\documentclass[12pt, a4paper]{article}

\usepackage{graphicx}
\usepackage{bmpsize}
\usepackage{amsmath}
\usepackage{amsfonts}
\usepackage{amssymb}
\usepackage{latexsym}
\usepackage{graphicx}
\usepackage{color}
\usepackage{mathrsfs}
\usepackage{slashed}
\usepackage{soul}
\usepackage{array}
\usepackage{afterpage}

\usepackage[draft,totoc]{listlbls}

\usepackage{placeins}
\usepackage{hyperref}
  
\usepackage{enumerate} 
\usepackage{multirow}
\usepackage[table]{xcolor} 
\usepackage{colortbl} 
\usepackage{tikz-feynman} 
\tikzfeynmanset{compat=1.0.0}

\usepackage{subcaption}

\usepackage{float}
\usepackage{braket}
\usepackage{xfrac}
\usepackage
[
version=4,
math-greek=default,
text-greek=default,
]{mhchem
}
\usepackage{booktabs}  

\usepackage[text={7in,10in}]{geometry}

\usepackage{comment}
\includecomment{toexclude} 
\excludecomment{addsections} 

\usepackage{pdflscape}

\usepackage[
  output-decimal-marker=.,
  separate-uncertainty=true,
]{siunitx}
\DeclareSIUnit\parsec{pc}
\DeclareSIUnit\erg{erg}

\usepackage{color}

\usepackage{soul}
 
\usepackage{slashed}
\usepackage{xfrac}

\usepackage{cite}

\usepackage{titlesec}
\titleformat{\paragraph} 
{\normalfont\normalsize\bfseries}{\theparagraph}{1em}{}
\titlespacing*{\paragraph}
{0pt}{3.25ex plus 1ex minus .2ex}{1.5ex plus .2ex}
 
\numberwithin{equation}{section}
\numberwithin{figure}{section}

\let\latexchii\chi
\makeatletter
\renewcommand\chi{\@ifnextchar_\sub@chi\latexchii}
\newcommand{\sub@chi}[2]{
  \@ifnextchar^{\subsup@chi{#2}}{\latexchii^{}_{#2}}%
}
\newcommand{\subsup@chi}[3]{
  \latexchii_{#1}^{#3}%
}
\makeatother

\newcommand{\latexchi}{T}

\newcommand{\e}{\mathrm e}
\newcommand{\lag}{\mathcal L}
\newcommand{\hlf}{\frac{1}{2}}
\newcommand{\hc}{\mathrm{h.c.}}
\newcommand{\tr}[1]{\mathrm{Tr}\left[#1\right]}
\newcommand{\nr}{N_R}
\newcommand{\chir}{T_R}
\newcommand{\sqt}{\frac{1}{\sqrt 2}}
\newcommand{\sqtf}[1]{\frac{#1}{\sqrt 2}}
\newcommand{\sqtsf}[1]{\sfrac{#1}{\sqrt 2}}
\newcommand{\cc}[1]{\left(#1\right)^c}
\newcommand{\zn}{z_N}
\newcommand{\ddd}[1]{\frac{\mathrm d {#1}}{\mathrm d \zn}}
\newcommand{\dd}[1]{\frac{\mathrm d {#1}}{\mathrm d \zn}}
\newcommand{\dy}[1]{\frac{\mathrm d Y_{#1}}{\mathrm d \zn}}
\newcommand{\dyz}[1]{\frac{\mathrm d Y_{#1}}{\mathrm d z}}

\newcommand{\dyvl}{\frac{\mathrm d Y_N^{VL}}{\mathrm d \zn}}

\newcommand{\yeq}[1]{Y_{#1}^{eq}}

\newcommand{\br}[1]{\mathrm {Br}_{#1}}

\newcommand{\yneq}{Y_N^{eq}}

\newcommand{\ynthA}{Y_{N_{\hat A}}^t}
\newcommand{\ynhA}{Y_{N_{\hat A}}}

\newcommand{\ynzB}{Y_{N_{B}}^z}

\newcommand{\ynVL}{Y_{N_{VL}}}
\newcommand{\yntVL}{Y_{N_{VL}}^t}
\newcommand{\ynzVL}{Y_{N_{VL}}^z}
\newcommand{\yntzVL}{Y_{N_{VL}}^{t,z}}

\newcommand{\besselk}[2]{\mathrm{K}_{#1}\!\left(#2\right)}
\newcommand{\mn}{M_{N}}
\newcommand{\gn}{g_{N}}
\newcommand{\mni}[1]{M_{N_{#1}}}
\newcommand{\gni}[1]{g_{N_{#1}}}

\newcommand{\mt}{M_\latexchi}
\newcommand{\meta}{M_\latexchi}
\newcommand{\gat}[1]{\gamma_{#1}}
\newcommand{\ms}{m_\sigma}
\newcommand{\ls}{\lambda_\sigma}
\newcommand{\mj}{m_J}
\newcommand{\geta}{g_{\latexchi}}

\newcommand{\be}[1]{\left[#1\right]}
\newcommand{\mtl}{\tilde m}
\newcommand{\mta}{\tilde m_S}
\newcommand{\mtaA}{\tilde m^{S}_{\hat A}}
\newcommand{\mtaB}{\tilde m^{S}_{\hat B}}

\newcommand{\norm}{Y_{N}^{0}}
\newcommand{\fnorm}{\frac{1}{\norm}}
\newcommand{\dall}{\delta_{\latexchi, \sigma, J}}

\newcommand{\gammaa}{\gamma_S}

\newcommand{\za}{z_S}
\newcommand{\zaA}{z_S^{\hat A}}
\newcommand{\zaB}{z_S^{\hat B}}
\newcommand{\zi}{z_I}
\newcommand{\zaeq}{z_{S}^{eq}}

\newcommand{\zaAeq}{z^{eq, \hat A}_{S}}
\newcommand{\zaBeq}{z^{eq, \hat B}_{S}} 

\newcommand{\zrho}{z_\rho}
\newcommand{\zrhoA}{z_\rho^A}
\newcommand{\zrhoB}{z_\rho^B}

\newcommand{\zalleq}{z^{(N,\sigma,J)}_{eq}}
\newcommand{\zieq}{z^{i}_{eq}}
\newcommand{\zneq}{z^{N}_{eq}}
\newcommand{\zvleq}{z^{VL}_{eq}}
\newcommand{\zseq}{z^{\sigma}_{eq}}
\newcommand{\zjeq}{z^{J}_{eq}}

\newcommand{\mtmaxap}{\hat{\tilde m}_{max}}
\newcommand{\mtmaxapA}{{\tilde m}^{\hat A}_{max}}
\newcommand{\mtmaxapB}{{\tilde m}^{\hat B}_{max}}
\newcommand{\mtmaxapAB}{{\tilde m}^{\hat A,\hat B}_{max}}
\newcommand{\mtmax}{{\tilde m}_{max}}
\newcommand{\mtmaxA}{{\tilde m}^{A}_{max}}
\newcommand{\mtmaxB}{{\tilde m}^{B}_{max}}
\newcommand{\mtmaxBt}{{\tilde m}^{B,t}_{max}}

\newcommand{\etaVL}{\eta_{VL}}
\newcommand{\etaA}{\eta_A}
\newcommand{\etaB}{\eta_B}
\newcommand{\etahA}{\eta_{\hat A}}
\newcommand{\etahB}{\eta_{\hat B}}
\newcommand{\etahAB}{\eta_{\hat A, \hat B}}
\newcommand{\etaAB}{\eta_{A, B}}

\newcommand{\etaVLt}{\eta_{VL}^t}
\newcommand{\etaVLz}{\eta_{VL}^z}
\newcommand{\etaVLtz}{\eta_{VL}^{t,z}}
\newcommand{\etaAt}{\eta_{A}^t}
\newcommand{\etaBt}{\eta_{B}^t}
\newcommand{\etaAz}{\eta_{A}^z}
\newcommand{\etaBz}{\eta_{B}^z}
\newcommand{\etahAt}{\eta_{{\hat A}}^t}
\newcommand{\etahAtz}{\eta_{{\hat A}}^{t,z}}
\newcommand{\etaAtz}{\eta_{{A}}^{t,z}}

\newcommand{\etahAz}{\eta_{{\hat A}}^z}

\newcommand{\etaVLzabs}{\left|\eta_{VL}^z\right|}

\newcommand{\etaBzabs}{\left|\eta_{B}^z\right|}

\newcommand{\gic}{\gn^{IA}}
\newcommand{\mic}{\mtl^{IA}}

\newcommand{\etaBzm}{\eta^-_B}
\newcommand{\etaBzp}{\eta^+_B}

\newcommand{\lasym}{{\lag}}
\newcommand{\basym}{{{\mathcal B}}}
\newcommand{\lnumber}{{L^\prime}}
\newcommand{\bnumber}{{B^\prime}}

\newcommand{\gaugeww}{{\latexchi, gauge}}

\begin{document}

\begin{center}
{\Huge
Leptogenesis in a Majoron~+~Triplet model
}
\\ [2.5cm]
{\large{\textsc{ 
Tim Brune\footnote{\textsl{tim.brune@tu-dortmund.de}}
}}}
\\[1cm]

\large{\textit{
Fakult\"at f\"ur Physik, Technische Universit\"at Dortmund,\\
44221 Dortmund, Germany
}}
\\ [2 cm]
{ \large{\textrm{
Abstract 
}}}
\\ [1.5cm]
\end{center}
\normalsize
We discuss leptogenesis in a majoron model extended by a right-handed $SU(2)_L$ triplet fermion. We study several different parameter assignments and find that the interactions of neutrinos with the new particles in the majoron+triplet model can significantly alter the way leptogenesis proceeds. We show that for large parts of the considered parameter space, it is essential to solve the set of coupled Boltzmann equations for the evolution of the neutrinos and the additional particles rather than solving the Boltzmann equations for the neutrino evolution only.

\def\thefootnote{\arabic{footnote}}
\setcounter{footnote}{0}
\pagestyle{empty}

\newpage
\pagestyle{plain}
\setcounter{page}{1}

\section{Introduction}  
The Standard Model (SM) of particle physics has been remarkably successfull on many levels. Despite its predictive power however, it fails to explain the origin of the Baryon Asymmetry of the Universe (BAU)\cite{Planck:2018vyg} and how the light neutrino masses arise \cite{KamLAND:2002uet,SNO:2002tuh,Super-Kamiokande:1998kpq}. \\
The type I Seesaw mechanism \cite{Minkowski:1977sc} is an elegant way of generating small neutrino masses. Extending the SM by heavy right-handed singlet neutrinos, the Seesaw mechanism naturally explains the smallness of the light neutrino masses as the consequence of a suppression by the mass scale of the heavy neutrinos. An inherent feature of the Seesaw mechanism is that it inhibits all ingredients necessary to explain the BAU via leptogenesis\cite{Fukugita:1986hr}. In the most simple version of the leptogenesis mechanism, often called vanilla leptogenesis (VL), CP violating out-of-thermal-equilibrium decays of a heavy neutrino generate a lepton asymmetry which is subsequently converted to a baryon asymmetry via Sphaleron transitions\cite{Khlebnikov:1988sr}. A compelling extension of the Seesaw mechanism is the singlet majoron model\cite{Chikashige:1980ui,Schechter:1981cv,Georgi:1981pg} where rather than including the masses of the Majorana neutrinos ad-hoc, their origin is based on the spontaneous symmetry breaking (SSB) of a global $U(1)_\lnumber$ symmetry where the index $\lnumber$ denotes lepton number. This SSB gives rise to a Goldstone boson, called majoron, that has been shown to be a viable dark matter (DM) candidate if it obtains a small mass\cite{Brune:2018sab,Frigerio:2011in,Hall:2009bx,Rothstein:1992rh,Berezinsky:1993fm}. 
As in the type I seesaw mechanism, a lepton asymmetry can be generated via neutrino decays \cite{Pilaftsis:2008qt,Gu:2009hn,AristizabalSierra:2014uzi}.\\
In this paper, we focus on the extension of the majoron model by a right-handed triplet fermion. This extension of the majoron model is originally motivated by the aim to prevent the appearance of cosmological domain walls as discussed in \cite{Lazarides:2018aev}. However, we note that in the absence of additional explicit baryon number violation on the Lagrangian level, $SU(2)_L$ instantons do not generate a Majoron potential \cite{Heeck:2019guh,FileviezPerez:2014xju, Anselm:1992yz, Anselm:1993uj, Csaki:2023ziz} and hence no domain walls appear. Nevertheless, the presence of the right-handed triplet fermion has interesting consequences on the dynamics of leptogenesis and additionally, it changes the $[SU(2)_L]^2\times U(1)_\lnumber$ anomaly factor associated with the Instanton transitions and therefore affects the conversion rate with which a lepton asymmetry is converted to a baryon asymmetry via Sphaleron transitions \cite{Brune:2022vzd}.
For simplicity, we restrict ourselves to a model where an additional $Z_2$ symmetry forbids a Yukawa coupling of the triplet fermion with the lepton doublet and the Higgs. We will refer to this setup as the "majoron+triplet model". In contrast to VL, neutrino interactions in the majoron+triplet model with the new particles introduce several additional parameters which can severly affect the dynamics that create the lepton asymmetry. Moreover, we find that the creation of the lepton asymmetry is independent from the presence of the triplet if the corresponding Yukawa coupling is sufficiently small.\\
We analyze how the parameters relevant in the majoron+triplet model affect the evolution of the lepton asymmetry compared to VL.
We find that the lepton asymmetry that can be generated depends significantly on the Majorana neutrino Yukawa coupling and the effective neutrino mass while the initial abundances of the new particles are only relevant for a specific subcategory of parameters. Depending on the parameters, a sizable lepton asymmetry can be generated while in other cases, the asymmetry is significantly diminished. \\
This paper is structured as follows. In Sec. \ref{sec:model}, we give an introduction to the basic ingredients of the majoron+triplet model. In Sec. \ref{sec:general} , we give an overview on the leptogenesis mechanism both in VL and in the majoron+triplet model, present the Boltzmann equations for the evolution of the particle abundances and the lepton asymmetry and introduce the parameters that we used to solve these Boltzmann equations. 
In Sec. \ref{sec:results}, we present our results for the efficiency factor that is relevant for leptogenesis and discuss the solutions of various scenarios in greater detail. 
In Sec. \ref{sec:dm}, we discuss dark matter constraints on the parameters of the model. 
In Sec. \ref{sec:basym}, we focus on a more specific realization of the model and discuss the lepton- and baryon asymmetries that can be realized in the majoron+triplet model.
In Sec. \ref{sec:summary}, we conclude with a summary. In App. \ref{sec:appendix}, we present useful formulae. 

\section{Model}
\label{sec:model}
In the majoron+triplet model, the SM is extended by a singlet complex scalar $\hat\sigma$, three right handed neutrinos $N_R^i$, and a fermion triplet $\chir$, 
\begin{align}
 \chir &= \begin{pmatrix} \sqtsf{\chir^0} && \chir^+ \\ \chir^- && -\sqtsf{\chir^0} \end{pmatrix}\,,
\end{align}
transforming under $(SU(3)_C\times SU(2)_L \times U(1)_Y)_\lnumber$ as 
\begin{align}
  \hat\sigma \sim (1,1,0)_{-2}\,, \qquad N_R^{1,2,3} \sim (1,1,0)_1\,,  \qquad \chir \sim (1,3,0)_1\,,
\end{align}
where the index $\lnumber$ denotes lepton number. 
The $U(1)_\lnumber$ invariant scalar potential is given by
\begin{align}
  V = -\mu_\sigma^2|\hat\sigma|^2 - \lambda_\sigma |\hat\sigma|^4-\mu_H^2|H|^2 - \lambda_H |H|^4 + 2\lambda_{mix}|\hat\sigma|^2|H|^2\,, \label{eq:potential}
\end{align}
where $H = \begin{pmatrix} \phi^+ \\ \phi^0 \end{pmatrix} $ is the Higgs doublet. Moreover, the relevant Yukawa and kinetic terms are given by 
\begin{align}
  \lag_{Yuk} &= -y_\nu \overline{L} \tilde H \nr - \hlf \gni{i} \overline{(\nr^i)^c}\nr^i\hat\sigma -\hlf g_\latexchi \tr{\overline{(\chir)^c}\chir \hat\sigma} + \hc\,,\label{lag:yukawa}\\
  \lag_{kin} &= i \overline{\nr}\slashed\partial \nr+ i \tr{\overline{\chir} \slashed D \chir}\,, \label{lag:kin}
\end{align} 
where $L$ are the lepton doublets and $\tilde H = i\sigma_2 H^*$.
In the Lagrangian above, we invoked an additional $Z_2$ symmetry under which $\chir$ is odd while all other particles are even.
This assumption simplifies the discussion for two main reasons: First, it prevents the triplet from mixing with the SM leptons via the Yukawa term $\overline{L}\tilde H \chir$. Second, as we discuss later, the absence of this term ensures that the only lepton number violating interactions in the Lagrangian are neutrino decays, thereby simplifing the Boltzmann equations.\\
At the Seesaw scale $f$, the $U(1)_\lnumber$ symmetry is broken as $\hat\sigma$ obtains its VEV with an expansion around its ground state given by  
\begin{align}
 \hat\sigma &= \sqt(f + \sigma + i J)\,,
\end{align}
where $J$ denotes the CP-odd majoron and $\sigma$ is CP-even. As $J$ is a Goldstone boson, it is massless. However, the majoron may obtain a small mass due to e.g.\ radiative corrections or gravitational Instanton processes and thereby become a dark matter candidate\cite{Brune:2018sab,Frigerio:2011in,Hall:2009bx,Rothstein:1992rh}. \\
For simplicity, we neglect $\lambda_{mix}$ so that the potentials for $\hat\sigma$ and $H$ decouple and we obtain 
\begin{align}
  V_\sigma &= \frac{m_\sigma^2}{2} \sigma^2 + k_\sigma J^2\sigma + k_\sigma \sigma^3 + \lambda_\sigma J^2\sigma^2 + \frac{\lambda_\sigma}{4}J^4+ \frac{\lambda_\sigma}{4}\sigma^4 + \mathrm{const}\, \label{eq:broken:potential}
\end{align}
for the singlet potential with 
\begin{align}
  f &= \frac{\mu_\sigma}{\sqrt{\lambda_\sigma}}\,,\qquad k_\sigma = \lambda_\sigma f\,, \qquad \ms = \sqrt{\lambda_\sigma} f\,.
\end{align} 
After SSB at the Seesaw scale, we define the heavy Majorana neutrinos as 
\begin{align}
  N_i \equiv N_R^i + \left(N_R^i\right)^c = \left(N_i\right)^c\,\qquad i=1,2,3\,,
\end{align}
with the corresponding Majorana masses given by $\mni{i} \equiv \frac{\gni{i}}{\sqrt 2}f$. 
Similarly, we write the triplet as 
\begin{align}
  \latexchi \equiv  \begin{pmatrix} \sqtsf{\latexchi^0} && \latexchi^+ \\ \latexchi^- && -\sqtsf{\latexchi^0} \end{pmatrix} = \chir + \cc{\chir}  = \latexchi^c
\end{align}
with  $M_\latexchi = \sqtf{g_\latexchi}f$ so that $\latexchi^0 = \chir^0 + \cc{\chir^0}$ is a Majorana fermion while the charged components are combined into a Dirac spinor as  
\begin{align}
  \latexchi^+ = \latexchi_R^+ + (\latexchi_R^-)^c = (\latexchi^-)^c\,.
\end{align}
Consequently, we can write the Yukawa terms \eqref{lag:yukawa} as
\begin{align}
\begin{split}
  \lag_{Yuk} &= \left(-y_\nu \overline{L} \tilde H N  + \hc \right)  -\hlf M_{N} \overline N N - \frac{1}{2\sqrt 2} g_N \overline{N} N \sigma \\&- \frac{1}{2\sqrt 2} g_N \overline{N}\gamma_5 N J - \hlf M_\latexchi \tr{\overline \latexchi \latexchi} - \frac{1}{2\sqrt 2} g_\latexchi \tr{\overline{\latexchi}\latexchi \sigma} - \frac{1}{2\sqrt 2} g_\latexchi \tr{\overline{\latexchi}\gamma_5\latexchi J} \,,  \label{lag:yukawa:broken}
\end{split}
\end{align}
while the kinetic terms become
\begin{align}
\begin{split}
  \lag_{kin} &=  \frac{i}{2} \overline{N}\slashed\partial N+ \frac{i}{2}  \tr{\overline{\latexchi} \slashed D \latexchi} =  \frac{i}{2} \overline{N}\slashed\partial N+  \frac{g}{2}\left(  \overline{\latexchi^0} W^+ \latexchi^- + \overline{\latexchi^0} W^- \latexchi^+ \right) \\&+ e \overline{\latexchi^+} A \latexchi^+ + \frac{g}{c_W} (1-s_W^2)\overline{\latexchi^+} Z \latexchi^+\,,
\end{split}
\end{align}
where we omitted generation indices for brevity. 
After electroweak symmetry breaking (EWSB), we have
\begin{align} 
  H =  \sqt \begin{pmatrix} 0 \\ v+h^0 \end{pmatrix}  \,,\\
\end{align}
where $v = \SI{246}{\giga\electronvolt}$ is the SM VEV and $h^0$ is the Higgs boson. Moreover, the neutrinos obtain a Dirac mass $m_D = \sqtf{y_\nu}v$ and we can write the neutrino mass terms as 
\begin{align}
 \lag_{mass}^\nu &= -\hlf \begin{pmatrix} \overline{\nu_L} && \overline{\cc{\nr}} \end{pmatrix} \underbrace{\begin{pmatrix} 0 && m_D \\ m_D^T && M_N\end{pmatrix}}_{\mathcal M} \begin{pmatrix} \cc{\nu_L} \\ \nr  \end{pmatrix} + \hc\,.
\end{align}
The Majorana mass matrix $\mathcal M$ can be diagonalized with a unitary matrix $U$ as 
\begin{align}
  U^T\mathcal{M}U = \mathrm{diag}(m_1, ..., m_6)\,,
\end{align}
where $m_1, ..., m_6$ are the physical neutrino masses. In the Seesaw limit $M_N \gg m_D$, block-diagonalization of $\mathcal {M}$ yields three light neutrinos with masses  of  order $-\frac{m_D m_D^T}{M_N}$ and three heavy neutrinos with masses of order $M_N$. Without loss of generality, we can assume that $M_N$ is diagonal and define the heavy neutrino masses as $m_{4,5,6} = \mni{1,2,3}$ while $m_{1,2,3}$ are the light neutrino masses. \\
In the following, we will assume that $\mni{1} \ll \mni{2,3}$ so that $N_1$ is the lightest of the heavy neutrino mass eigenstates. For brevity, we will generally omit the index $1$ and stress that unless otherwise specified, $\gn, \mn$ refer to the Yukawa coupling and the mass of $N \equiv N_1$. 

\section{Leptogenesis}
In this section, we give a brief overview of leptogenesis. For an overview of the basic formulae and conventions, we refer the reader to App \ref{sec:interactions}. The relevant Feynman diagrams are shown in App. \ref{sec:app:feynman}, the relevant cross sections are given in App. \ref{sec:app:rates} and a more detailed derivation of the Boltzmann equations in the majoron+triplet model is presented in App. \ref{sec:app_boltzmann}.
\label{sec:general} 
\subsection{Overview}
In the VL scenario, the first term given in \eqref{lag:yukawa:broken} is already sufficient to produce a lepton asymmetry via CP violating out-of-equilibrium neutrino decays $N\to LH, \overline L \overline H$ with a decay rate given by
\begin{align}
  \Gamma_D &= \frac{\left(y_\nu^\dagger y_\nu\right)_{11} \mn}{8 \pi}= \frac{\mn^2}{8 \pi v^2}\mtl\,,
\end{align}
where $\mtl = \left(y_\nu^\dagger y_\nu\right)_{11} v^2\mn^{-1}$ is the effective neutrino mass \cite{Plumacher:1996kc}. 
In our computations, we also include scattering processes involving third generation quarks, induced via the top yukawa coupling $y_t$, $Q_3 U_3 \leftrightarrow NL$ and $L \overline{Q_3} \leftrightarrow N \overline{U_3}$, while neglecting gauge boson contributions. 
The corresponding feynman diagrams can be found in Fig. \ref{fig:feynman:vl}.
In the one flavor approximation, the Boltzmann equation for the neutrino abundance can then be written as \cite{Kolb:1979qa}
\begin{align}
  s H \zn  \dyvl &= -\left( \delta_N-1 \right)\gamma_D -2\left( \delta_N-1 \right)\gamma_Q \,, \label{eq:dyn_vl} 
\end{align}
where the abundance $Y_i$ is defined as the number density $n_i$ normalized to the entropy density $s$ \eqref{eq:entropy} so that $ Y_i = \sfrac{n_i}{s}$ and $ \delta_i \equiv \sfrac{ Y_i}{Y_i^{eq}}$ where $Y_i^{eq}$ is the equilibrium abundance. Moreover, we defined $\zn \equiv \frac{\mn}{T}$\footnote{Here, $T$ is the temperature. } while $\gamma_D$ and $\gamma_Q\equiv \gat{Q_3U_3 NL} + \gat{L\overline{Q_3} N\overline{U_3}}+ \gat{L\overline{U_3} N\overline{Q_3}}$ are the thermal rates for neutrino decays and quark scatterings, respectively. \\
The CP violation in VL is a result of the interference of the tree- and one-loop-level diagrams of $N\to LH, \overline L \overline H$ presented in Fig. \ref{fig:cp:diagrams} and in the case of strongly hierarchical heavy neutrinos with $\mni{2,3} \gg \mni{1}$, it is given by \cite{Covi:1996wh}
\begin{align}
  \varepsilon &= \frac{\Gamma\left(N \to L H \right)-\Gamma\left(N \to \bar{L} \bar{H} \right)}{\Gamma\left(N \to L H \right)+\Gamma\left(N \to \bar{L} \bar{H} \right)} \eqsim \frac{1}{8\pi} \frac{1}{\left(y_\nu^\dagger y_\nu\right)_{11}}\sum_{i = 2,3} \mathrm{Im}\left[\left(y_\nu^\dagger y_\nu\right)_{1i}^2\right]f\left( \frac{\mni{i}^2}{\mni{1}^2} \right)\,, 
\end{align}
where
\begin{align}
  f(x) &= \sqrt x \left[ \frac{x-2}{x-1}-(1+x) \log\left(\frac{1+x}{x}\right) \right]\,.
\end{align}
Using the Davidson-Ibarra bound \cite{Davidson:2002qv} and assuming $\sfrac{\mni{2,3}}{\mni{1}} \to\infty$, an upper limit on the CP violation can be given as 
\begin{align}
  |\varepsilon^{DI}| &\leq \frac{3}{16\pi} \frac{\mn }{v^2} (m_3-m_1)\,.\label{eq:di}
\end{align}
If the light neutrinos follow a normal hierarchy with $m_1 \ll m_2 \ll m_3$, \eqref{eq:di} can be written as 
\begin{align}
  |\varepsilon^{DI}| &\leq \frac{3}{16\pi} \frac{\mn}{v^2} \sqrt{\Delta m_{atm}^2}\,,
\end{align}
where $\Delta m_{atm}^2$ is the atmospheric neutrino mass splitting with \cite{Salas}
\begin{align}
  \Delta m_{atm}^2 &= m_3^2 - m_2^2 \approx \SI{2.55e-3}{\electronvolt\squared}\,.
\end{align}
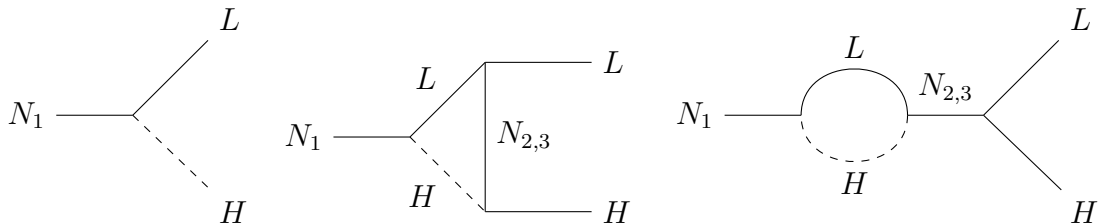
\begin{figure}[H]
  \centering
  \begin{subfigure}{0.2\textwidth}
  \centering
  \begin{tikzpicture}
  \begin{feynman}
  \vertex (a1) {\(N_1\)};
  \vertex[right=1.4cm of a1] (a2);
  \vertex[above right=1.4cm of a2] (b1){\(L\)};
  \vertex[below right=1.4cm of a2](b2){\(H\)};
  \diagram*{
    (a1) --(a2) , 
    (a2) -- (b1),
     (a2) -- [scalar](b2),
  };
  \end{feynman}
  \end{tikzpicture}
  \end{subfigure}
  \begin{subfigure}{0.31\textwidth}
  \centering
  \begin{tikzpicture}
  \begin{feynman}
  \vertex (a1) {\(N_1\)};
  \vertex[right=1.4cm of a1] (a2);
  \vertex[above right=1.4cm of a2] (b1);
  \vertex[below right=1.4cm of a2](b2);
  \vertex[right=1.4cm of b1] (c1){\(L\)};
  \vertex[right=1.4cm of b2](c2){\(H\)};
  \diagram*{
    (a1) --(a2) , 
    (a2) --[edge label=\(L\)] (b1),
     (a2) --[scalar,edge label'=\(H\)] (b2),
     (b1) -- (c1),
     (b2) -- (c2),
     (b1) -- [edge label=\(N_{2,3}\)](b2),
  };
  \end{feynman}
  \end{tikzpicture}
  \end{subfigure}
  \begin{subfigure}{0.35\textwidth}
  \centering
  \begin{tikzpicture}
  \begin{feynman}
 \vertex (a1) {\(N_1\)};
  \vertex[right=1.4cm of a1] (a2);
  \vertex[right=1.4cm of a2] (b1);
  \vertex[right=1cm of b1] (b2);
  \vertex[above right=1.4cm of b2] (c1){\(L\)};
  \vertex[below right=1.4cm of b2](c2){\(H\)};
  \diagram*{
    (a1) --(a2) , 
    (a2) -- [half left, edge label=\(L\)](b1) -- [half left,scalar, edge label=\(H\)](a2),
     (b1) -- [edge label=\(N_{2,3}\)](b2),
     (b2) -- (c1),
     (b2) -- (c2),
  };
  \end{feynman}
  \end{tikzpicture}
  \end{subfigure}
  \caption{Feynman diagrams contributing to the CP violating $N$ decay.}
  \label{fig:cp:diagrams}
\end{figure} 
\noindent
It is convenient to express the lepton asymmetry $Y_\lasym \equiv Y_\lnumber-Y_{\overline{\lnumber}}$ in terms of an efficiency factor $\eta$ \cite{Barbieri:1999ma}, 
\begin{align}
   Y_\lasym(\zn) = \varepsilon  Y_N^0  \eta(\zn)\,, \label{eq:lasymmetry}
\end{align}
where $Y_N^0 \equiv Y_N(\zn \to \infty)$ is a normalization factor, yielding a Boltzmann equation for the efficiency independent from $\varepsilon$,
\begin{align}
    s H \zn\dd{\eta} &= \fnorm\left(\delta_N-1 \right)\gamma_D - \frac{\eta}{\yeq{L}}\left( \frac{\gamma_D}{2} + \delta_N\gat{Q_3U_3 NL} + 2\gat{L\overline{Q_3} N\overline{U_3}}\right)\label{eq:eta}\,.
\end{align}
The final lepton asymmetry is given by $Y_\lasym(\zn \to\infty) = \varepsilon  Y_N^0  \eta$ where we defined the final efficiency as $\eta\equiv \eta(\zn\to\infty)$. 
In \eqref{eq:eta}, the first term is the source term that produces the lepton asymmetry while the term proportional to $\eta$ is called the washout (WO) term which reduces the efficiency via inverse decays and quark scatterings. \\
Compared to VL, the interactions given in \eqref{lag:yukawa}, \eqref{lag:kin} and \eqref{eq:broken:potential} induce a vast number of new interactions between the additional particles as shown in Figs. \ref{fig:feynman:N}  and \ref{fig:feynman:diagrams:rest}.
While these interactions affect the neutrino evolution, the only lepton number violating processes are still the (inverse) neutrino decays and quark scatterings and consequently, the corresponding Boltzmann equation for the efficiency is the same as in VL. We also assume that no additional CP violation compared to VL is present in the majoron+triplet model. \\
Clearly, the lepton asymmetry is directly linked to the neutrino evolution and consequently, the most relevant processes are these that change the neutrino abundance (see Fig. \ref{fig:feynman:N}). It is therefore convenient to combine these processes to a summed thermal scattering rate $\gamma_S$ as 
\begin{align}
	\gamma_S \equiv&  \gat{NN \latexchi\latexchi}+ \gat{NN JJ}+ \gat{NN \sigma\sigma}+ \gat{NN  \sigma J} + \br{\sigma, NN}\gat{\sigma,NN}\,, \label{eq:gammaA} 
\end{align} 
where we introduced the short-hand notations
\begin{align}
\gamma(12\leftrightarrow34)=:\gat{1234}\,,\qquad
\gamma(1\leftrightarrow23)=:\gat{1,23}\,,
\end{align}
for the thermal rates $\gamma$ and $\br{\sigma, NN} \equiv \frac{\Gamma_{\sigma \to NN}}{\sum_{i = N, \latexchi, J} \Gamma_{\sigma \to ii}}$ is the branching ratio of $\sigma \to NN$. 
In contrast to the VL scenario, we expect that a large scattering rate $\gamma_S$ effectively thermalizes the neutrinos, thus allowing the generation of a lepton asymmetry only once the scattering processes decouple. This effect is similar to type III leptogenesis \cite{AristizabalSierra:2010mv,Hambye:2012fh,Hambye:2003rt,Strumia:2008cf,Zhuridov:2012hb,AristizabalSierra:2012js} where the lepton asymmetry is produced via the decays of a $SU(2)_L$ triplet fermion.
An appealing feature of type III leptogenesis is that gauge interactions thermalize the triplet, rendering leptogenesis independent from the triplets initial abundance while in type I leptogenesis, the asymmetry is independent from the initial neutrino abundance only in the so called strong washout regime where $\mtl \gtrsim \SI{e-3}{\electronvolt}$. 
In contrast to type III leptogenesis though, the triplet in the model discussed here does not directly contribute to the lepton asymmetry due to the additional $Z_2$ symmetry and further, we can treat the couplings $\gn, \geta, \ls$ as free parameters and explore the effects of changes in these couplings on the generation of the asymmetry. For example, if $\gamma_S$ is small, neutrinos might not be thermalized as efficiently as in type III leptogenesis. One complicating factor when calculating the lepton asymmetry however are the abundances of $\latexchi, \sigma$ and $J$. Their evolution depends on the same couplings as the evolution of the neutrinos which suggests that a coupled set of Boltzmann equations for $\latexchi, \sigma, J$ and $N$ needs to be solved. However, in order to explore qualitative implications, it is convenient to begin with the discussion of a simplified scenario where we assume that $\latexchi, \sigma$ and $J$ are in thermal equilibrium, i.e.\ $\dall = 1$, and consequently, the Boltzmann equation for $N$ is independent from their abundance. \\
In the next section, we present the full set of Boltzmann equations relevant for the evolution of $N, \latexchi, \sigma$ and $J$ and the Boltzmann equation for $Y_N$ in the simplified scenario where $\dall = 1$. Afterwards, we specify the parameters we used to solve the Boltzmann equations.

\subsection{Boltzmann Equations}
The full set of coupled Boltzmann equations for the evolution of $Y_N, Y_\latexchi, Y_\sigma$ and $Y_J$ is given by 
\begin{align}
\begin{split}
  s H \zn  \dy{N} &= \dyvl  -2\left(\delta_N^2 - \delta_\latexchi^2\right) \gat{NN \latexchi\latexchi}-2\left(\delta_N^2 - \delta_J^2\right) \gat{NN JJ}-2\left(\delta_N^2 -\delta_\sigma^2\right) \gat{NN \sigma\sigma}\\&-2\left(\delta_N^2 - \delta_J \delta_\sigma\right) \gat{NN  \sigma J}+  2 \delta_{sub}\gat{\sigma,NN}  \\
    &\equiv -\left( \delta_N-1 \right)\gamma_D -2\left( \delta_N-1 \right)\gamma_Q - 2(\delta_N^2-1)\gamma_S - 2\rho \label{eq:ben} 
  \end{split} \\
  \begin{split}   
  s H \zn  \dy{\latexchi} &= -2 \left(\delta_\latexchi^2  - \delta_N^2\right) \gat{\latexchi\latexchi  NN}
  -2 \left(\delta_\latexchi^2  - \delta_J^2\right) \gat{\latexchi\latexchi  JJ}-2\left(\delta_\latexchi^2  -\delta_\sigma^2\right) \gat{\latexchi\latexchi  \sigma\sigma}\\&-2 \left(\delta_\latexchi^2  - \delta_\sigma\delta_J\right) \gat{\latexchi\latexchi  \sigma J}-2 \left(\delta_\latexchi^2  -1\right) \gat{\gaugeww}+  2 \delta_{sub}\gat{\sigma , \latexchi\latexchi}\,,\label{be:chi}
  \end{split}\\  
  \begin{split}
      s H \zn\dy{\sigma} &= -2 \left(\delta_\sigma^2 - \delta_N^2\right) \gat{\sigma\sigma  NN}-2 \left(\delta_\sigma^2 - \delta_\latexchi^2 \right) \gat{\sigma\sigma  \latexchi\latexchi}-2 \left(\delta_\sigma^2 - \delta_J^2\right) \gat{\sigma\sigma  JJ}\\&-\left(\delta_\sigma \delta_J - \delta_\latexchi^2 \right) \gat{\sigma J  \latexchi\latexchi}-\left(\delta_\sigma \delta_J - \delta_N^2\right) \gat{\sigma J  NN} -\left(\delta_N \delta_\sigma- \delta_N \delta_J\right) \gat{N\sigma  N J}\\&-\left(\delta_\latexchi \delta_\sigma- \delta_\latexchi \delta_J\right) \gat{\latexchi\sigma  \latexchi J}- \left(\delta_\sigma- \delta_N^2\right) \gat{\sigma,  N N}- \left(\delta_\sigma -\delta_\latexchi^2 \right) \gat{\sigma,  \latexchi \latexchi}\\&- \left(\delta_\sigma- \delta_J^2\right) \gat{\sigma,  J J}\,,\label{be:s}
  \end{split}\\
  \begin{split}
      s H \zn\dy{J} &= -2 \left(\delta_J^2 - \delta_N^2\right) \gat{JJ  NN}-2 \left(\delta_J^2 -\delta_\sigma^2\right) \gat{JJ  \sigma\sigma}-2 \left(\delta_J^2 - \delta_\latexchi^2 \right) \gat{JJ  \latexchi\latexchi} \\&-\left(\delta_J \delta_\sigma - \delta_N^2\right) \gat{J\sigma  NN}-\left(\delta_J \delta_\sigma - \delta_\latexchi^2 \right) \gat{J\sigma  \latexchi\latexchi}- \left(\delta_J \delta_N - \delta_\sigma \delta_N\right) \gat{JN \sigma N}\\&- \left(\delta_J \delta_\latexchi - \delta_\sigma \delta_\latexchi\right) \gat{J\latexchi \sigma \latexchi} + 2\delta_{sub}\gat{\sigma, JJ}\,,\label{be:j}
  \end{split}
\end{align}
where we wrote the Boltzmann equation for $\latexchi$ as the sum of the three triplet components $n_\latexchi \equiv n_{\latexchi^0}+n_{\latexchi^+}+n_{\latexchi^-}$, the summed scattering rate $\gammaa$ is given by \eqref{eq:gammaA} and we defined 
\begin{align}
  \begin{split}
  \rho \equiv&\left( 1-\delta_J^2 \right)\left(\gat{NNJJ}- \br{\sigma, JJ}\gat{\sigma, NN}\right) + \left( 1-\delta_\latexchi^2 \right)\left(\gat{NN\latexchi\latexchi}- \br{\sigma, \latexchi\latexchi}\gat{\sigma, NN}\right)\\ &+ \left( 1-\delta_\sigma^2 \right)\gat{NN \sigma \sigma}+\left( 1-\delta_\sigma\delta_J \right)\gat{NN \sigma J} +\left(1- \delta_\sigma\right)\gat{\sigma,NN} \\
  =&\left( 1-\delta_J^2 \right)\gat{NNJJ} + \left( 1-\delta_\latexchi^2 \right)\gat{NN\latexchi\latexchi}+ \left( 1-\delta_\sigma^2 \right)\gat{NN \sigma \sigma}+\left( 1-\delta_\sigma\delta_J \right)\gat{NN \sigma J} \\&+\left[\left(1- \delta_\sigma\right) - \left( 1-\delta_J^2 \right)\br{\sigma,JJ}- \left( 1-\delta_\latexchi^2 \right)\br{\sigma,\latexchi\latexchi}\right]\gat{\sigma,NN} \,\label{eq:rho}
\end{split}
\end{align}
in order to isolate the terms in the Boltzmann equation for the neutrino evolution that explicitly contain $\dall$.\footnote{We stress however that $\dall$ and $\delta_N$ are clearly not independent.}
Moreover, 
\begin{align}
   \delta_{sub} &= \delta_\sigma-\delta_J^2\br{\sigma, JJ}-\delta_N^2\br{\sigma, NN}-\delta_\latexchi^2 \br{\sigma, \latexchi\latexchi} 
\end{align}
appears due to the substraction of on-shell scattering as explained in App. \ref{sec:app_boltzmann}
and $\gat{\gaugeww}$ is the reaction density of the triplet from interactions involving gauge bosons. \\
In the simplified scenario where $\dall = 1$, only the Boltzmann equation for the neutrinos remains which simplifies to 
\begin{align}
  s H \zn  \dy{N} &= -\left( \delta_N-1 \right)\gamma_D -2\left( \delta_N-1 \right)\gamma_Q - 2\left( \delta_N^2-1 \right)\gamma_S\,. \label{eq:dyn_ad} 
\end{align}
Note that we recover \eqref{eq:dyn_vl}for $\rho, \gamma_A \to 0$ in \eqref{eq:ben} while we have \eqref{eq:ben}$\to$\eqref{eq:dyn_ad} for $\rho \to 0$.   

\subsection{Parameters}
\begin{figure}[H]
  \centering
  \includegraphics[width=0.45\textwidth]{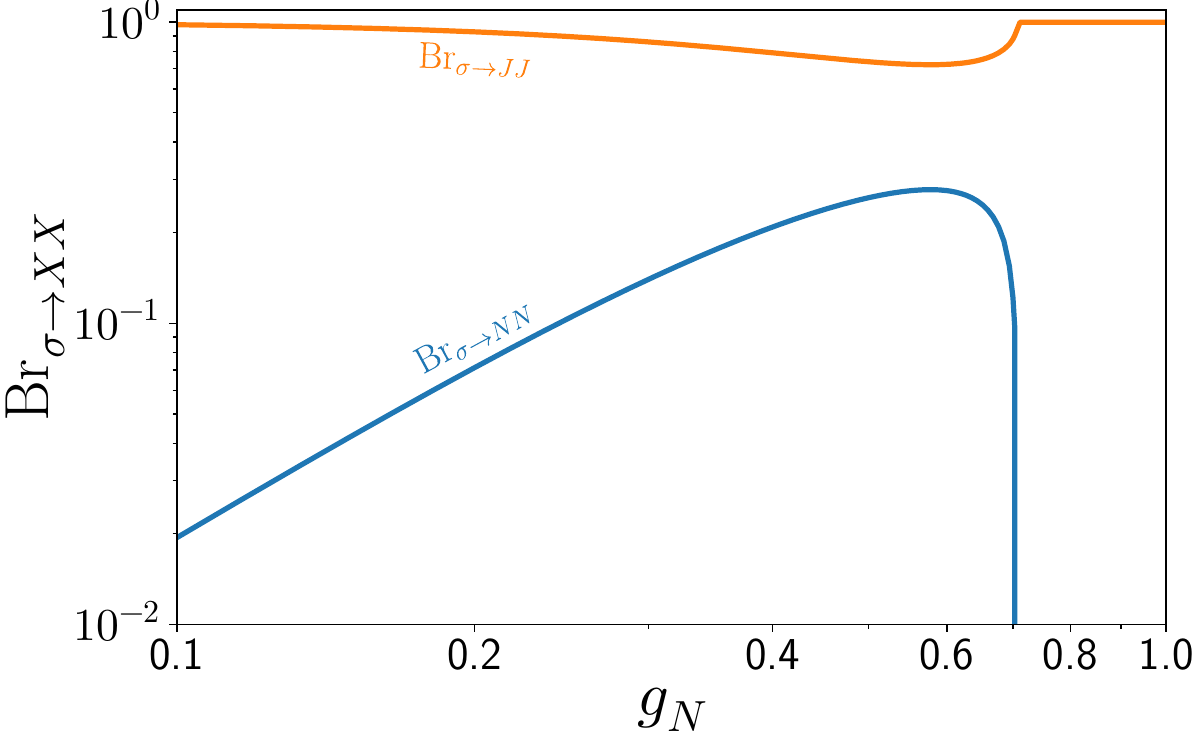}
  \caption{Branching ratios of $\sigma\to NN$ and $\sigma \to JJ$ as functions of $\gn$. We stress that in cases $A, \hat A$ the decay $\sigma \to \latexchi\latexchi$ is kinematically forbidden while in cases $B, \hat B$, the branching ratio $\br{\sigma \to \latexchi\latexchi}$ is of order $\num{e-13}$ and can therefore be neglected. As a result, the branching ratios $\br{\sigma\to JJ}$ and $\br{\sigma\to NN}$ are practically the same in all considered scenarios. Moreover, we note that $\sigma$-decays to a pair of majorons are clearly the dominant decay channel over the full $\gn$ range. Above $\gn=0.7$, $\sigma$ decays to a pair of neutrinos are kinematically forbidden, resulting in $\br{\sigma \to JJ} = 1$. }
  \label{fig:branchingPlot}
\end{figure}

Before discussing the solutions of the Boltzmann equations in detail, a short examination of the parameters appearing in the Boltzmann equations is in order. 
In VL, the thermal rates $\gamma_D, \gamma_Q$ govern the neutrino evolution and are mainly determined by the effective neutrino mass $\mtl$ while the thermal rates in the majoron+triplet model depend on four additional parameters: The VEV $f$ and the couplings $\gn,\geta,\ls$. Solving the Boltzmann equations introduces additional parameters in form of the initial conditions. 
We will therefore distinguish between various different cases, depending on the couplings, the initial conditions and whether the simplified scenario or the full set of Boltzmann equations is considered. 
As it is reasonable to assume that changes in $\geta$ and $\ls$ have comparably small effects on the asymmetry compared to changes in $\gn$, we vary $\gn$ in the range $\gn \in [0.1,1]$ in each case. We then distinguish between cases $A, \hat A$ where $\geta = \ls = 1$ and $B, \hat B$ where $\geta = \num{e-7}, \ls = 1$ where $A, B$ correspond to the case where the full set of Boltzmann equations is considered while $\hat A, \hat B$ correspond to the simplified scenario. 
In each case, we vary $\mtl$ in the range $\mtl \in [0.5\times10^{-4},0.1]\si{\electronvolt}$\footnote{In case of normal ordering, neutrino data suggest $m_1 \leq \mtl \lesssim m_3 \sim \SI{0.05}{\electronvolt}$, although a larger $\mtl$ is in principle possible if cancellations occur.\cite{Buchmuller:2003gz} } and fix the VEV at $f = \SI{e10}{\giga\electronvolt}$, neglecting any temperature dependence.\footnote{We discuss the effect of the VEV on the efficiency briefly in Sec. \ref{sec:basym}.} \\
In contrast to VL, the lepton asymmetry in the majoron+triplet model is produced only after the $U(1)_\lnumber$ symmetry is broken at $f$ and thus the relevant range for the asymmetry is $\zn \geq z_I \equiv \sfrac{\gn}{\sqrt 2}$. For simplicity, we solve the Boltzmann equations for the particle abundances in that range as well and hence specify the initial condtions at $\zn = \zi$. We then broadly distinguish between the two limiting scenarios where either all relevant particles have a thermal initial abundance or a vanishing initial abundance at $\zi$. 
Thus, for cases $\hat A, \hat B$, the neutrinos have either a thermal initial abundance, 
\begin{align}
  \hat A_t, \hat B_t\,\, \text{initial abundance:}\qquad Y_N(\zi) = Y_N^{eq}(\zi)\,,
\end{align}
denoted by a subscript $t$, or a vanishing initial abundance,   
\begin{align}
  \hat A_z, \hat B_z\,\, \text{initial abundance:}\qquad Y_N(\zi) = 0\,,
\end{align}
denoted by a subscript $z$. 
In cases $A, B$, we assume for simplicity that the initial abundances of $N, \sigma, J, \latexchi$ are the same and similarly solve the Boltzmann equations for their evolution for a scenario where the new fields have either a thermal initial abundance,
\begin{align}
  A_t, B_t\,\, \text{initial abundance:}\qquad Y_{N,\sigma, J,\latexchi}(\zi) = Y_{N,\sigma, J,\latexchi}^{eq}(\zi)\,,
\end{align}
denoted with a subscript $t$, or a vanishing initial abundance, 
\begin{align}
  A_z, B_z\,\, \text{initial abundance:}\qquad Y_{N,\sigma, J,\latexchi}(\zi) = 0\,,
\end{align}
denoted with a subscript $z$. 
Assuming that no other mechanisms generated an lepton asymmetry for $\zn < \zi$, we solve the Boltzmann equation for the efficiency with the initial condtion $\eta(\zi) =0 $. \\
For comparison, we also solve the Boltzmann equations in the VL scenario. We choose the same initial conditions as cases $\hat A, \hat B$ with the distinction that the initial conditions are given at $\zi^{VL}\equiv\zn = 0$. 
The parameter assignments are summarized in Tab. \ref{tab:parameters}.
Even though it does not affect the numerical results, we explicitly included a small but non-zero majoron mass $\mj \sim \SI{e-3}{\giga\electronvolt}$ inspired by previous works on majoron DM \cite{Brune:2018sab}. \\
The sizes of the couplings $\gn, \geta, \ls$ are particularly relevant when considering the decays $\sigma \to NN, \latexchi\latexchi, JJ$ with decay rates given by
\begin{align}
	\Gamma_{\sigma \to NN} = \frac{\gn^2}{8\pi}\sqrt{\frac{\ms^2}{4}- \mn^2}\,, \quad \Gamma_{\sigma \to TT} = 3\frac{\geta^2}{8\pi}\sqrt{\frac{\ms^2}{4}- \mt^2}\,,\quad \Gamma_{\sigma\to JJ} = \frac{\pi k_\sigma^2}{16\pi^2 \ms^2}\sqrt{\frac{\ms^2}{4}-\mj^2}\,.
\end{align}
Clearly, $\sigma \to NN$ and $\sigma \to \latexchi\latexchi$ are kinematically allowed only when $\gn \leq 0.7$ and $\geta \leq 0.7$, respectively. This implies that $\sigma \to \latexchi\latexchi$ is forbidden in cases $A, \hat A$ and allowed in cases $B, \hat B$. However, we find $\br{\sigma \to \latexchi\latexchi} \sim \num{e-13}$ for $\geta = \num{e-7}$ so that we can safely neglect $\sigma \to \latexchi\latexchi$ even in cases $B, \hat B$. As $\mj \ll \ms$ independent of the couplings, $\sigma \to JJ$ is always allowed.  In Fig. \ref{fig:branchingPlot}, we show $\br{\sigma \to JJ, NN}$ as a function of $\gn$. We clearly see that $\sigma\to JJ$ is the preferred decay channel while $\br{\sigma \to NN} \ll \br{\sigma \to NJJ}$ peaks at $\gn = \frac{1}{\sqrt 3}$. \\
We further note that we distinguish the thermal rates where $\latexchi$ appears in the initial or final state for cases $A, \hat A$ and cases $B, \hat B$ while the other thermal rates do not depend on $\geta$ and are therefore identical for all cases.
\begin{table}[H]
  \centering
  \begin{tabular}{c c c c c c c c c}
  \toprule
  {Case} & $\gn$& $\geta$ & $\ls$ & $f/\si{\giga\electronvolt}$ & $\mtl/\si{\electronvolt}$& $ Y_N(\zi) $& $ Y_{\sigma, J, \latexchi}(\zi) $ & $\dall(\zn) = 1$ \\
  \midrule
  $\hat A_z$ & $[0.1,1]$ & $1$ & $1$ & $\num{e10}$ & $[\num{5 e-5},\num{e-1}]$ & 0 &   & yes\\
  $\hat A_t$ & $[0.1,1]$ & $1$ & $1$ & $\num{e10}$ & $[\num{5 e-5},\num{e-1}]$ & $Y_{N}^{eq}(\zi)$ &   & yes\\
  $\hat B_z$ & $[0.1,1]$ & $\num{e-7}$ & $1$ & $\num{e10}$ & $[\num{5 e-5},\num{e-1}]$ & 0 &   & yes\\
  $\hat B_t$ & $[0.1,1]$ & $\num{e-7}$ & $1$ & $\num{e10}$ & $[\num{5 e-5},\num{e-1}]$ & $Y_{N}^{eq}(\zi)$ &   & yes\\
  \midrule
  $A_z$ & $[0.1,1]$ & $1$ & $1$ & $\num{e10}$ & $[\num{5 e-5},\num{e-1}]$ & 0 & $0$  & no \\
  $A_t$ & $[0.1,1]$ & $1$ & $1$ & $\num{e10}$ & $[\num{5 e-5},\num{e-1}]$ & $Y_{N}^{eq}(\zi)$ & $Y_{\sigma, J, \latexchi}^{eq}(\zi)$  & no \\
  $B_z$ & $[0.1,1]$ & $\num{e-7}$ & $1$ & $\num{e10}$ & $[\num{5 e-5},\num{e-1}]$ & 0 & $0$  & no \\
  $B_t$ & $[0.1,1]$ & $\num{e-7}$ & $1$ & $\num{e10}$ & $[\num{5 e-5},\num{e-1}]$ & $Y_{N}^{eq}(\zi)$ & $Y_{\sigma, J, \latexchi}^{eq}(\zi)$  & no \\
  \midrule
  $\mathrm{VL}_z$ &  &  &  &  & $[\num{5 e-5},\num{e-1}]$ & 0 &  &  \\
  $\mathrm{VL}_t$ &  &  &  &  & $[\num{5 e-5},\num{e-1}]$ & $Y_{N}^{eq}(\zi)$ &   &  \\
  \bottomrule
\end{tabular}
\caption{Parameters and initial condtions used to solve to Boltzmann equations \eqref{eq:dyn_vl}, \eqref{eq:eta}, \eqref{eq:ben}, \eqref{be:chi}, \eqref{be:s} \eqref{be:j} and \eqref{eq:dyn_ad}. Cases $\hat A, \hat B$ correspond to the simplified scenario where $\dall = 1$ while in cases $A, B$, the full set of Boltzmann equations for $N, \latexchi, \sigma, J$ is considered. The index $t(z)$ denotes that the corresponding Boltzmann equations are solved with thermal (vanishing) initial particle abundances. Note that $\mn$ does not have an effect on $\eta_{VL}$. }
\label{tab:parameters}
\end{table}

\section{Results}
\label{sec:results}
We numerically solve the Boltzmann equations given in \eqref{eq:dyn_vl}, \eqref{eq:eta}, \eqref{eq:ben}, \eqref{be:chi}, \eqref{be:s}, \eqref{be:j} and \eqref{eq:dyn_ad} with the parameters given in Tab. \ref{tab:parameters}. As our main interest is the effect of the different parameter assignments on the final efficiency, we will begin with a brief overview of the corresponding results before discussing the evolutions of the particle abundances and the efficiency in the respective cases in greater detail. 

\subsection{Overview}
\label{sec:comparison}
\begin{figure}[h!]
    \centering
    \begin{subfigure}{\textwidth}
    \includegraphics[width=\textwidth]{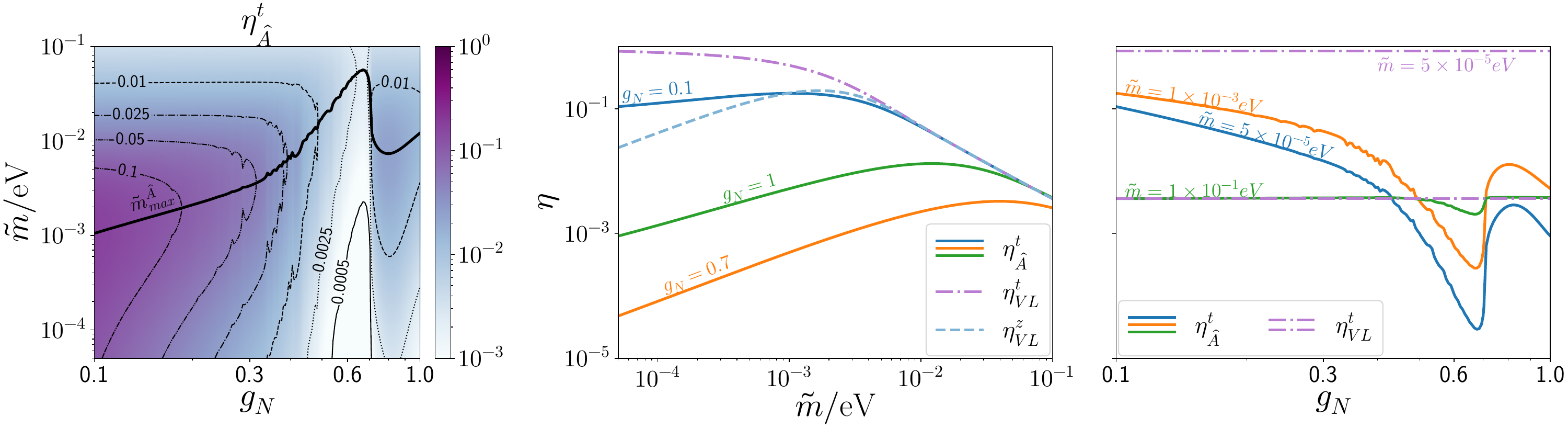}
    \end{subfigure}\\
    \centering
    \begin{subfigure}{\textwidth}
    \includegraphics[width=\textwidth]{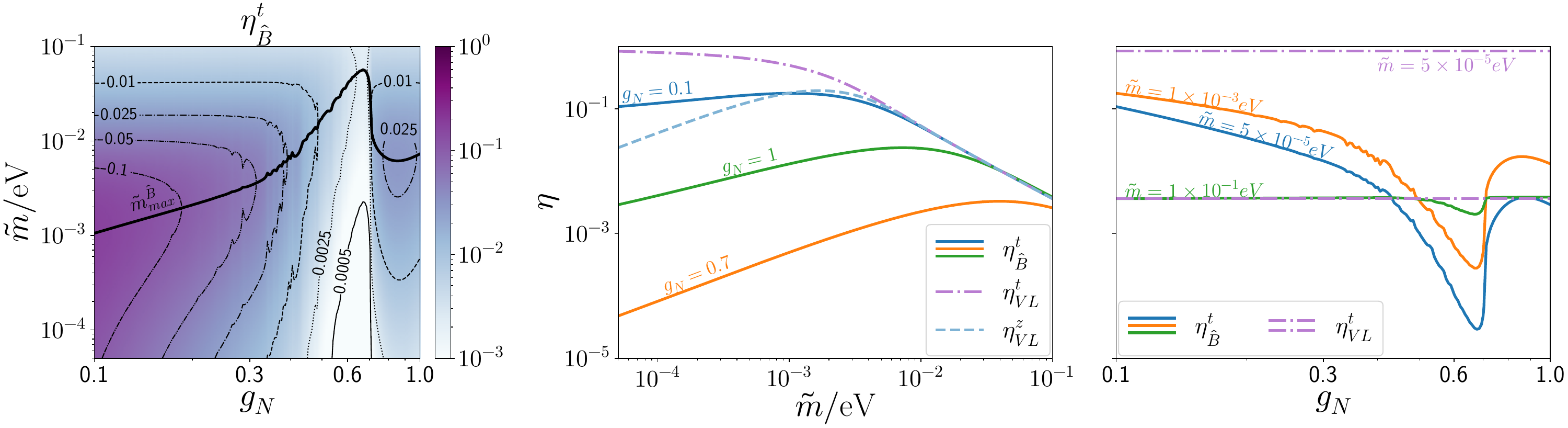}
    \end{subfigure}
    \caption{
    \textit{Left:} Density plots of $\etahA$ (upper plot) and $\etahB$ (lower plot) in the $\gn - \mtl$ plane. The black lines indicates at which $\mtl$ the efficiency reaches its maximum. \\
    \textit{Middle:}  $\etahA$ (upper plot) and $\etahB$ (lower plot) as functions of $\gn$.\\  
    \textit{Right:}  $\etahA$ (upper plot) and $\etahB$ (lower plot) as functions of $\mtl$.\\ 
    See text for discussion. }
    \label{fig:density:hatA:hatB}
\end{figure}

\begin{figure}[h!]  
    \centering
    \begin{subfigure}{\textwidth}
    \includegraphics[width=\textwidth]{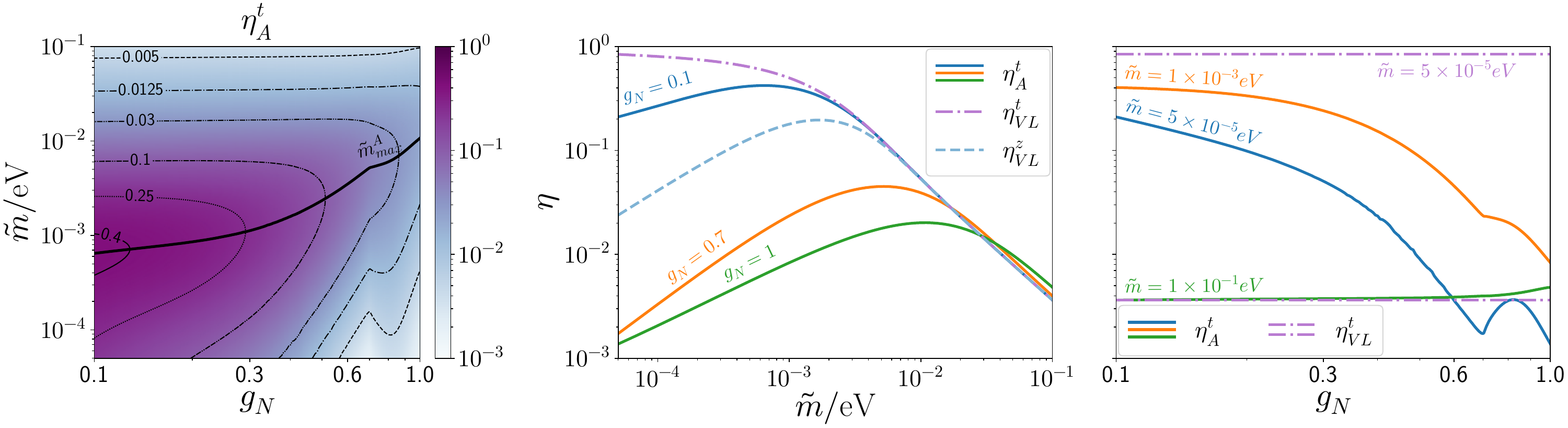}
    \end{subfigure}\\
    \centering
    \begin{subfigure}{\textwidth}
    \includegraphics[width=\textwidth]{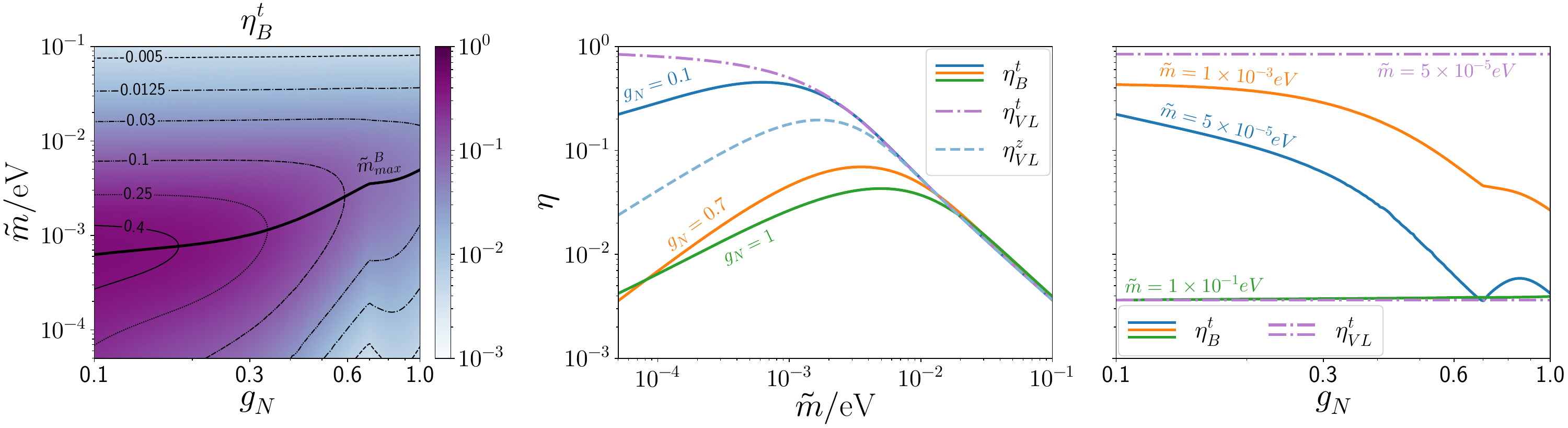}
    \end{subfigure}\\
    \centering
    \begin{subfigure}{\textwidth}
    \includegraphics[width=\textwidth]{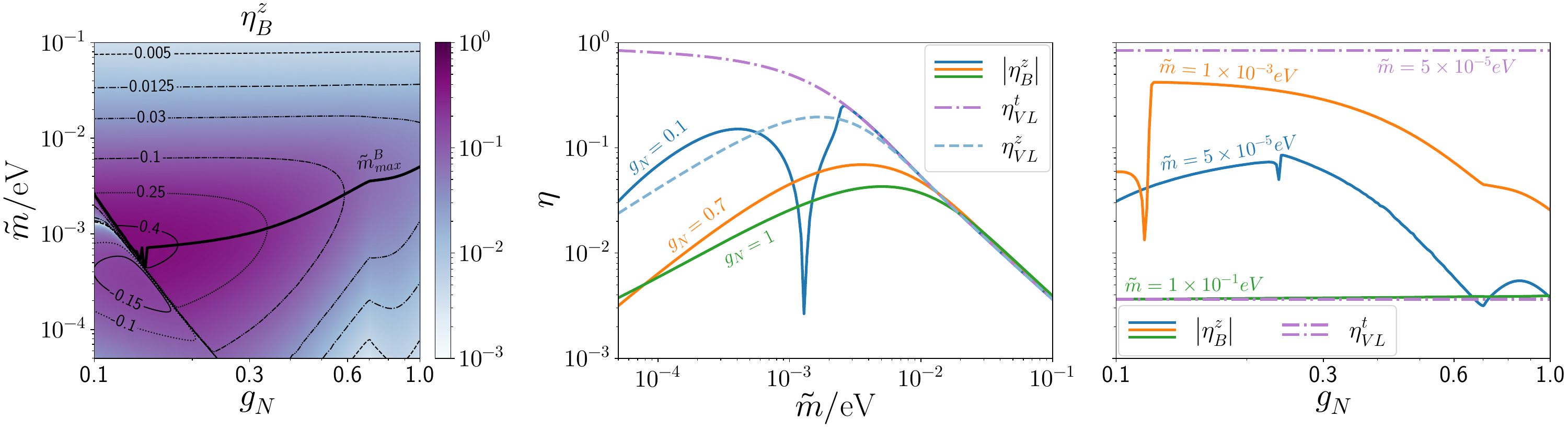}
    \end{subfigure}
    \caption{
    \textit{Left:} Density plots of $\etaA$ (top column), $\etaBt$ (middle column) and $\etaBz$ (bottom column) in the $\gn - \mtl$ plane. The black lines indicates at which $\mtl$ the efficiency reaches its maximum. \\
    \textit{Middle:}  $\etaA$ (top column), $\etaBt$ (middle column) $\etaBz$ (bottom column) as functions of $\gn$.\\  
    \textit{Right:}  $\etaA$ (top column), $\etaBt$ (middle column) $\etaBz$ (bottom column) as functions of $\mtl$.\\ 
    See text for discussion. }
    \label{fig:density:A:B}
\end{figure}

In Figs. \ref{fig:density:hatA:hatB},\ref{fig:density:A:B}, we show density plots of the efficiency $\eta$ in the $\gn-\mtl$ plane(left panels) and $\eta(\mtl)$(middle panels), $\eta(\gn)$(right panels) for exemplary values of $\gn$ and $\mtl$, respectively. \\
Note that we find that $\eta$ is independent from the initial conditions in cases $A, \hat A, \hat B$ while the efficiency in case $B$ depends on the initial conditions. Therefore, we will not distinguish between the initial conditions in the discussion of the final efficiency for cases $A, \hat A, \hat B$, i.e.\ 
\begin{align}
	\etahA \equiv \etahA^{t,z}\,,\qquad	\etahB \equiv \etahB^{t,z}\,,\qquad	\etaA \equiv \etaA^{t,z} \,.
\end{align}
 \begin{figure}[h!]
  \centering
    \includegraphics[width=0.72\textwidth]{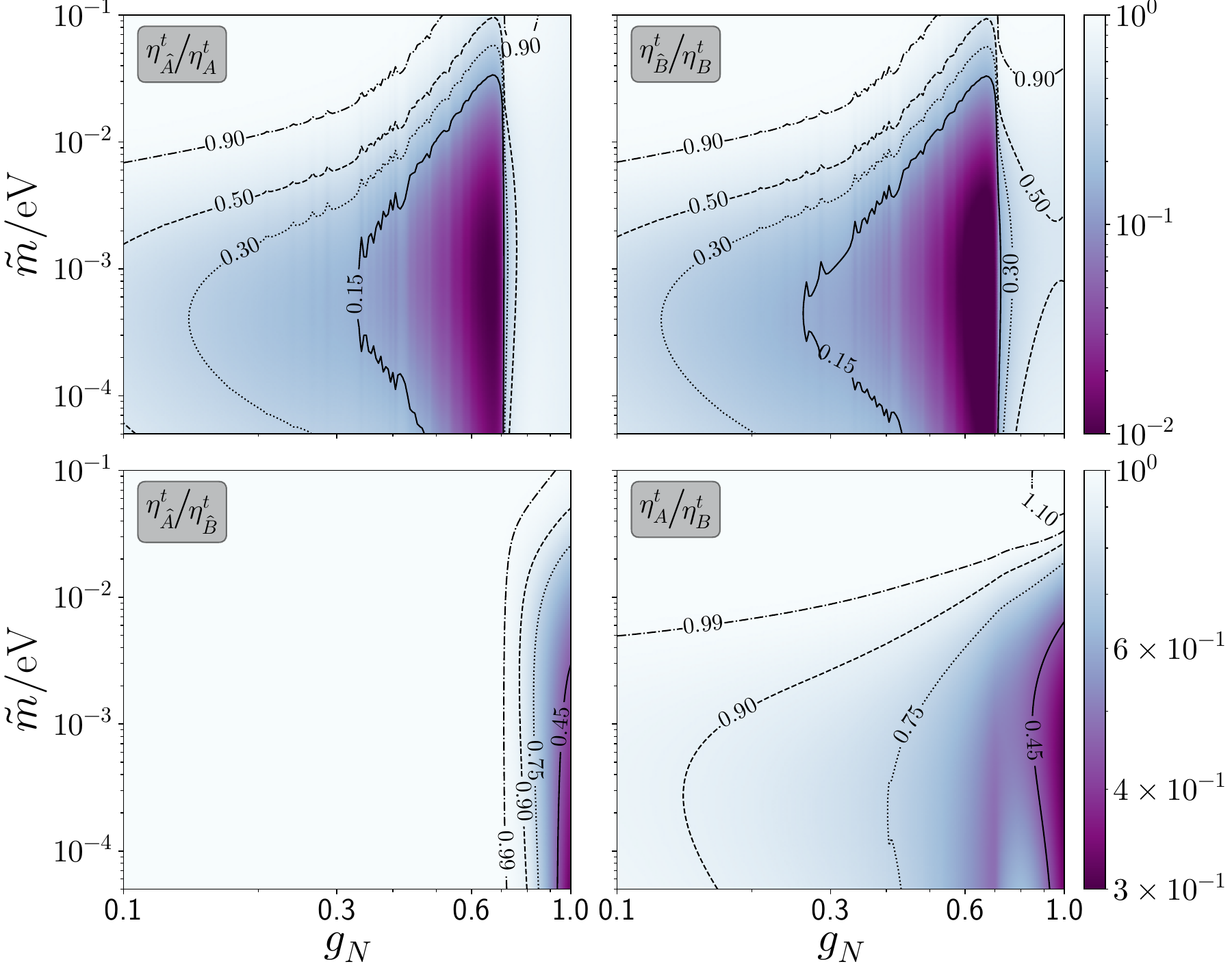}
    \caption{
    \textit{Upper plots: }Density plots of $\sfrac{\etahA}{\etaA}$ (left) and $\sfrac{\etahB}{\etaBt}$ (right) in the $\gn-\mtl$ plane. \\
    \textit{Lower plots: }Density plots of $\sfrac{\etahA}{\etahB}$ (left) and $\sfrac{\etaA}{\etaBt}$ (right) in the $\gn-\mtl$ plane. 
    See text for discussion.
    }
    \label{fig:densitycomp}
\end{figure}

\begin{figure}[h!]
  \centering
    \includegraphics[width=0.8\textwidth]{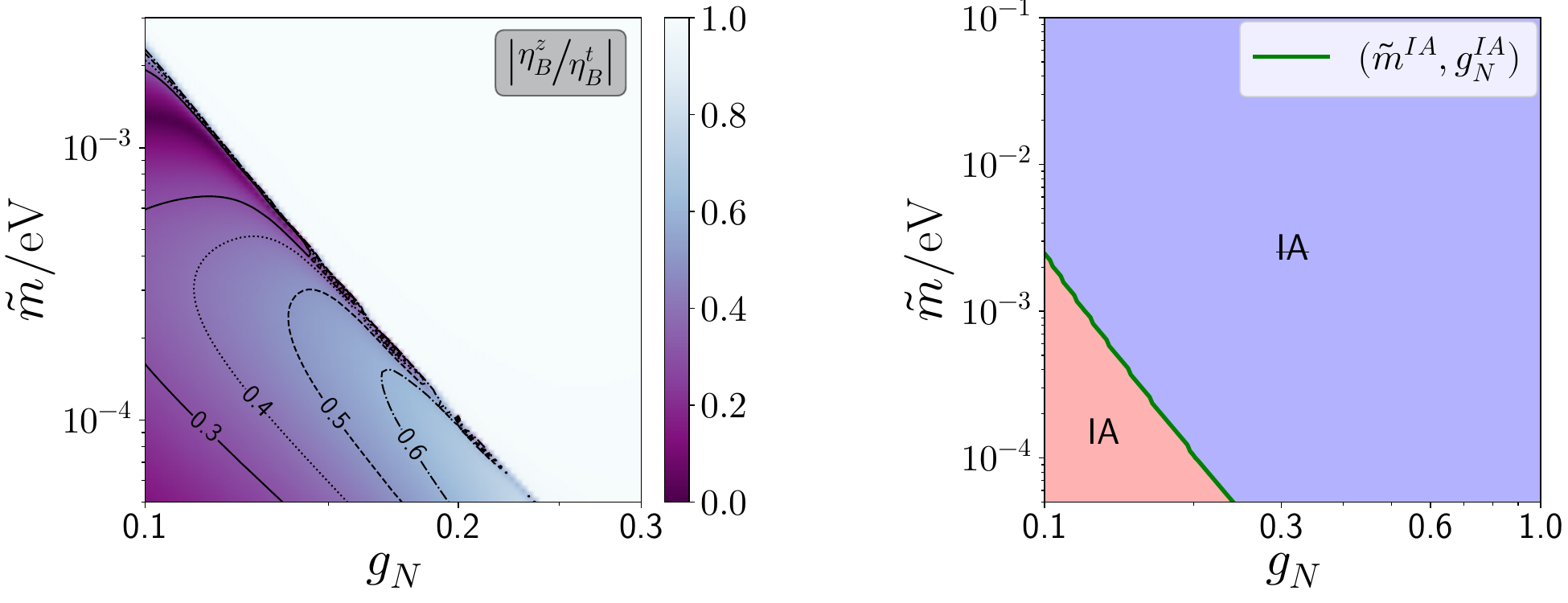}
    \caption{\textit{Left:} Density plot of $\left|\sfrac{\etaBz}{\etaBt}\right|$ in the $\gn-\mtl$ plane.  \\
    \textit{Right:} Schematic density plot of IA, \st{IA} and $(\gic, \mic)$.\\
    See text for discussion.
    }
    \label{fig:densitycomp2} 
\end{figure}
\noindent 
First, we dicuss cases $A, \hat A, \hat B, B_t$ as they display some similar features. 
Concerning the dependence on $\mtl$, the efficiencies $ \etahA, \etaA, \etahB, \etaBt$ initially increase with $\mtl$ until they reach a maximum at some $\mtmax$ and subsequently decrease. For $\mtl > \mtmax$, we find that similarly to what happens in VL, all efficiencies eventually align and obtain a common efficiency we define as $\eta^{WO}$. This holds except for two notable exceptions: In cases $\hat A, \hat B$ with $\gn \approx 0.7$, the efficiency stays below $\eta^{WO}$ while in case $A$ with $\gn \approx 1$, the efficiency slightly exceeds $\eta^{WO}$.\\
For $\mtl \lesssim \mtmax$, we find that the efficiencies for cases $A,B$ exceed the efficiencies reached in the cases $\hat A, \hat B$. 
This is highlighted in Fig. \ref{fig:densitycomp} (upper panels) where we show density plots of $\sfrac{\etahA}{\etaA}(\gn, \mtl)$ (left) and $\sfrac{\etahB}{\etaB}(\gn, \mtl)$ (right).
However, the efficiencies in the considered scenarios of the majoron+triplet model for $\mtl \lesssim \mtmax$ are generally smaller than $\etaVLt$ and unless $\gn$ is small, they are also smaller than $\etaVLz$ (see Figs. \ref{fig:density:hatA:hatB},\ref{fig:density:A:B}). \\
Considering the dependence on $\gn$, we find that the efficiencies in the simplified scenario $\etahAB(\gn)$ depend strongly on $\gn$ and irrespective of $\mtl$, they reach a striking minimum around $\gn \sim 0.7$. Overall, we note that for $\gn \ll 0.7$ and $\gn \gg 0.7$, $\etahAB(\gn)$ decrease with $\gn$. 
Similarly, the efficiencies obtained from the full set of Boltzmann equations $\eta_{A,B}^t(\gn)$ decrease with $\gn$ for $\gn \ll 0.7$ and $\gn \gg 0.7$. 
On the other hand, the minimum of $\eta_{A,B}^t$ around $\gn \sim 0.7$ is significantly less striking compared to the simplified scenario and even disappears for $\mtl \to \SI{e-1}{\electronvolt}$. 
Thus, as can be seen in Fig. \ref{fig:densitycomp} (upper panels), the relative deviations $\sfrac{\etahA}{\etaA}(\gn,\mtl)$ and $\sfrac{\etahB}{\etaB}(\gn,\mtl)$ are maximal around $\gn \sim 0.7$. 
Finally, as can be seen in Fig. \ref{fig:densitycomp} (lower panels) where we show density plots of $\sfrac{\etahA}{\etahB}(\gn, \mtl)$(left) and $\sfrac{\etaA}{\etaB}(\gn, \mtl)$(right), the efficiencies $\etahA, \etahB$ are close to identical for $\gn <0.7$ while $\etahB$ slightly exceeds $\etahA$ for $\gn >0.7$. Similarly, the relative deviation $\sfrac{\etaA}{\etaB}(\gn, \mtl)$ is maximal for $\gn >0.7$ but in contrast to the simplified scenario, $\etaB$ slightly exceeds $\etaA$ already for $\gn <0.7$. \\
Next, let us discuss $\etaBz$. 
In contrast to the previous cases, we find that $\etaBz$ is negative in a triangular region spanning from $0.1 \leq \gn \lesssim 0.24$ and $\SI{5e-5}{\electronvolt} \leq \mtl \lesssim \SI{2.5 e-3}{\electronvolt}$ while outside of this region, $\etaBz$ is positive and we have $\etaBt \approx \etaBz$ as highlighted in Fig. \ref{fig:densitycomp2}(left) where we show $\sfrac{\etaBz}{\etaBt}(\gn, \mtl)$. 
Thus, we seperate $|\etaBz(\mtl, \gn)|$ into an IA (initial abundance) regime where the initial abundances of $N, \sigma, J, \latexchi$ are relevant and into an \st{IA} regime where the inital abundances have no significant effect so that $\etaBt \approx \etaBz$. This is displayed in Fig. \ref{fig:densitycomp2}(right) where we also introduced 
 $(\gic,\mic)$ in order to quantify when the transition from IA to the \st{IA} regime occurs.\footnote{We will properly define  $(\gic,\mic)$ in Sec. \ref{sec:case:b}} \\
To summarize, we find that while the additional scattering processes generally diminish the efficiency that can be reached compared to VL, certain sets of parameters still allow for a sizable efficiency. 
Moreover, solving the full Boltzmann equations incluing $\latexchi, \sigma$ and $J$ can have significant effects on the final efficiency. 
Further, we note that solving the full Boltzmann equations can also result in a dependence on the initial abundances if $\gn, \geta$ and $\mtl$ are small while for most of the parameters sets we considered, the initial abundances are irrelevant. In the following, we will discuss the dynamics of the respective cases in greater detail. 

\subsection{Cases $\hat A, \hat B$: $\dall = 1$} 
\label{sec:approximation}
In this section, we discuss the cases $\hat A, \hat B$ in more detail. Recall that we found in the previous section that the initial abundance of neutrinos does not affect the final efficiency in this scenario. Additionally, we found that the final efficiencies in both cases differ only slightly. \\
We will begin with a discussion of the relevant thermal rates and how they depend on $\mtl$ and $\gn$. We then proceed with a discussion of the neutrino evolution and how this affects the efficiency. Note that for brevity, we will only distinguish between case $\hat A$ and case $\hat B$ when necessary and introduce the convention to refer to generic quantities of the simplified scenario with a hat so that e.g.\ $\hat \eta = \eta_{\hat A, \hat B}$.

\subsubsection{Discussion of $\gammaa$}
\begin{figure}[h!]
    \centering
    \includegraphics[width=0.5\textwidth]{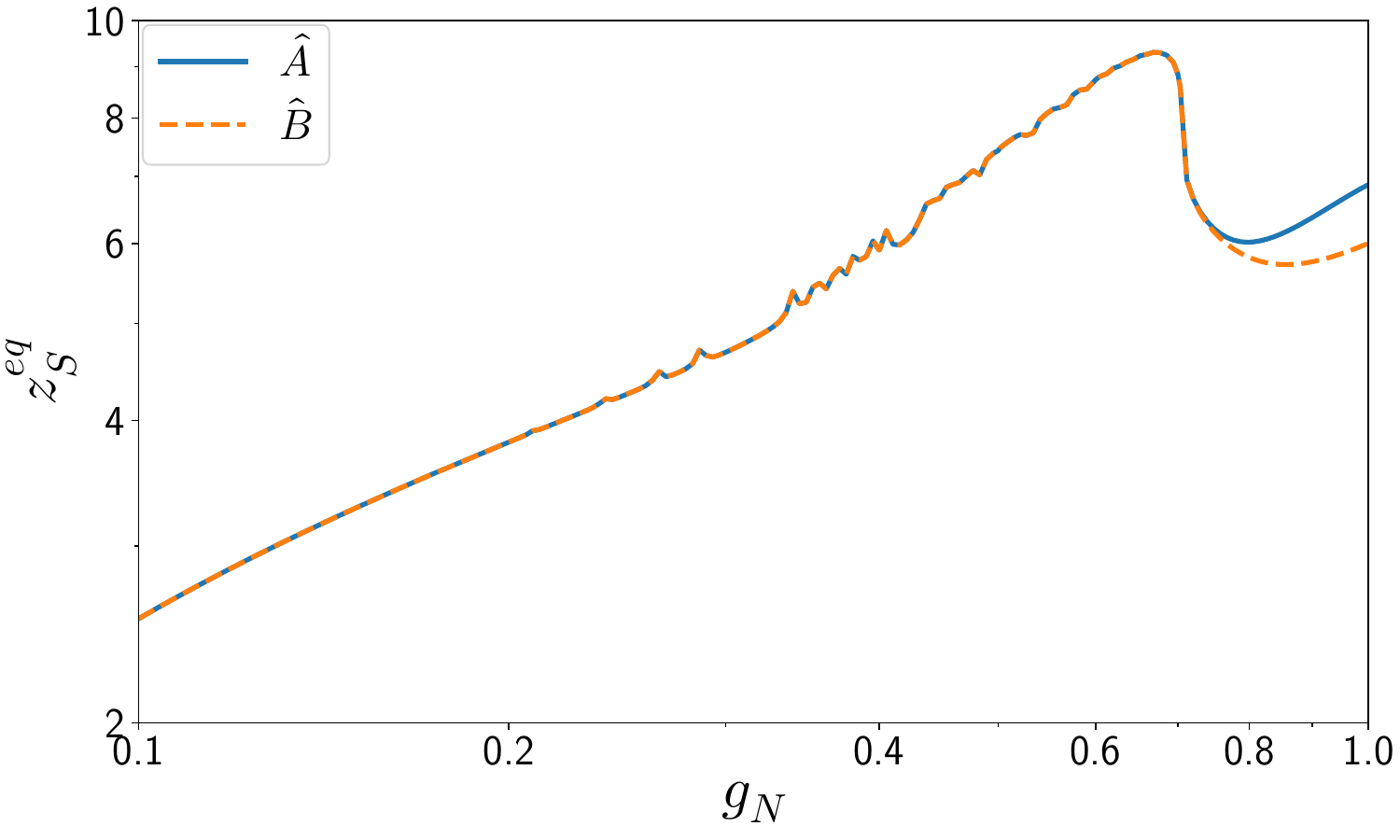}
    \caption{$\zaeq(\gn)$ as defined in \protect\eqref{eq:zeq:123} with the blue solid line corresponding to case $\hat A$ und the orange   dashed line corresponding to case $\hat B$. 
    For $\gn \ll 0.7$ and $\gn > 0.7$, both $\zaAeq$ and $\zaBeq$ increase with $\gn$ due to the Boltzmann suppression of the relevant thermal rates appearing in $\gammaa$ being damped as $\gn$ increases. 
    Moreover, for $\gn < 0.7$, $\zaeq(\gn)$ is identical for cases $\hat A$ and $\hat B$ as the processes dominating $\gammaa$ are independent from $\geta$.
    At $\gn \sim 0.7$, both $\zaAeq$ and $\zaBeq$ rapidly drop with can be attributed to $\gat{\sigma,NN}$ becoming forbidden, thus decreasing the overall magnitudes of $\gammaa^{\hat A}$ and $\gammaa^{\hat B}$(see Fig. \ref{fig:case1_gamman}). 
    For $\gn > 0.7$, we have $\zaAeq > \zaBeq$ as the summed scattering rate $\gammaa$ in case $\hat B$ decouples faster than in case $\hat A$.  
    } 
    \label{fig:gammaA:zeq}
\end{figure}

\begin{figure}[h!]
  \centering
  \includegraphics[width=\textwidth]{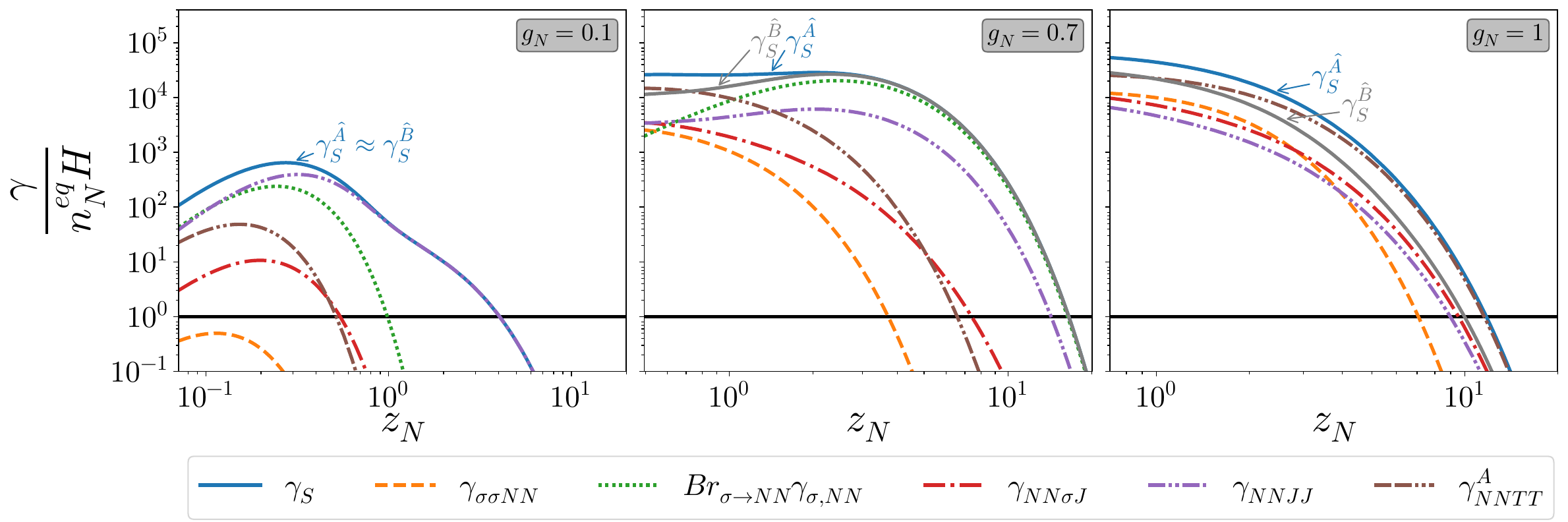}
  \caption{Thermal rates of scattering processes that appear in the majoron+triplet model for $\gn=0.1$ (left), $\gn=0.7$(middle) and $\gn=1$. Note that we show $\gat{NN\latexchi\latexchi}^A=\gat{NN\latexchi\latexchi}^{\hat A}$ while $\gat{\sigma\sigma NN}, \gat{NN\sigma J}, \gat{NNJJ}$ and $\gat{\sigma,NN}$ are identical for cases $\hat A$ and $\hat B$. For $\gn =0.1$, $\gat{NN\latexchi\latexchi}^{\hat A}$ is tiny, resulting in $\gammaa^{\hat A} \approx \gammaa^{\hat B}$. As $\gn$ increases, the contribution of $\gat{NN\latexchi\latexchi}^{\hat A}$ to $\gammaa^{\hat A}$ grows so that eventually $\gammaa^{\hat A} > \gammaa^{\hat B}$. 
  This is especially striking for $\gn=1$ where $\gammaa^{\hat B} \approx \gat{\sigma\sigma NN} + \gat{NN\sigma J} + \gat{NNJJ}$ whereas $\gammaa^{\hat A}$ is dominated by $\gat{NN\latexchi\latexchi}^{\hat A}$ while the contributions from $\gat{\sigma\sigma NN}, \gat{NN\sigma J}$ and $\gat{NNJJ}$ are subdominant. Moreover, as $\gn$ increases, the Boltzmann suppression of the thermal rates is shifted to larger $\zn$ and consequently, the corresponding interactions are coupled to the plasma longer.
  We further stress that we used $\zi$ as the lower limit for the $\zn$ range presented and consequently, the respective plot range changes with $\gn$. 
  }
  \label{fig:case1_gamman}
\end{figure}

First, let us discuss the overall behaviour of the summed scattering rate $\gammaa$. 
In Fig. \ref{fig:case1_gamman}, we show $\gammaa$ and the relevant scattering rates that contribute to $\gammaa$ for exemplary values of $\gn$. 
It is apparent that for small $\zn$, we have 
\begin{align}
  \frac{\gammaa}{n_N^{eq}H}(\zn) >1
\end{align}
and thus the scattering interactions thermalize the neutrinos.  
In fact, the scattering processes are so fast around $\zn = \zi$ that neutrinos immediatly reach thermal equilibrium even if $\hat Y_N(\zi) = 0$. For convenience, we will therefore not distinguish between $\hat Y_N^t(\zn)$ and $\hat Y_N^z(\zn)$ for the remainder of this section. We note however that $\sfrac{\gammaa}{n_N^{eq}H}(\zn)$ is Boltzmann suppressed and therefore the scatterings decouple when
\begin{align}
  \frac{\gammaa}{n_N^{eq}H}(\zn)\leq 1 \label{eq:zeq:123}\,,
\end{align}
i.e.\ scattering processes keep neutrinos close to thermal equilibrium until $\zn = \zaeq$ defined via 
\begin{align}
  \frac{\gammaa}{n_N^{eq}H}(\gn, \zn= \zaeq)= 1 \label{eq:zeq2:123}\,.
\end{align}
In Fig. \ref{fig:gammaA:zeq}, we show $\zaeq(\gn)$ as a function of $\gn$. 
We can easily see that for $\gn < 0.7$, $\zaAeq = \zaBeq$ and additionally, both reach a peak at $\gn \lesssim 0.7$ before rapidly decreasing. 
After reaching a local minumum, both $\zaAeq$ and $\zaBeq$ start to increase again with $\zaAeq$ growing faster than $\zaBeq$. This behaviour can be explained by considering the individual terms that appear in $\gammaa$.\\
\noindent
Both for case $\hat A$ and case $\hat B$, we can see in Fig. \ref{fig:case1_gamman} that $\gammaa$ is dominated by $\gamma_{\sigma ,NN}$ and $\gamma_{NNJJ}$ when $\gn \leq 0.7$, thus explaining why $\zaeq$ is identical for case $\hat A$ and case $\hat B$ in that range. On the other hand, for $\gn > 0.7$, we note that in case $\hat A$, $\gammaa$ is dominated by $\gat{NNTT}^{\hat A}$ while in case $\hat B$, the dominant processes in $\gammaa$ are $\gat{NNJJ}$ and $\gat{NN\sigma J}$. The change of behaviour around $\gn \sim 0.7$ can be traced back to $\sigma \to NN$ being forbidden for $\gn > 0.7$ and a changing Boltzmann suppression due to the change in $\gn$. \\
In particular, let us consider $\gat{NNJJ}$ and $\gat{\sigma,NN}$ in the range $\gn \leq 0.7$. As $\gn$ increases, the overall magnitudes of $\gat{NNJJ}$ and $\gat{\sigma,NN}$ increase and as a consequence, $NN\leftrightarrow JJ$ scatterings and $\sigma\leftrightarrow NN$ (inverse) decays are thermalized longer, translating to an increasing $\zaeq$. Moreover, the Boltzmann suppression of $\sfrac{\gat{\sigma,NN}}{n_N}$ does change notably with $\gn$. In general, we have
\begin{align} 
  \frac{\gat{\sigma,NN}}{n_N^{eq}} \propto \frac{n_\sigma^{eq}}{n_N^{eq}}\,,
\end{align}
where $n_\sigma^{eq}$ and $n_N^{eq}$ are given by \eqref{eq:numberdensity} and depend on $z_\sigma = \sfrac{\ms}{T}$\footnote{Here, $T$ is the temperature. } and $\zn$, respectively. With 
\begin{align}
z_\sigma = \frac{\ms}{\mn}\zn = \frac{\sqrt 2}{\gn}\zn\,,
\end{align}
we can easily see that the minimal value $z_\sigma$ can take is given by $z_\sigma^{min} = \frac{\sqrt 2}{\gn}\zi^{min} = 1$, i.e.\ $n_\sigma^{eq}$ is Boltzmann suppressed over the full range of interest, irrespective of $\gn$. 
On the other hand, we have $z_N^{min} = z_I(\gn) < 1$, meaning that $n_N^{eq}$ is not initially Boltzmann suppressed and we need to distinguish between $\zn <1$ and $\zn >1$, i.e.\
s\begin{align}
  \frac{\gat{\sigma,NN}}{n_N^{eq}} \propto\begin{cases} \left(\frac{\sqrt 2}{\gn}\right)^\frac{3}{2} \e^{-\frac{\sqrt 2}{\gn}\zn}\,,&\quad \zn <1\,, \\  \frac{\ms}{\mn} \e^{-\left(\frac{\sqrt 2}{\gn}-1\right)\zn}\,, &\quad \zn >1\,. \end{cases} 
\end{align}
In both regimes however, we find that the overall Boltzmann suppression of $ \sfrac{\gat{\sigma,NN}}{n_N^{eq}}$ decreases with increasing $\gn$. Additionally, it is evident from Fig. \ref{fig:gammaA:zeq} that $\Gamma_{\sigma,NN}$ increases with $\gn$ until it peaks at $\gn = \sfrac{1}{\sqrt 3}$. Putting all these observations together, it is clear that $\zaeq$ increases with $\gn$ while $\gn \leq 0.7$. \\
On the other hand, when $\gn > 0.7$, the previously dominant process $\gat{\sigma,NN}$ is forbidden and $\zaeq$ is determined by the processes $\gat{NNTT}^{\hat A}$ (case $\hat A$) and $\gat{NN\sigma J},\gat{NNJJ},\gat{\sigma \sigma NN}$ (case $\hat B$) which have an overall smaller magnitude, thereby explaining the rapid drop in $\zaeq$.
\subsubsection{Neutrino Abundance}
\begin{figure}[h!]
  \centering
  \includegraphics[width=\textwidth]{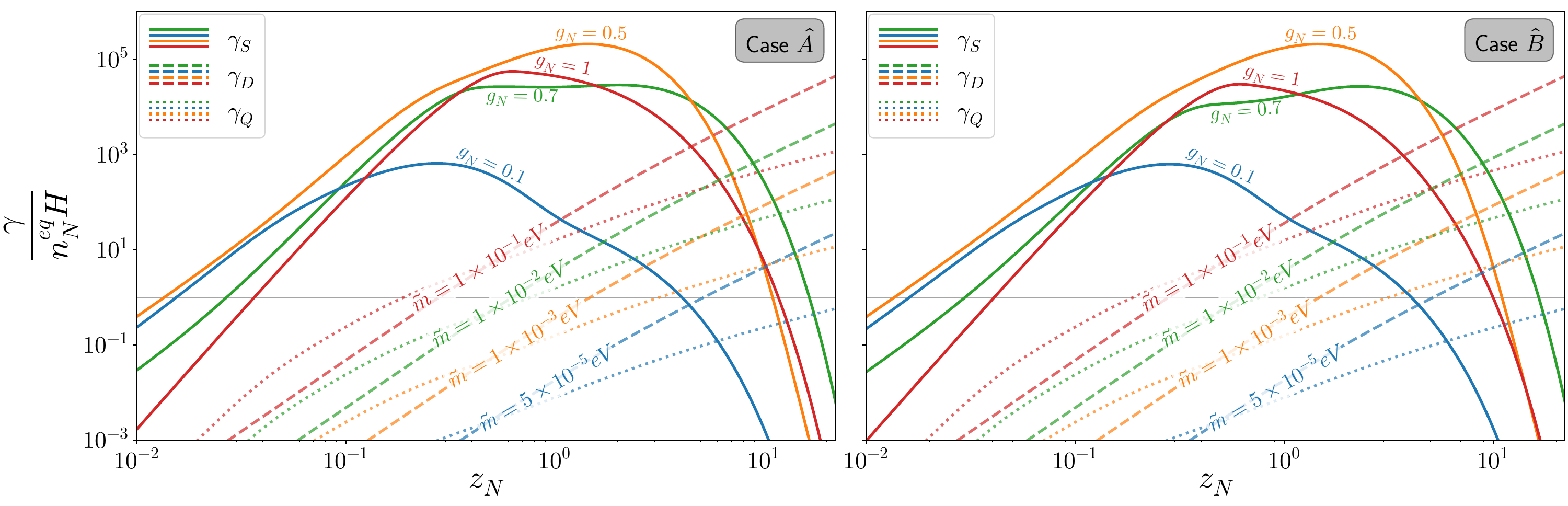}
  \caption{Thermal rates relevant for the evolution of $Y_N$ \eqref{eq:dyn_ad} for $\gn = [0.1,0.5,0.7,1]$ and 
  $\mtl = [\num{5e-5}, \num{e-3},\num{e-2},\num{e-1}]\si{\electronvolt}$ for case $\hat A$(left) and case $\hat B$(right). Note that $\sfrac{\gammaa}{n_N^{eq}H}$ exceeds $\sfrac{\gamma_D}{n_N^{eq}H}$ until it drops due to being Boltzmann suppressed and subsequently, $\sfrac{\gamma_D}{n_N^{eq}H}$ becomes dominant while $\gamma_Q$ is always subdominant to either $\gammaa$ or $\gamma_D$.}
  \label{fig:gammaBSM_N}
\end{figure}

\begin{figure}[h!]
  \centering
  \begin{subfigure}{\textwidth}
  \includegraphics[width=\textwidth]{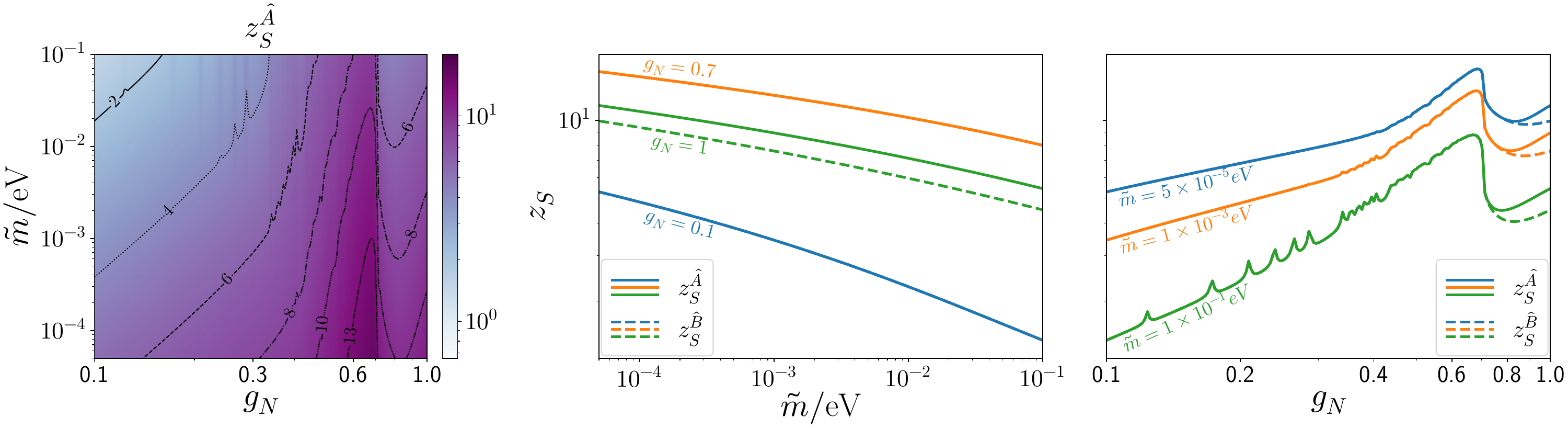}
   \end{subfigure}\\
   \centering
  \begin{subfigure}{\textwidth}
  \includegraphics[width=\textwidth]{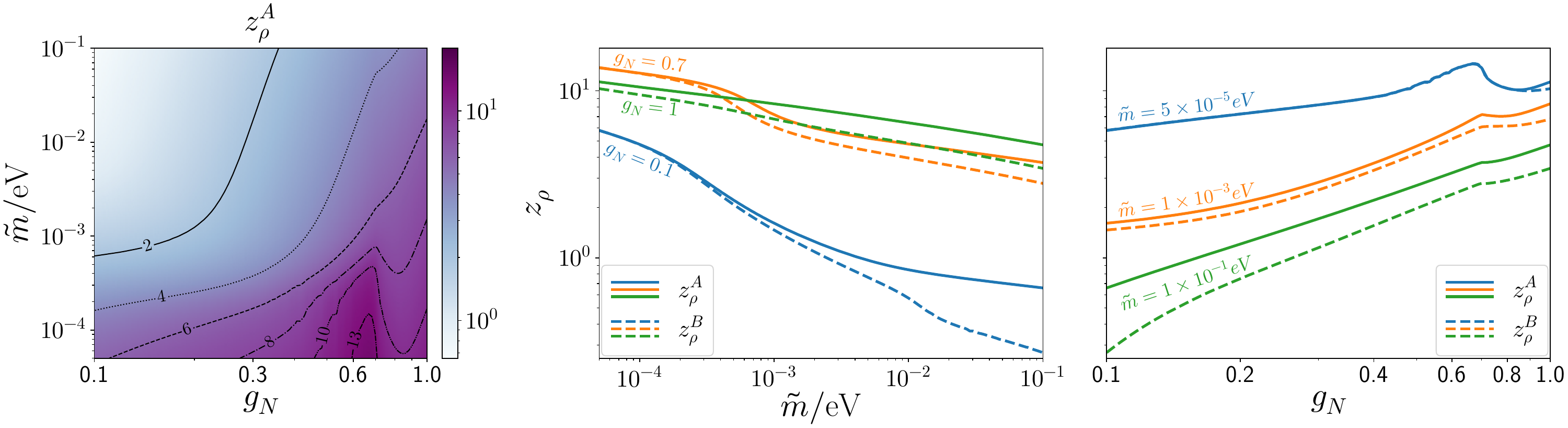}
   \end{subfigure}
    \caption{\textbf{Upper plots: }\textit{Left: }Density plot of $\zaA$ in the $\gn-\mtl$ plane. 
    \textit{Middle: } $\zaA$(solid) and $\zaB$(dashed) as functions of $\mtl$. 
    \textit{Right: } $\zaA$(solid) and $\zaB$(dashed) as functions of $\gn$.  Note that for $\gn \leq 0.7$, we have $\zaA \sim \zaB$. We stress the striking similarity in the behaviour of $\za$ compared to $\zaeq$ (see Fig. \ref{fig:gammaA:zeq}). \\
  \textbf{Lower plots: }\textit{Left: }Density plot of $\zrhoA$ in the $\gn-\mtl$ plane. 
    \textit{Middle: } $\zrhoA$(solid) and $\zrhoB$(dashed) as functions of $\mtl$. 
    \textit{Right: } $\zrhoA$(solid) and $\zrhoB$(dashed) as functions of $\gn$. 
    }
  \label{fig:density:za:1}
\end{figure}

\begin{figure}[h!]
    \centering
    \includegraphics[width=\textwidth]{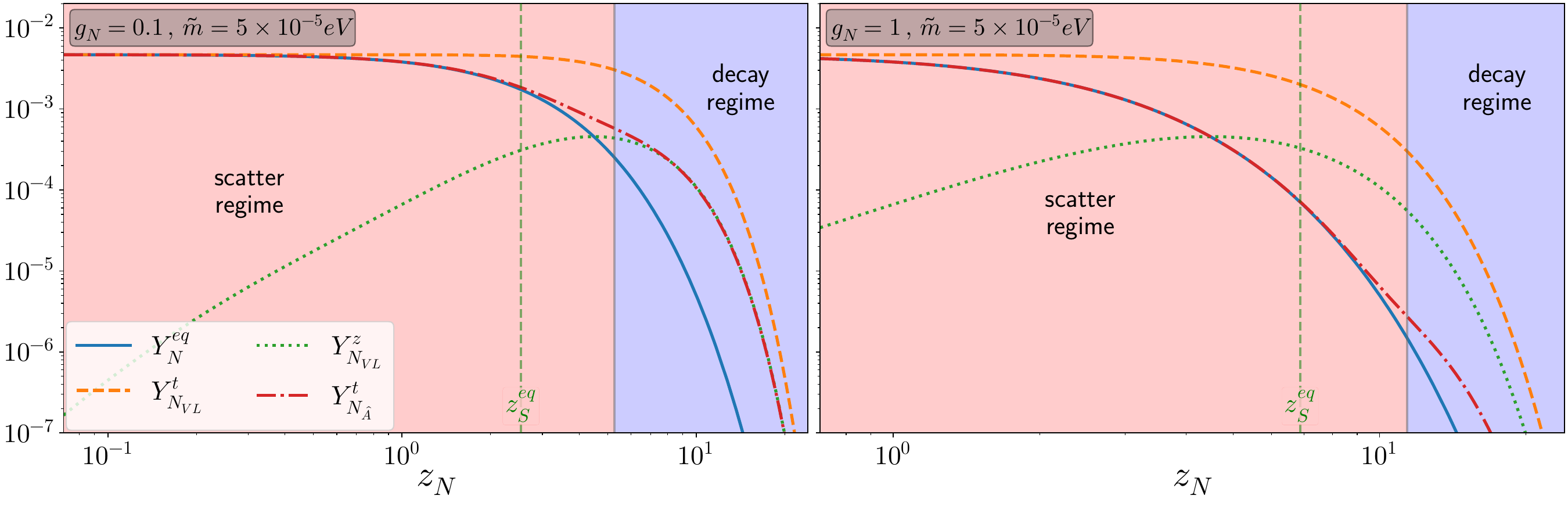}
    \caption{Evolution of $\ynthA(\zn)$ and  $\yntzVL(\zn)$, compared to $Y_{N}^{eq}$. The red region denotes the scatter regime where the neutrino evolution in the majoron+triplet model is dominated by scattering processes while the blue region denotes the decay regime where (inverse) neutrino decays are dominant. The vertical line denotes $\zaeq$ where the scattering processes decouple. 
    Note that the lower limits of the $\zn$ range are given by $\zi$ and hence depend on $\gn$.  }
    \label{fig:yPlots:ZA:1:1:1:194}
\end{figure}

Next, we will compare the behaviour of $\gammaa$ with respect to $\gamma_D$ in more detail.
In Fig. \ref{fig:gammaBSM_N}, we show $\sfrac{\gamma_{S,D,Q}}{n_N^{eq}H}$ which are the relevant quantities for the neutrino evolution \eqref{eq:dyn_ad}. \\
\noindent
It is apparent that $\sfrac{\gammaa}{n_N^{eq}H}$ initially significantly exceeds $\sfrac{\gamma_D}{n_N^{eq}H}$ until $\sfrac{\gammaa}{n_N^{eq}H}$ drops quickly due to Boltzmann suppression setting in. As $\sfrac{\gamma_D}{n_N^{eq}H}$ is not Boltzmann suppressed, it eventually exceeds $\sfrac{\gammaa}{n_N^{eq}H}$. On the other hand, quark scatterings never dominate the neutrino evolution. Thus, neutrinos initially scatter with $\latexchi, \sigma$ and $J$ before they decay until these scattering processes become inefficient and neutrinos can freely decay via $\gamma_D$. 
Consequently, we can distinguish between two regions: In the \textit{scatter} regime, scattering processes dominate the neutrino evolution while in the \textit{decay} regime, the neutrino  evolution is driven by (inverse) neutrino decays. 
From \eqref{eq:dyn_ad}, one can deduce that scattering processes dominate the neutrino evolution as long the scatter-term exceeds the decay-term, i.e.
\begin{align}
 4\gammaa > \gamma_D \,.\, \label{eq:cond_ad}
\end{align} 
Thus, we can make the approximation 
\begin{align}
s H \zn  \ddd{\hat Y_{N}} \approx  \begin{cases}   
s H \zn  \ddd{\hat Y^{scatter}_{N}} = -\left( \hat\delta_{N}-1 \right) 4\gammaa\,, &\quad \zn  \lesssim \za\,,\\
s H \zn  \ddd{\hat Y^{decay}_{N}} = -\left( \hat\delta_{N}-1 \right) \gamma_D\,, &\quad \zn \gtrsim \za\,,
\end{cases} \label{eq:za:def}
\end{align}
where $\za(\mtl, \gn)$ is defined via 
\begin{align}
 4\gammaa(\zn = \za, \mtl) = \gamma_D(\zn = \za, \gn)\,.
\end{align} 
Note that while $\gammaa$ is independent from $\mtl$, we have $\gamma_D \sim \mtl$ and conclusively, we expect that $\za$ increases with $\mtl$. On the other hand, $\gamma_D$ is independent of $\gn$ and the dependence of $\za$ on $\gn$ is dictated by $\gammaa$. If $\sfrac{\gammaa}{n_N^{eq}H}$ is already Boltzmann suppressed and consequently drops quickly around $\za$, we expect that $\za$ depends similarly on $\gn$ than $\zaeq$. In Fig. \ref{fig:density:za:1}(upper panels), we show a density plot of $\zaA(\gn, \mtl)$(left) and $\za(\mtl)$(middle) and $\za(\gn)$(right) for exemplary values of $\gn$ amd $\mtl$, respectively, which clearly display these dependencies. However, recalling the previous discussion, note that scattering processes keep neutrinos close to thermal equlibrium only for $\zn < \zaeq$. 
If $\za \gg \zaeq$, this implies that for $\zaeq < \zn < \za$ , scattering processes dominate the neutrino evolution over decays but are not fast enough to ensure that neutrinos stay in thermal equilibrium. As a result, the neutrino abundance at $\za$ exceeds the thermal abundance at $\za$. On the other hand, $\za \lesssim \zaeq$ results in a thermal neutrino abundance at $\za$. 
This behaviour is also highlighted in Fig. \ref{fig:yPlots:ZA:1:1:1:194} where we show $\ynhA(\zn)$ and $\ynVL(\zn)$ for $\gn = 0.1$(left) and $\gn = 1$(right). It is apparent that $\ynthA(\zn)$ follows $\yneq(\zn)$ closely until scattering processes fall out of thermal equilibrium around $\zaeq$ while $\yntVL(\zn)$ deviates from thermal equilibrium significantly stronger and already for smaller $\zn$. 
Note that for $\gn=0.1$, we have $\ynhA \approx\ynzVL$ around $\za$ so that $\ynhA$ approaches $\ynzVL$ in the decay regime, indicating that the neutrino evolution in both cases is determined by $\gamma_D$. In the right panel with $\gn=1$, the neutrino abundance in case $\hat A$ at the transition from the scatter to the decay regime is significantly smaller than $\ynzVL$ and thus $\ynhA$ does not approach $\ynzVL$ in the $\zn$ range relevant for leptogenesis. 
 \begin{figure}[!htbp]
  \centering
  \includegraphics[width=\textwidth]{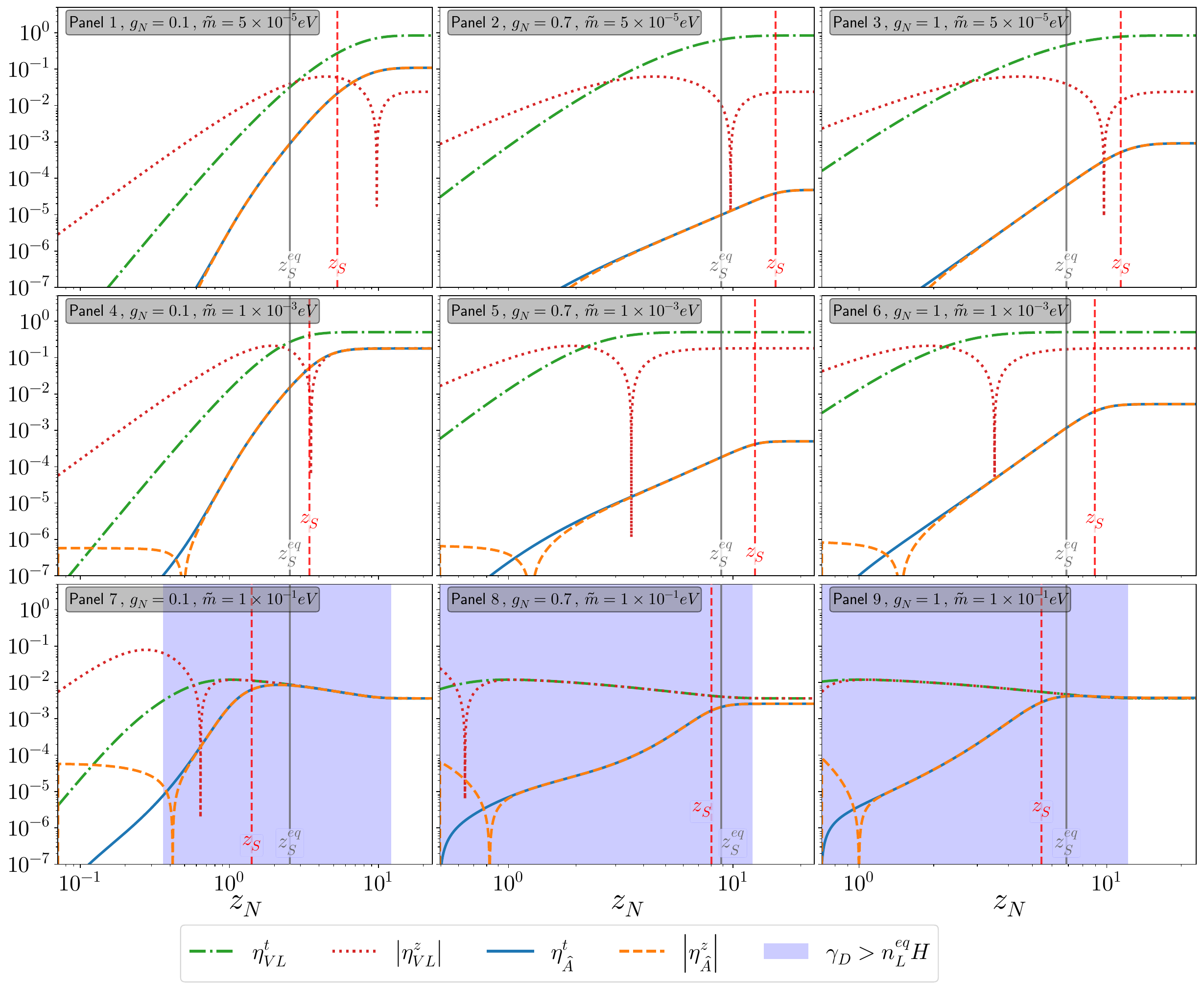}
    \caption{Evolution of the efficiencies $\etaVLtz$ and $\etahAtz$ for $\gn = [0.1, 0.7, 1]$ and $\mtl = [\num{5e-5}, \num{e-3}, \num{e-1}]\si{\electronvolt}$. The blue region in panels 7--9 indicates where inverse neutrino decays are in thermal equilibrium, the solid vertical line denotes $\zaeq$ and the dashed vertical line denotes $\za$.
    Note that the lower limits of the $\zn$ range are given by $\zi$ and hence depend on $\gn$.  
    As the efficiency in VL does not depend on $\gn$, $\etaVLtz$ are the same in each column. \\
    The efficiencies $\etahAz(\zn)$ and $\etaVLz(\zn)$ are initially negative and change sign at $\zn^{\pm}$ and $z_{VL}^{\pm}$ (located at the striking singularities), respectively. 
    In VL, neutrino production proceeds only via lepton number violating inverse neutrino decays and quark scatterings while in case $\hat A$, the scattering processes in $\gammaa$ are significantly more effective at populating the plasma with neutrinos, resulting in $\mathrm{Max}\left(|\etaVLz(\zn < z_{VL}^{\pm})|\right) \gg \mathrm{Max}\left(|\etahAz(\zn < \zn^{\pm})|\right)$. As $\mtl$ increases, neutrino production via inverse decays and quark scatterings becomes more effective, hence $\mathrm{Max}\left(|\etahAz(\zn < \zn^{\pm})|\right)$ increases with $\mtl$ (see e.g.\ panels $1$, $4$ and $5$) while scattering processes with $\sigma, J$ and $\latexchi$ are more effective as $\gn$ increases, i.e.\ $\mathrm{Max}\left(|\etahAz(\zn < \zn^{\pm})|\right)$ decreases with $\gn$ (see e.g.\ panels 4--6).
    Comparing $\etahAt$ and $\etaVLt$, the suppression of $\etahAt$ while $\zn < \za$ is striking. In particular, even if inverse decays are in thermal equilibrium (panels 7--9), they are mostly ineffective at diminishing $\etaVLt$ until the suppression is lifted at $\za$. 
    }
  \label{fig:yPlots:1:1:63:162:1:194paper}
\end{figure}
\subsubsection{Efficiency}
\begin{figure}[h!] 
  \centering
   \includegraphics[width=0.5\textwidth]{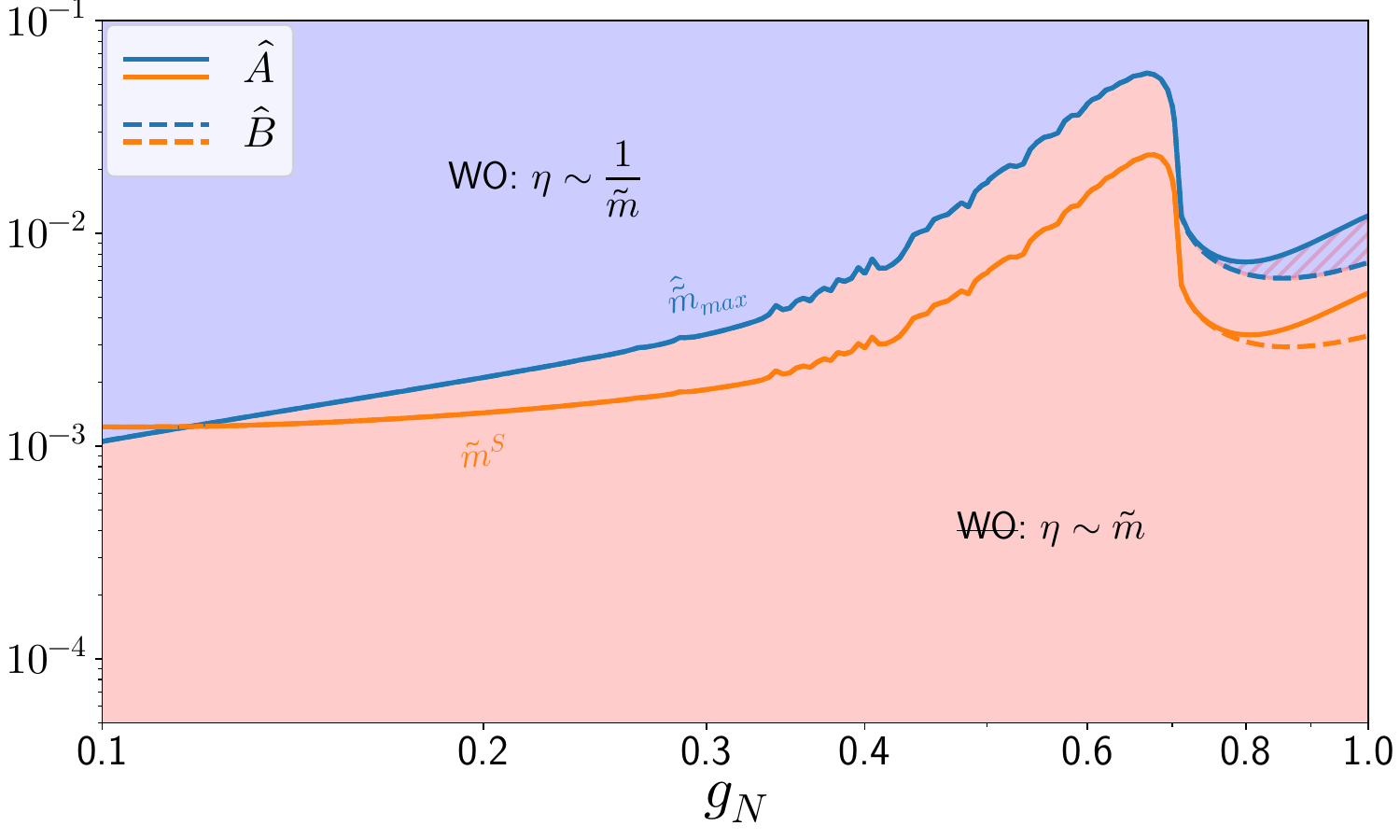}
  \caption{
  Comparison of $\mta$ and $\mtmaxap$ for case $\hat A$ (solid) and case $\hat B$ (dashed). In the red region, the final efficiency is determined by scattering processes and neutrino decays and thus proportional to $\mtl$. On the other hand, the final efficiency in the blue regions is determined by washout processes, resulting in $\etahAB \sim \sfrac{1}{\mtl}$. The hatched region corresponds to the WO regime in case $\hat A$ and to the \st{WO} regime in case $\hat B$. \\
  We can easily see how the strong suppression via scattering processes around $\gn \sim 0.7$ translates to peaks in $\mta$ and $\mtmaxap$. With the suppresion in case $\hat B$ being slightly weaker for $\gn> 0.7$, we also find that this results in $\mtaA\gtrsim \mtaB$ and $\mtmaxapA\gtrsim\mtmaxapB$. We also note that although $\mta$ slightly underestimates $\mtmaxap$, it does display the same qualitative features and thus is in reasonably well agreement with $\mtmaxap$. 
  }
  \label{fig:mtildegplots:cases:1:11:approx}
\end{figure}

Let us now discuss the evolution of the efficiency. To that end, we show $\etahA^{t,z}(\zn)$ and $\etaVL^{t,z}(\zn)$ for $\gn = [0.1, 0.7, 1]$ and $\mtl = [\num{5e-5}, \num{e-3}, \num{e-1}]\si{\electronvolt}$ in Fig. \ref{fig:yPlots:1:1:63:162:1:194paper}.\\
Let us first discuss the case of a vanishing initial neutrino abundance. 
In VL, it is well known that at first, a negative efficiency is produced via lepton number violating inverse decays and quark scatterings until neutrinos are thermalized at $\zvleq$ so that 
\begin{align}
 \ynzVL(\zn = \zvleq) = \yneq(\zn = \zvleq)\,. 
\end{align}
For $\zn > \zvleq$, a positive efficiency is produced and the final efficiency can be written as \cite{Buchmuller:2004nz}
\begin{align}
  \etaVLz = \eta^-_{VL}\big|_0^{\zvleq} + \eta^{+}_{VL}\big|_{\zvleq}^\infty\,.
\end{align}
As $\eta^+_{VL} > |\eta^-_{VL}|$, $\etaVLz(\zn)$ eventually changes sign at some $z_{VL}^{\pm}$ so that the final efficiency $\etaVLz$ is positive.  
In case $\hat A$ in the majoron+triplet model on the other hand, neutrinos are thermalized mainly via the scattering processes in $\gammaa$ and only a fraction of neutrinos is produced via lepton number violating inverse decays and quark scatterings. Moreover, neutrinos are thermalized so quickly that inverse decays can only create a negative efficiency around $\zn \gtrsim \zi$ while for $\zn > \zn^{\pm}\gtrsim \zi$, the efficiency $\etahAz(\zn)$ is positive. Thus, the negative efficiency created around $\zi$ is significantly less sizable compared to the negative efficiency created in VL and  $\etahA^z(\zn)$ approaches $\etahAt(\zn)$ already for small $\zn$ (see Fig. \ref{fig:yPlots:1:1:63:162:1:194paper}). \footnote{From the panels $8$ and $9$ in Fig. \ref{fig:yPlots:1:1:63:162:1:194paper}, we can deduce that for these parameters sets, $\etaVLz$ changes sign for smaller $\zn$ compared to case $\etahAz$. We stress however that this is merely due to the different $\zi$ used in the Boltzmann equations: Recall that in VL, we used $\zi^{VL} = 0$ while in the majoron+triplet model, leptogenesis can only take place for $\zneq \geq\zi$. In other words, $\etaVLz$ changes sign for smaller $\zn$ compared to $\etahAz$ as the leptogenesis era begins earlier.} Hence, the initial conditions are never relevant and for the remainder of this section, we will focus on $\etahAt(\zn)$. The same considerations apply for case $\hat B$. \\
\begin{figure}[h!]
  \centering
  \includegraphics[width=0.5\textwidth]{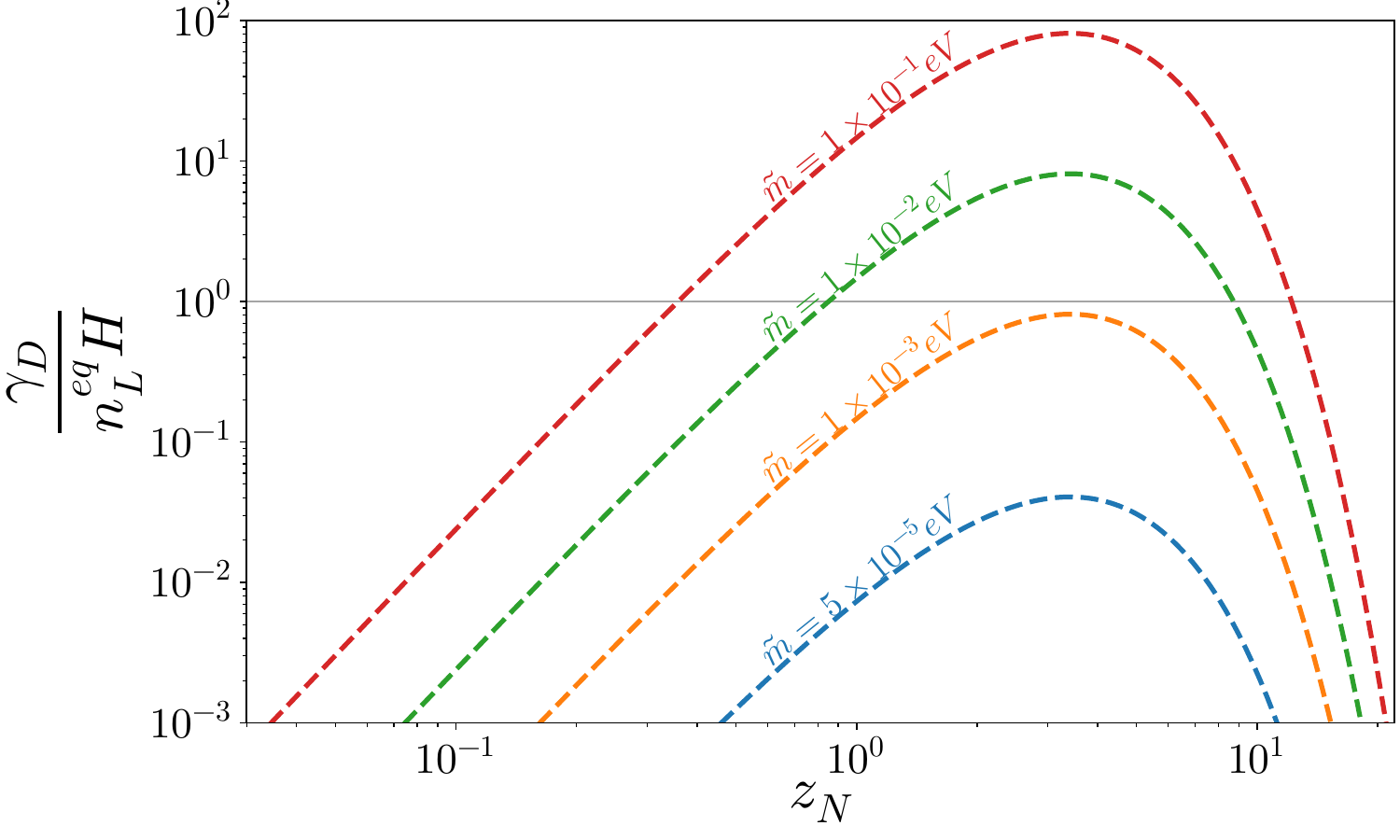}
  \caption{Thermal rate $\sfrac{\gamma_D}{n_L^{eq}H}$ relevant for \eqref{eq:eta}. We can easily see that inverse decays are thermalized only for $\mtl > \SI{e-3}{\electronvolt}$. }
  \label{fig:gammaL}
\end{figure}
\noindent
As can be seen in Fig. \ref{fig:gammaL}, the inverse decays $\sfrac{\gamma_D}{n_L^{eq}H}$ that are relevant for washout of the efficiency are Boltzmann suppressed and additionally, they are never thermalized when $\mtl$ is small. In this case, washout processes are not effective and we can write the Boltzmann equation for the efficiency \eqref{eq:eta} as 
\begin{align}
    s H \zn\ddd{\hat\eta} &= \fnorm\left(\hat\delta_{N}-1 \right)\gamma_D\,. \label{eq:eta:hatA}
\end{align}
In order to examine the evolution of the efficiency in the scatter and decay regimes, we insert \eqref{eq:za:def} in \eqref{eq:eta:hatA}, yielding
\begin{align}
\ddd{\hat \eta} \approx  \begin{cases}   
- \fnorm \frac{\gamma_D}{4\gammaa} \ddd{\hat Y_{N}}\,, &\quad \zn  \lesssim \za\,,\\
-\fnorm \ddd{\hat Y_{N}}\,, &\quad \zn \gtrsim \za\,,
\end{cases} \label{eq:eta:hat:approx}
\end{align}
and  the final efficiency can be written as 
\begin{align}
  \hat\eta(\zn\to\infty) &= \fnorm\left( \hat Y_{N}(\za)-\int_{z_I}^{\za} \mathrm{d}z' \frac{\gamma_D}{4\gammaa} \ddd{\hat Y_{N}} \right) \,. \label{eq:l_a} 
\end{align}
It is apparent that the efficiency consists of two parts: 
For $\zn < \za$, the efficiency is suppressed by $4\gammaa$ as neutrinos dominantly scatter while decays are ineffective. 
On the other hand, for $\zn > \za$, the suppression is lifted as neutrinos dominantly decay via $\gamma_D$ and thus the main portion of the efficiency is dictated by the neutrino abundance around $\za$. However, the efficiency created around $\za$ is still indirectly suppressed as the scattering processes diminish the neutrino abundance that is left to decay at $\za$. 
We can therefore deduce that the final efficiency evolves similarly to $\sfrac{1}{\za}$: 
For increasing $\za$, $\hat\eta(\zn)$ is suppressed over a larger range and as the neutrino abundance decreases with $\zn$, it reduces the neutrino abundance at $\za$. 
Recalling our previous discussion, this implies that $\hat\eta(\mtl)$ increases with $\sfrac{1}{\za(\mtl)}\sim\mtl$ while $\hat\eta(\gn)$ displays a similar behaviour as $\sfrac{1}{\za(\gn)}\sim\sfrac{1}{\zaeq(\gn)}$, resulting in a dip in the efficiency around $\gn\sim 0.7$. This behaviour is striking when comparing Fig. \ref{fig:density:hatA:hatB} with Fig. \ref{fig:density:za:1} (top panels).
In particular, we stress how $\zaB < \zaA$ for $\gn > 0.7$ translates to $\etahB > \etahA$. \\
This picture changes drastically when washout effects become relevant for larger $\mtl$. 
In VL, $\eta_{VL}(\zn)$ is diminished by inverse decays and quark scatterings soon after they reach thermal equilibrium, resulting in $\eta_{VL} \sim \sfrac{1}{\mtl}$ for $\mtl \gtrsim \SI{e-3}{\electronvolt}$. 
Due to the additional scattering processes however, inverse decays being thermalized does not necessarily imply that washout processes reduce the efficiency $\hat\eta$.  \\
Instead, recall that $\hat\eta$ is suppressed as long as $\zn < \za$. 
As the washout terms are proportional to the efficiency, this implies that washout is mostly ineffective even in thermal equilibrium if $\zn < \za$ and only relevant once a sizable efficiency is created for $\zn > \za$. 
More precisely, if inverse decays are in thermal equilibrium, i.e.\ $\sfrac{\gamma_D}{n_L^{eq}H} >1$ in the range $z_{D_1} < \zn <z_{D_2}$, the region where inverse decays can have a sizable effect on the efficiency is reduced to $\mathrm{Max}[z_{D_1}, \za] < \zn <z_{D_2}$ and may even completely vanish if $\za > z_{D_2}$. 
This can easily be seen in Fig. \ref{fig:yPlots:1:1:63:162:1:194paper}(panels 7--9) where the blue region indicates where inverse decays are in thermal equilibrium.  
It is apparent that $\hat\eta$, unlike $\eta_{VL}$, is affected by WO effects only for $\zn \gtrsim \za$ even if inverse decays are in thermal equilibrium.
Consequently, for a given $\gn$, we can find the smallest $\mtl=:\mta$ for which the interval  $[\mathrm{Max}[z_{D_1}, \za], z_{D_2}]$ is not empty as
\begin{align}
  \frac{\gamma_D}{n_L^{eq}H}(\mta, \za) =1\,. \label{eq:maza}
\end{align}
This implies that WO processes are relevant only for $\mtl > \mta$, resulting in $\eta_{\hat A} \sim \sfrac{1}{\mtl}$ ("WO regime"), while for $\mtl < \mta$, WO processes can be neglected and therefore $\hat\eta\sim \mtl$ ("\st{WO} regime"). 
Thus, $\mta$ seperates the \st{WO} regime where the efficiency depends strongly on the additional scattering processes from the WO regime where leptogenesis in the majoron+triplet model proceeds very similar to the VL scenario with initially thermalized neutrinos. 
We stress however that this is only a rough lower bound since the WO term is generally Boltzmann suppressed for $\zn > \za$ and thus for $\mtl \gtrsim \mta$, it is thermalized too briefly to have a notable effect, i.e.\ we expect that WO becomes effective for $\mtl$ slightly larger than $\mta$. Moreover, we neglected the effect of quark scatterings in this definition of $\mta$.\footnote{This treatment is similar to the discussions in \cite{AristizabalSierra:2010mv,Hambye:2012fh} regarding type III leptogenesis. We note however that the results differ fundamentally due to the different particle content of the models.} \\
Nevertheless, these considerations imply that $\hat\eta(\mtl)$ reaches a maximum around $\mtl \gtrsim \mta$. From the definition of $\mta$ it is also clear that $\mta$ depends similarly on $\gn$ as $\za$. Using our results for $\etahAB(\gn, \mtl)$, we further determine $\mtl =: \mtmaxapAB$ where $\etahAB(\gn, \mtl)$ actually become maximal. In Fig. \ref{fig:mtildegplots:cases:1:11:approx}, we show $\mta$ and $\mtmaxap$ and we find indeed $\mtmaxap \gtrsim \mta \sim \za$. As expected from the previous discussion, we also find $\mtaA = \mtaB$ for $\gn < 0.7$ amd $\mtaA > \mtaB$ for $\gn >0.7$. 
Thus, the suppression due to the scattering processes significantly diminishes the efficiency which in turn severly restricts the $\mtl$ range in which WO processes have an effect on the efficiency compared to VL. This is especially striking for $\gn \sim 0.7$ where the suppression is so strong that $\etahAB < \eta^{WO}$ and WO barely has an effect. \\
To summarize, we find that similarly to type III leptogenesis, the initial neutrino abundance is irrelevant as neutrinos are rapidly thermalized due to the presence of the new scattering processes. However, we find that over a significant parameter space, scattering processes not only effectively thermalize the neutrinos but also stop them from decaying up to late $\zn$, thereby preventing the creation of a sizable efficiency. 
In the next section, we will explore how the efficiency is affected when we drop the assumtion that $\sigma, J, T$ are in thermal equlibrium. \\
 
\subsection{Case $A$: $\dall \neq 1\,, \geta = \ls = 1$}
In this section, we discuss the efficiency for case $A$ in more detail.
Compared to the simplified scenarios $\hat A, \hat B$, the Boltzmann equations are significantly more complex and depend delicately on the effectiveness with which $N,\sigma,J$ and $\latexchi$ are thermalized. For example, $\delta_{\latexchi} > \delta_N$ implies that $\gat{NN\latexchi\latexchi}$ enhances the neutrino abundance while $\delta_{\latexchi} < \delta_N$ implies that the neutrino abundance is diminished. 
In the following, we therefore only highlight some fundamental features that have notable effects on the neutrino evolution and stress that the details are highly non-trivial.

\subsubsection{Thermal Rates}
\begin{figure}[h!]
    \centering
    \includegraphics[width=\textwidth]{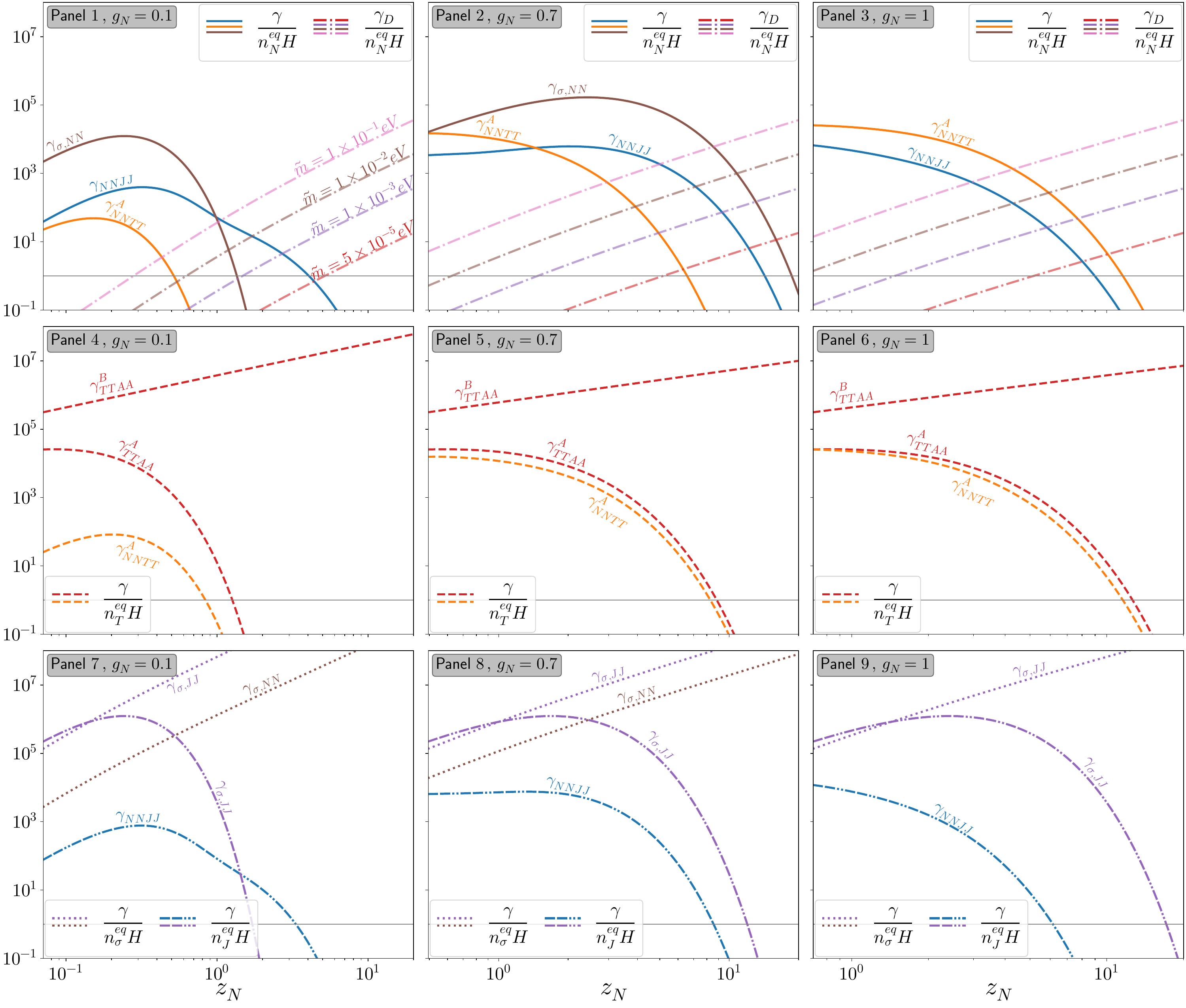}
    \caption{Thermal rates relevant for the evolution of $N$ \eqref{eq:ben} (panels 1--3), $\latexchi$ \eqref{be:chi} (panels 4--6), $\sigma$ \eqref{be:s}(panels 7--9) and $J$ \eqref{be:j}(panels 7--9) for $\gn = 0.1$ (left), $\gn=0.7$ (middle) and $\gn = 1$ (right). 
    The thermal rates without $\latexchi$ in the initial or final state are indentical for cases $A$ and $B$. Note that the lower limits of the $\zn$ range are given by $\zi$ and hence depend on $\gn$. See text for details.  
    }
    \label{fig:gammai}
\end{figure}

In Fig. \ref{fig:gammai}, we show a subset of the thermal rates appearing in the Boltzmann equations \eqref{eq:ben}-\eqref{be:j}, normalized to the respective number density, for $\gn \in [0.1,0.7,1]$. For clarity, we do not present all thermal rates but focus in these that are the most relevant for the overall perspective. 
For example, $\latexchi$ interacts dominantly via gauge scatterings $\gat{\latexchi\latexchi AA}^A$ and even if $Y_{\latexchi_A}(\zi) = 0$, these gauge scatterings instantly thermalize $\latexchi$. They are however Boltzmann suppressed with 
\begin{align}
  \frac{\gat{\latexchi\latexchi AA}^A}{n_\latexchi^{eq}} \sim \left(\frac{\gn}{\zn}\right)^\frac{3}{2} \exp^{-\frac{\zn}{\gn}}
\end{align}
and consequently, $\latexchi$ freezes out at $z_\latexchi = \sfrac{\zn}{\gn} \sim 10$ (see Fig. \ref{fig:yPlots:1:1:1:162:194abundancies}). Clearly, $\latexchi$ freezes out at larger $\zn$ as $\gn$ increases. \\
At $\zn \sim \zi$, interactions between $N, \sigma, J$ and $\latexchi$ are so strong that $N, \sigma$ and $J$ are instantly thermalized as well. Thus, as in the simplified case, we find that the initial abundances of $N, \sigma, J$ and $\latexchi$ are irrelevant, i.e.\ $Y_{{(N,\sigma,\latexchi,J)}_A}^z(\zn>\zi) \sim Y_{{(N,\sigma,\latexchi,J)}_A}^t(\zn>\zi)$.  
Moreover,  $J$ interacts dominantly via $\sfrac{\gat{\sigma,JJ}}{n_J^{eq}H}$ and $\sfrac{\gat{NNJJ}}{n_J^{eq}H}$ which are Boltzmann suppressed as 
\begin{align}
  \frac{\gat{\sigma,JJ}}{n_J^{eq}} &\sim \frac{n_\sigma^{eq}}{n_J^{eq}} \sim \left(\frac{\ms \zn}{\mn}\right)^\frac{3}{2} \exp^{-\frac{\sqrt{2}}{\gn}\zn}\,,\\
  \frac{\gat{NNJJ}}{n_J^{eq}} &\sim \frac{{n_N^{eq}}^2}{n_J^{eq}}\sim
    \begin{cases}
        \left(\frac{\mn}{\zn}\right)^3\,,\quad &\zn < 1\,,\\
        \mn^3\exp^{-2\zn} \,,\quad &\zn > 1\,,
    \end{cases}
\end{align}
i.e.\ the respective Boltzmann suppressions are damped as $\gn$ grows and $J$ freezes out at larger $\zn$. As already discussed in Sec. \ref{sec:approximation}, the suppression of the scattering processes dominating the neutrino evolution is similarly less pronounced for larger $\gn$. 
On the other hand, the evolution of $\sigma$ is determined by the decays $\sfrac{\gat{\sigma,JJ}}{n_\sigma^{eq}},\sfrac{\gat{\sigma,NN}}{n_\sigma^{eq}}$ which are not Boltzmann suppressed. This has drastic consequences for the evolution of $N$ as we demonstrate in the following. 

\subsubsection{Abundances}
\begin{figure}[!htbp]
  \centering
  \includegraphics[width=\textwidth]{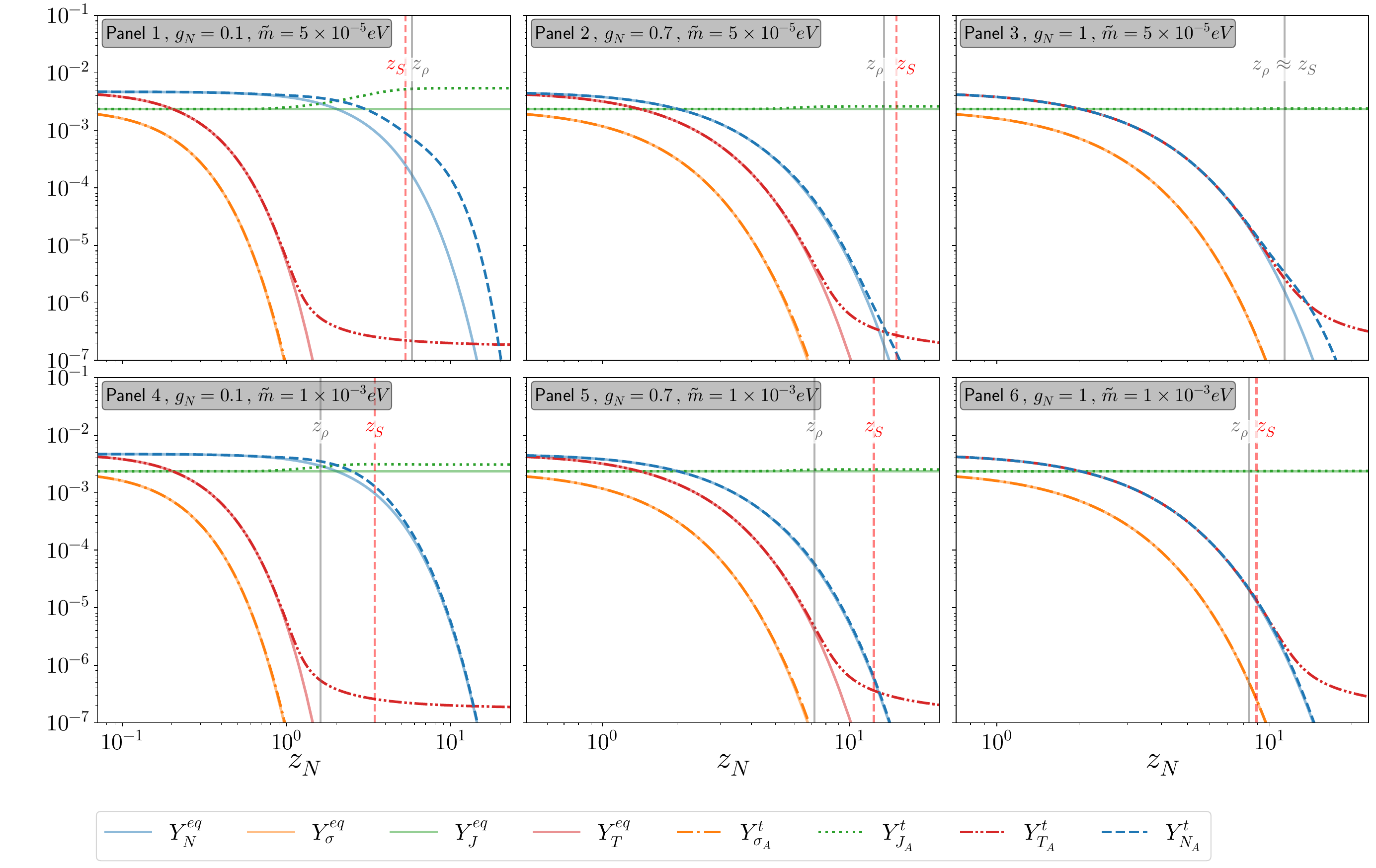}
  \caption{
  Evolution of the abundances of $N, \sigma, \latexchi$ and $J$ for $\gn = [0.1,0.7,1]$ and $\mtl = [\num{5e-5}, \num{e-2}]\si{\electronvolt}$. In the right plot, we have $\gn = \geta$ and therefore $Y_{N_A}^{eq} = Y_\latexchi^{eq}$. Note that the lower limits of the $\zn$ range are given by $\zi$ and hence depend on $\gn$. \\
  Scattering processes and inverse decays initially keep $N, \latexchi, \sigma$ and $J$ close to thermal equilibrium. 
  While $\sigma$ never deviates significantly from thermal equilibrium, $\latexchi$ freezes out and eventually obtains a constant abundance once gauge interactions decouple at $z_\latexchi = \sfrac{\zn}{\gn} \sim 10$, independtly from $\mtl$. The evolutions of both neutrinos and majorons on the other hand are significantly affected by $\gn$ and $\mtl$ and if $\gn$ and $\mtl$ are small, their abundances deviate significantly from thermal equilibrium. In particular, $J$ freezes out once interactions with $\sigma$ and $N$ decouple and for e.g.\ $\gn = 0.1$ and $\mtl = \SI{5e-5}{\electronvolt}$ (panel 1), the majoron abundance after freeze out notably exceeds the corresponding thermal abundance. 
  }
  \label{fig:yPlots:1:1:1:162:194abundancies}
\end{figure}

Similarly to the discussion in Sec. \ref{sec:approximation}, let us first distinguish between a scattering regime where the neutrino evolution is determined by scattering processes and a decay regime were neutrino decays are the dominant processes, i.e.\ 
\begin{align}
  sH\zn\ddd{Y_{N_A}} \approx sH\zn
  \begin{cases} 
    \ddd{Y_{N_A}^{scatter}} = -2(\delta_{N_A}^2-1)\gamma_S-2\rho_A \,,\quad &\zn < \zrhoA\,, \label{eq:scatter}\\
    \ddd{Y_{N_A}^{decay}} = -(\delta_{N_A}-1)\gamma_D \,,\quad &\zn > \zrhoA \,,
  \end{cases}
\end{align} 
where at $\zrhoA$ the transistion from scatter to decay regime occurs and we neglected quark scatterings. Note the similarities to \eqref{eq:za:def}. 
In the scatter regime, the first term proportional to $\gamma_S$ in \eqref{eq:scatter} generally reduces the neutrino abundance as $\delta_{N_A}, \gamma_S > 1$. On the other hand, the sign of the $\rho_A$-term is not immediatly clear due to the substractions of on-shell scatterings (see \eqref{eq:rho}). We do however find that irrespective of $\gn$ and $\mtl$, $\rho_A(\zn)$ is always negative. 
Concerning the neutrino evolution, this implies that $-2\rho_A(\zn)$ gives a positive contribution to $\ddd{Y_{N_A}}(\zn)$, i.e.\ the deviations from thermal equilibrium of $\sigma, J$ and $\latexchi$ can enhance the neutrino abundance if $\rho_A(\zn)$ is sufficiently sizable. We stress however that for the most part, we have $\left[(\delta_{N_A}^2-1)\gamma_S\right](\zn) > \left|\rho_A(\zn)\right|$ so that $\ddd{Y_{N_A}}(\zn) < 0$, i.e.\ the neutrino abundance overall decreases. 
In the decay regime, it is apparent that the neutrino evolution is mainly determined by the same Boltzmann equation as in the simplified scenario and thus we expect that the difference between the neutrino evolution in both scenarios is mainly due to the differing neutrino abundances at $\za$ and $\zrhoA$, respectively. \\
\begin{figure}[h!]
    \includegraphics[width=\textwidth]{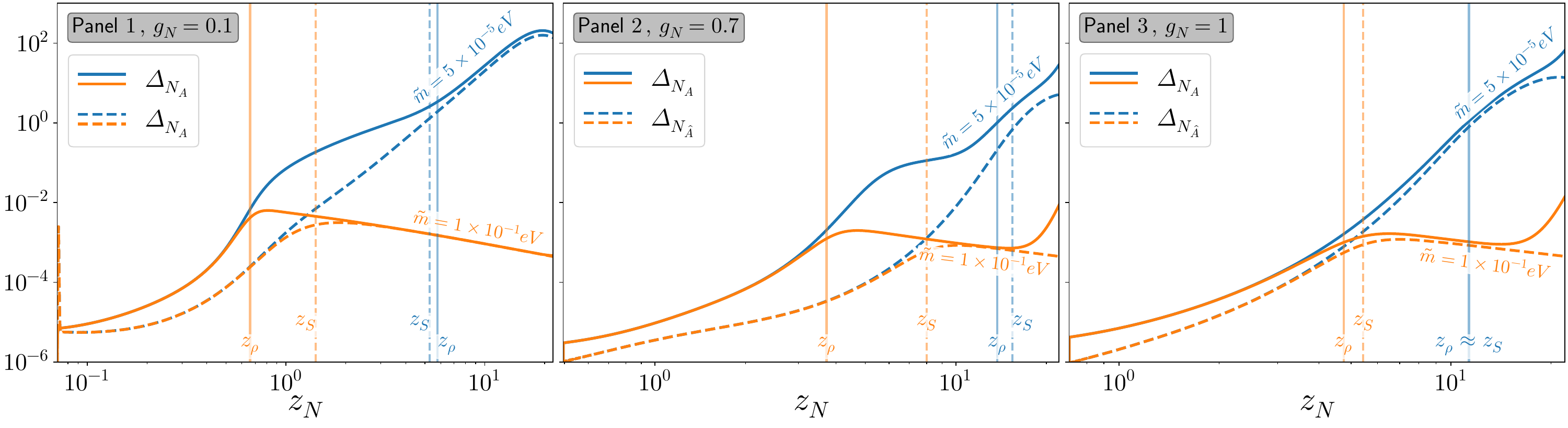}
    \caption{
    Evolution of $\Delta_{N_A} = \delta_{N_A}-1$ (solid lines) and $\Delta_{N_{\hat A}} = \delta_{N_{\hat A}}-1$ (dashed lines) for $\mtl = [\num{5e-5}, \num{e-1}]\si{\electronvolt}$ and $\gn = 0.1$ (panel $1$), $\gn=0.7$ (panel $2$), $\gn=1$ (panel $3$). The vertical solid lines indicate $\zrhoA$ while the vertical dashed lines denote $\zaA$. Note that the lower limits of the $\zn$ range are given by $\zi$ and hence depend on $\gn$. \\
    For $\zn < \zrhoA$ and $\zn < \zaA$, respectively, the neutrino evolution is determined by scattering processes with $\latexchi, \sigma$ and $J$ and thus $\Delta_{N_A}(\zn <\zrhoA)$ and $\Delta_{N_{\hat A}}(\zn <\zaA)$ are independent from $\mtl$. 
    In case $A$ for $\zn< \zrhoA, \zaA$, neutrinos deviate significantly more from thermal equilibrium compared to the simplified scenario $\hat A$, i.e.\ $\Delta_{N_A}(\zn <\zrhoA, \zaA)\gg\Delta_{N_{\hat A}}(\zn <\zrhoA, \zaA)$.
    In panel $1$, $\Delta_{N_A}(\zn)$ approaches $\Delta_{N_{\hat A}}(\zn)$ for $\zn > \zrhoA$ as the neutrino evolution in this regime is entirely driven by inverse neutrino decays. In panels $2$ and $3$, $\Delta_{N_A}(\zn)$ intially approaches $\Delta_{N_{\hat A}}(\zn)$ for $\zn > \zrhoA$ but eventually starts to increase again due to freeze out of $\latexchi$. 
    }
    \label{fig:deltaN}
\end{figure}
In order to examine the neutrino evolution in the scatter and decay regimes in more detail, it is convenient to parametrize the deviation from thermal equilibrium of a particle $i$ as 
\begin{align}
  \Delta_i \equiv \delta_i-1\,
\end{align}
and discuss the evolutions of $\Delta_{N_{A, \hat A}}(\zn)$ rather than $\ddd{Y_{N_{A, \hat A}}}(\zn)$ as shown in Fig. \ref{fig:deltaN} for $\gn = [0.1,0.7,1]$ and $\mtl = [\num{5e-5},\num{e-1}]\si{\electronvolt}$. 
Moreover, in Fig. \ref{fig:density:za:1}(lower panels), we show a density plot of $\zrhoA(\mtl, \gn)$ (left) and $\zrhoA(\mtl)$ (middle), $\zrhoA(\gn)$(right) for exemplary values of $\gn$ and $\mtl$, respectively. \\
First, let us discuss the effect of $\mtl$ on the neutrino evolution in the scatter regime. As the thermal rates that dominate in the scatter regime do not depend on $\mtl$, it is clear that  $\Delta_{N_A}(\zn)$ is independent from $\mtl$ as long as $\zn < \zrhoA$, see Fig. \ref{fig:deltaN}. Additionally, as $\gamma_D \sim \mtl$, we conclude that $\zrhoA$ decreases with $\mtl$, similarly to $\zaA$. Comparing $\zaA$ and $\zrhoA$, we find however that $\zrhoA$ decreases faster with $\mtl$ than $\zaA$ (see left panels in Fig. \ref{fig:density:za:1}) which can be traced back to effects of $\rho_A$ on the transition from the scatter to the decay regime. More precisely, the transition occurs when $\left|\ddd{Y_{N_A}^{scatter}}\right| = \left|\ddd{Y_{N_A}^{decay}}\right|$ which yields the condition
\begin{align}
  \frac{\gamma_D}{4\gamma_S} = 1+ \frac{\rho_A}{2\gamma_S (\delta_{N_A}-1)} + \epsilon\,,\quad \zn = \zrhoA\,, 
\end{align}
where we defined $\delta_{N_A} =: 1+ \epsilon$ with $\epsilon \ll 1$. For most of the parameter space, $|\frac{\rho_A}{2\gamma_S (\delta_{N_A}-1)}| \gg \epsilon$ holds and we can write the condition for the transition as 
\begin{align}
  \gamma_D\approx  4\gamma_S  +  \frac{2\rho_A}{ (\delta_{N_A}-1)} \lesssim 4\gamma_S\,,\quad \zn = \zrhoA\,.
\end{align}
This implies that $\zrhoA$ can easily be smaller than $\zaA$ and in particular, it results in $\zrhoA$ so small that the Boltzmann suppression of $\gammaa$ is not yet very effective, resulting in a significantly stronger depenence on $\mtl$ compared to $\zaA$. \\
The effects of $\gn$ on the neutrino evolution in the scatter regime are more evolved.
For $\gn \lesssim 0.7$, we find that $\Delta_{N_A}(\zn) \gg \Delta_{N_{\hat A}}(\zn)$, i.e.\ neutrinos deviate significantly more from thermal equilibrium compared to the simplified case (panels 1 and 2 in Fig. \ref{fig:deltaN}). While the details are highly non-trivial, two essential ingredients are that interactions of $N, \sigma$ and $J$ with $\latexchi$ are too weak to thermalize them effectively (recall that $\latexchi$ is efficiently thermalized via gauge interactions) while fast $\sigma$ decays create non-thermal $N$ and $J$ abundances as demonstrated in Fig. \ref{fig:yPlots:1:1:1:162:194abundancies}. Moreover, note that with increasing $\gn$, $\Delta_{N_A}(\zn)$ deviates from thermal equilibrium up to larger $\zn$, as one would expect due to the shift of the Boltzmann suppression to larger $\zn$.
For $\gn > 0.7$ on the other hand, interactions with $\latexchi$ are more efficient at thermalizing $N, \sigma$ and $J$. Additionally, $\sfrac{\gat{\sigma,JJ}}{n_JH}$ is fast up to larger $\zn$ compared to $\gn < 0.7$ while $\sigma \leftrightarrow NN$ is kinematically forbidden. 
As a consequence, the neutrino abundance is not significantly enhanced in the scatter regime and we have  $\Delta_{N_A}(\zn)\gtrsim\Delta_{N_{\hat A}}(\zn)$ for $\zn < \zrhoA$ (panel $3$ in Fig. \ref{fig:deltaN}). \\
Next, let us discuss the neutrino evolution in the decay regime. 
For $\gn < 0.7$, we find that irrespective of the value of $\mtl$, $\Delta_{N_A}(\zn)$ eventually approaches $\Delta_{N_{\hat A}}(\zn)$ for $\zn\gg\zrhoA,\zaA$, i.e.\ the neutrino evolution is determined only by $\gamma_D$ while scattering processes with the new particles are irrelevant, similarly to the simplified scenario (panel $1$ in Fig. \ref{fig:deltaN}). As discussed above, this is to be expected given that the evolution of the neutrinos in the decay regimes in cases $\hat A$ and $A$ is governed by the same Boltzmann equation.
This simple picture changes however for $\gn \geq 0.7$. We find that initially, $\Delta_{N_A}(\zn)$ still tracks $\Delta_{N_{\hat A}}(\zn)$, although $\Delta_{N_A}(\zn)$ slightly exceeds $\Delta_{N_{\hat A}}(\zn)$ even for $\mtl \to \SI{e-1}{\electronvolt}$. Eventually, $\Delta_{N_A}(\zn)$ begins to increase while $\Delta_{N_{\hat A}}(\zn)$ keeps decreasing as for $\gn<0.7$ (panels 2 and 3 in Fig. \ref{fig:deltaN}). This is especially striking for $\mtl \to \SI{e-1}{\electronvolt}$. The reason for this behaviour is that although decay processes dominate the neutrino evolution for $\gn \geq 0.7$, the neutrino abundance is slightly enhanced due to the freeze out of $\latexchi$: In the decay regime, neutrinos are thermalized mainly via $\gamma_D$ and as $\sfrac{\gat{\latexchi\latexchi AA}^A}{n_\latexchi^{eq}}$ drops, $\latexchi$ couples weaker to the plasma than the neutrinos, resulting in $\delta_{\latexchi_A}(\zn) > \delta_{N_A}(\zn)$ and consequently, the respective term in the Boltzmann equation for the neutrino evolution gives a positive contribution to the neutrino abundance. As a result, $\left|\ddd{Y_{N_A}}\right|$ drops slower than in a scenario where only decays are relevant, thereby enhancing $\Delta_{N_A}(\zn)$. Just before $\latexchi$ completely freezes out and decouples, $\delta_{\latexchi_A}(\zn)$ begins to increase significantly as $Y_\latexchi^{eq}(\zn)$ becomes Boltzmann suppressed while $Y_{\latexchi_A}(\zn)$ becomes constant, resulting in a short but striking enhancement of $\Delta_{N_A}(\zn)$. 
Thus, despite decay processes being dominant for $\zn > \zrhoA$, scattering processes are still relevant and affect the neutrino evolution. \\
To summarize, we find that neutrinos can deviate significantly more from thermal equilibrium compared to the simplified case as the (inverse) decays and neutrino scatterings involving $\sigma,J$ and $\latexchi$ introduce a source of non-equilibration which is redistributed amongst $N, \sigma, J$ and $\latexchi$. Although the distinction between a scatter and a decay regime is not entirely accurate for $\gn > 0.7$, it does grasp the overall qualitative features of the neutrino evolution. In the following, we will discuss how this affects the evolution of the efficiency. 

\subsubsection{Efficiency}
\begin{figure}[h!]
  \centering  
   \includegraphics[width=0.6\textwidth]{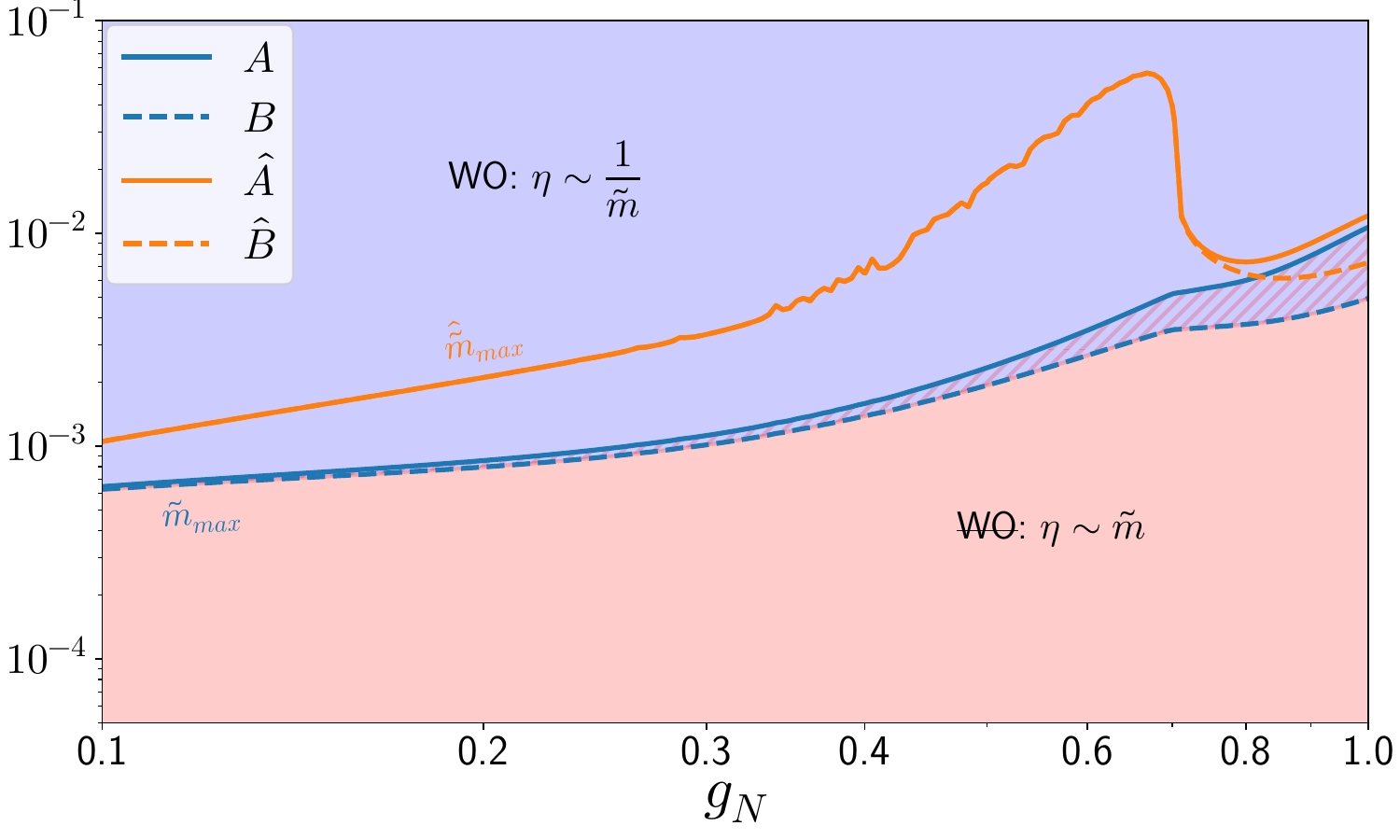}
  \caption{
  Comparison of $\mtmax$ and $\mtmaxap$ for case $\hat A, A$ (solid) and case $\hat B, B$ (dashed). In the red region, the final efficiency in cases $A,B$ is determined by scattering processes and neutrino decays and thus proportional to $\mtl$. On the other hand, the final efficiencyy in the blue regions is determined by washout processes, resulting in $\etaAB \sim \sfrac{1}{\mtl}$. The hatched region corresponds to the WO regime in case $A$ and to the \st{WO} regime in case $B$. \\
  We can immediatly see that the strong suppression of the efficiency in the simplified scenario that results in a peak around $\gn \sim 0.7$ in $\mtmaxap$ is absent in cases $A,B$. Addtionally, neutrinos are thermalized weaker in cases $A,B$ compared to the simplified scenario, translating to $\mtmax < \mtmaxap$. Even more precisely, in case $B$, the absence of interactions with $\latexchi$ renders the scattering processes less efficient in thermalizing the neutrinos compared to case $A$, resulting in $\mtmaxB < \mtmaxA$. 
   }
  \label{fig:mtildegplots:cases:1}
\end{figure}

\begin{figure}[!htbp]
  \centering
  \includegraphics[width=\textwidth]{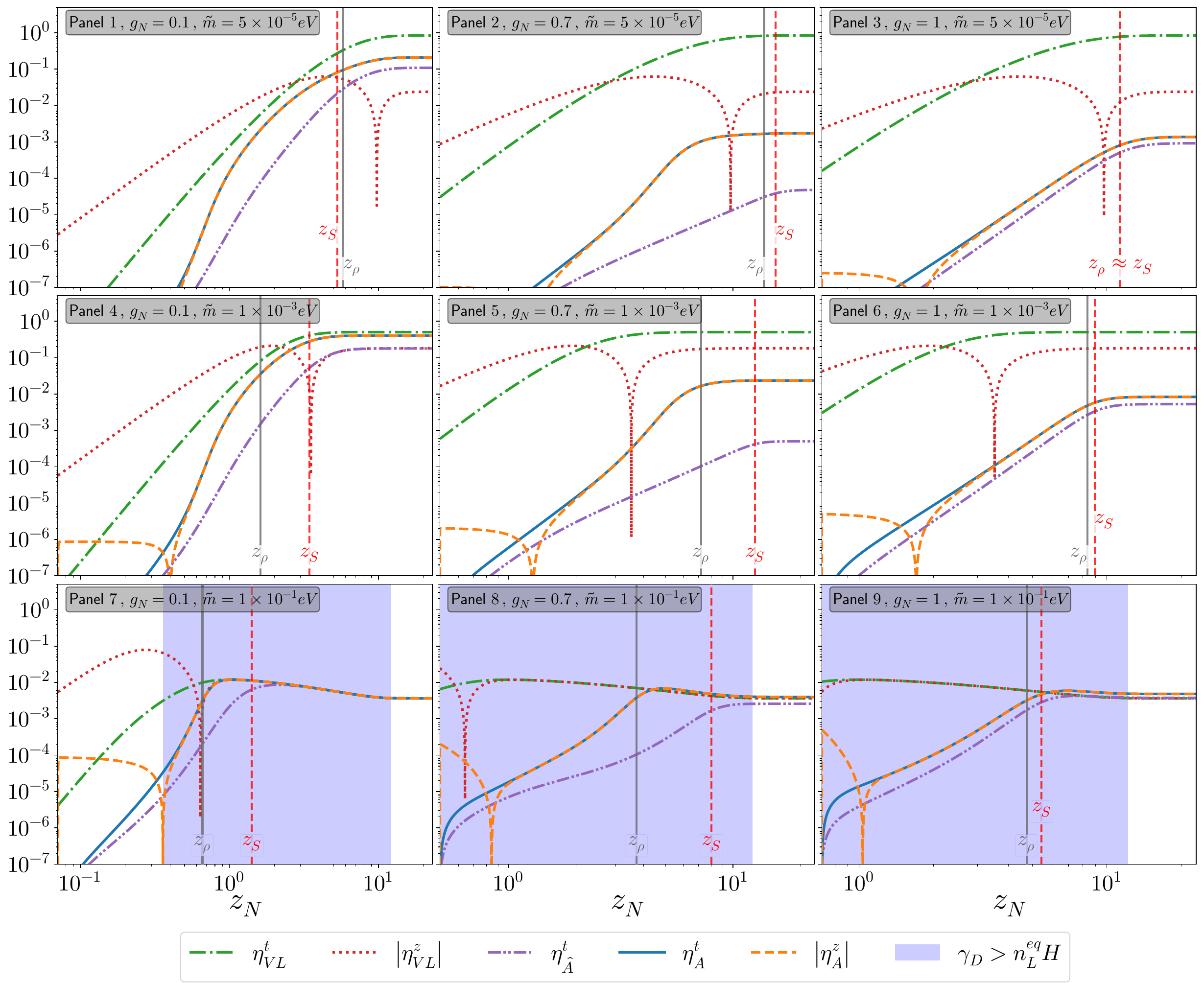}
  \caption{Efficiencies $\etaAtz$, $\etahAt$ and $\etaVLtz$ as functions of $\zn$. The blue region indicates where inverse decays are in thermal equilibrium, the solid vertical line denotes $\zrho$ and the dashed vertical line denotes $\za$.
    Note that the lower limits of the $\zn$ range are given by $\zi$ and hence depend on $\gn$.  As the efficiency in VL does not depend on $\gn$, $\etaVLtz$ are the same in each column. 
    In case $A$, neutrinos are coupled weaker to the plasma compared to case $\hat A$, resulting in a weaker suppression of $\etaAt$ compared to $\etahAt$. In panels 7--9, it is apparent that WO processes diminish $\etaA$ for smaller $\zn$ compared to $\etahA$. 
    As the initial abundances of $\latexchi, \sigma$ and $J$ vanish in case $A_z$ (in contrast to case $\hat A_z$ where they have thermal abundances), a larger portion of the neutrino production is via lepton number violating inverse decays and quark scatterings compared to case $\hat A$ (see Fig. \ref{fig:yPlots:1:1:63:162:1:194paper}). Hence, $\mathrm{Max}\left(|\etaAz(\zn < \zn^{\pm})|\right) \gg \mathrm{Max}\left(|\etahAz(\zn < \zn^{\pm})|\right)$. Nevertheless, $\etaAz$ eventually approaches $\etaAt$, irrespective of $\mtl$ and $\gn$. 
    }
  \label{fig:yPlots:1:1:63:162:1:162:194eta}
\end{figure}

In Fig. \ref{fig:yPlots:1:1:63:162:1:162:194eta}, we show $\eta_A(\zn)$ for exemplary values of $\mtl$ and $\gn$. 
Similarly to the simplified scenario, we find that in the case of vanishing initially abundances of $N, \sigma, \latexchi$ and $J$, the intially negative efficiency $\etaAz$ quickly changes sign as neutrinos are mainly produced via scattering processes involving $\sigma, \latexchi$ and $J$. However, neutrino production via lepton number violating inverse decays and quark scatterings plays a larger role in case $A$ compared to the simplified scenario as in the former case, the initial abundances of $\latexchi, \sigma$ and $J$ vanish as well. 
Nevertheless, $\etaAz$ approaches $\etaAt$ very and and for simplicity, we will focus the following discussion on $\etaAt$. \\
Let us at first focus on a regime where $\mtl$ is small so that WO can be neglected.
The Boltzmann equations in the scatter and decay regimes, respectively, can be then written as (neglecting quarks)
\begin{align}
  \ddd{\eta_A} = \frac{1}{sH\zn}(\delta_{N_A}-1)\gamma_D \fnorm\approx 
  \begin{cases} 
      -\fnorm \frac{\gamma_D}{4\gamma_S}\left(\ddd{Y_{N_A}} + \frac{2\rho_A}{sH\zn}\right) \,, \quad &\zn<\zrhoA\,, \\ 
     -\fnorm \ddd{Y_{N_A}}\,, \quad &\zn > \zrhoA\,,
  \end{cases} \label{eq:etaA:scatter:decay}
\end{align}
and we can write the final efficiency as
\begin{align}
  \eta_A(\infty) = \eta_A(\zrhoA) + \fnorm Y_{N_A}(\zrhoA)\,.
\end{align}
We can easily see that as in the simplified scenario, the efficiency in the scatter regime is suppressed by $\sfrac{\gamma_D}{4\gamma_S}$. 
However, neutrinos deviate significantly more from thermal equilibrium in the scatter regime compared to the simplified scenario which counteracts the suppression due to scattering processes. 
Thus, while the main portion of the efficiency in the simplified scenario is created at the transition from the scatter to the decay regime at $\zaA$, this picture changes drastically in case $A$ as the enhanced neutrino abundance allows the creation of an substantial part of the final efficiency already in the scatter regime. This is especially striking for  $\mtl = \SI{5e-5}{\electronvolt}$ in Fig. \ref{fig:yPlots:1:1:63:162:1:162:194eta}(panels 1--3) where $\zrhoA \sim \zaA$ and $\eta_A(\zrhoA) \gg \eta_{\hat A}(\zrhoA)$. 
Moreover, as the transition from the scatter to the decay regime in case $A$ generally occurs for smaller $\zn$ than the transition in the simplified scenario, it is reasonable to assume that $Y_{N_A}(\zrhoA)$ exceeds $Y_{N_{\hat A}}(\zaA)$ which additionally enhances the efficiency that can be generated in the decay regime of case $A$ compared to the simplified scenario. Thus, as we already saw in Fig. \ref{fig:density:A:B}(right), the efficiency in case $A$ for small $\mtl$ is suppressed similarly to the simplified scenario but the suppression is overall much less pronounced. \\
As in the simplified scenario, increasing $\mtl$ damps the suppression of the efficiency and while WO effects can be neglected, we have $\etaA\sim \mtl$ until a maximum is reached at $\mtmax$ and WO processes subsequently dominate, resulting in $\etaA \sim \sfrac{1}{\mtl}$, see Fig. \ref{fig:density:A:B}(middle).
However, the regime where WO processes are effective in case $A$ differs significantly from the simplified scenario. 
As can be seen in Fig. \ref{fig:yPlots:1:1:63:162:1:162:194eta}(panels 7--9), the suppression of $\etaA(\zn)$ is significantly weaker compared to $\etahA(\zn)$ but still sufficiently strong in order to render WO processes ineffective up to $\zn \sim \zrhoA$, similarly to the discussion in Sec. \ref{sec:approximation}. As is demonstrated in Fig. \ref{fig:mtildegplots:cases:1}, this results in $\mtmaxA \sim \zrhoA \ll \mtmaxapA \sim \za$ and in particular, the striking peak around $\gn \sim 0.7$ in $\mtmaxapA$ is missing in $\mtmaxA$. \\
Finally, let us shortly comment on the efficiency for $\gn \to 1$ and $\mtl \to \SI{e-1}{\electronvolt}$. From Fig. \ref{fig:density:A:B}, recall that the final efficiency in this case slightly exceeds the "common" efficiency in the strong WO regime $\eta^{WO}$. In light of our previous discussion, this unexpected behaviour can be explained as a combination of two effects. For $\gn \sim 1$, the suppression of $\eta_A(\zn)$ in the scatter regime is maximal and thus WO processes effectively diminish the efficiency only for $\mtl \gtrsim \SI{e-2}{\electronvolt}$(see Fig. \ref{fig:mtildegplots:cases:1}). As can be seen in Fig. \ref{fig:yPlots:1:1:63:162:1:162:194eta} (panel $9$) though, even for $\mtl = \SI{e-1}{\electronvolt}$, washout is effective only for $\zn \gtrsim 6$. Just after that, $\latexchi$ begins to freeze out which enhances the neutrino abundance and therefore counteracts the washout of the efficiency. 
Thus, even though WO processes dictate the overall evolution of $\etaA(\mtl)$ in the WO regime, leptogenesis is not entirely driven by VL interactions. 
Note that the significant enhancement $\Delta_{N_A}$ around $\zn \sim 20$ we observed in Fig. \ref{fig:deltaN} does not have a significant effect as due to the strong Boltzmann suppression of $Y_N(\zn)$ around $\zn \sim 20$, even a sizable deviation from thermal equilibrium in this regime only corresponds to a tiny enhancement of the neutrino abundance. 

\subsection{Case $B$: $\dall \neq 1\,, \geta =\num{e-7}\,, \ls = 1 $}
\label{sec:case:b}
In this section, we discuss the efficiency in case $B$ where in contrast to case $A$, we have $\geta = \num{e-7}$ which results in a dependence of the efficiency on the initial conditions. We begin with a short discussion of the thermal rates before seperately analyzing the evolution of the efficiency for the distinct initial conditions. In the discussion of case $B_z$, we focus in particular on the regime where the initial conditions are relevant for the final efficiency. 
\subsubsection{Thermal Rates}
In Fig. \ref{fig:gammai}, we show thermal rates relevant for the Boltzmann equations \eqref{eq:ben}-\eqref{be:j} and as in case $A$, we focus only on the most relevant ones. In contrast to the thermal rates in case $A$, the interactions of $\latexchi$ with $N, \sigma$ and $J$ in case $B$ are practically irrelevant while gauge interactions thermalize $\latexchi$ over the full range of $\zn$ that is of interest for leptogenesis. In particular, $\latexchi$ does not freeze out during the leptogenesis era. On the other hand, the interactions between $N, \sigma$ and $J$ that do not involve $\latexchi$ are not affected and identical to $A$. As a result, the evolutions of $N, \sigma$ and $J$ are decoupled from the evolution of $\latexchi$.  
\subsubsection{Discussion of case $B_t$}
\begin{figure}[h!]
    \includegraphics[width=\textwidth]{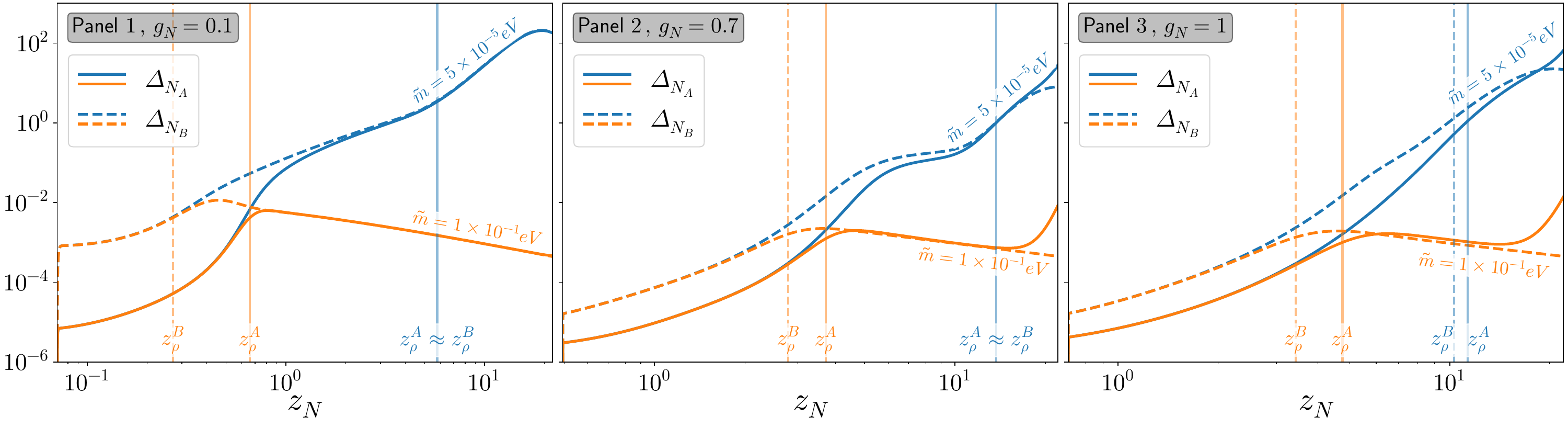}
    \caption{
        Evolution of $\Delta_{N_A} = \delta_{N_A}-1$ (solid lines) and $\Delta_{N_B} = \delta_{N_B}-1$ (dashed lines) for different values of $\mtl = [\num{5e-5}, \num{e-1}]\si{\electronvolt}$ and $\gn = 0.1$ (panel $1$), $\gn=0.7$ (panel $2$), $\gn=1$(panel $3$). The vertical solid lines indicate $\zrhoA$ while the vertical dashed lines belong to $\zrhoB$. Note that the lower limits of the $\zn$ range are given by $\zi$ and hence depend on $\gn$. 
        For $\zn \lesssim \zrhoB$, the neutrino abundance in case $B$ deviates more from thermal equilibrium than in case $B$ as interactions with $\latexchi$ are absent. For $\gn \geq 0.7$, the neutrino evolution in case $B$ is entirely driven by inverse neutrino decays, hence $\Delta_{N_B}(\zn \gg \zrho) < \Delta_{N_A}(\zn \gg \zrho)$.  
    }
    \label{fig:deltacasesN}
\end{figure}

After discussing the efficiency for cases $\hat A, \hat B$ and $A$, applying the findings to case $B_t$ is straightforward. 
First, note that gauge interactions keep $\latexchi$ close to thermal equilibrium, independently of $\gn$ and $\mtl$. On the other hand, due to the absence of notable interactions with $\latexchi$, we find that in the scatter regime, $N, \sigma$ and $J$ deviate from thermal equilibrium slightly more compared to case $A$. This can be seen in Fig. \ref{fig:deltacasesN} where we show $\Delta_{N_A}(\zn)$ and $\Delta_{N_B}(\zn)$ for different values of $\gn$ and $\mtl$. Clearly, this translates to a larger final efficiency compared to case $A$. 
Additionally, as is highlighted in Fig \ref{fig:density:za:1}, the transition from the scatter to the decay regime occurs for a slightly smaller $\zn$ compared to case $A$, i.e.\ $z_{\rho_B} \lesssim z_{\rho_A}$, as the absence of interactions with $\latexchi$ slightly reduces the overall magnitude of scattering processes involving neutrinos.
This is particularly striking for $\gn >0.7$ where in case $A$, $\gat{NN\latexchi\latexchi}$ is the dominating process which is absent in case $B$, similarly to our discussion regarding $\hat A, \hat B$. 
Further, as can be seen in Fig. \ref{fig:mtildegplots:cases:1}, the increased efficiency implies that the transition from the \st{WO} to the WO regime takes place at smaller $\mtl$ compared to case $A$ and therefore $\mtmaxBt < \mtmaxA$.
\subsubsection{Discussion of case $B_z$}
\begin{figure}
  \centering
  \includegraphics[width=\textwidth]{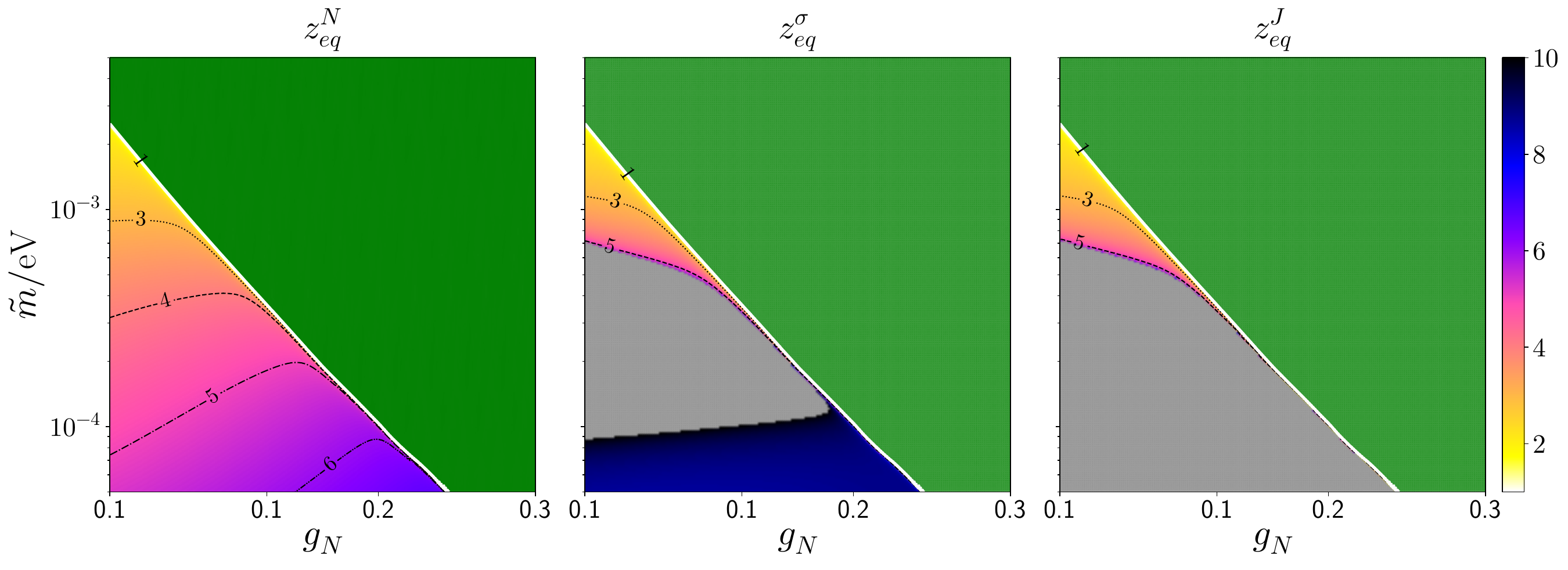}
  \caption{Density plots of $\zalleq$ in the $\gn-\mtl$ plane. In the green region, thermalization proceeds rapidly so that $\zieq \ll 1$ while in the grey region, the respective particle is not thermalized until $\zn = 10$. They white line indicate $\zieq = 1$. Note that we do not show the full $\gn-\mtl$ plane. Seet text for details.}
  \label{fig:heatmaps:zeq:11}
\end{figure}

\begin{figure}[!htbp]
  \centering
  \includegraphics[width=\textwidth]{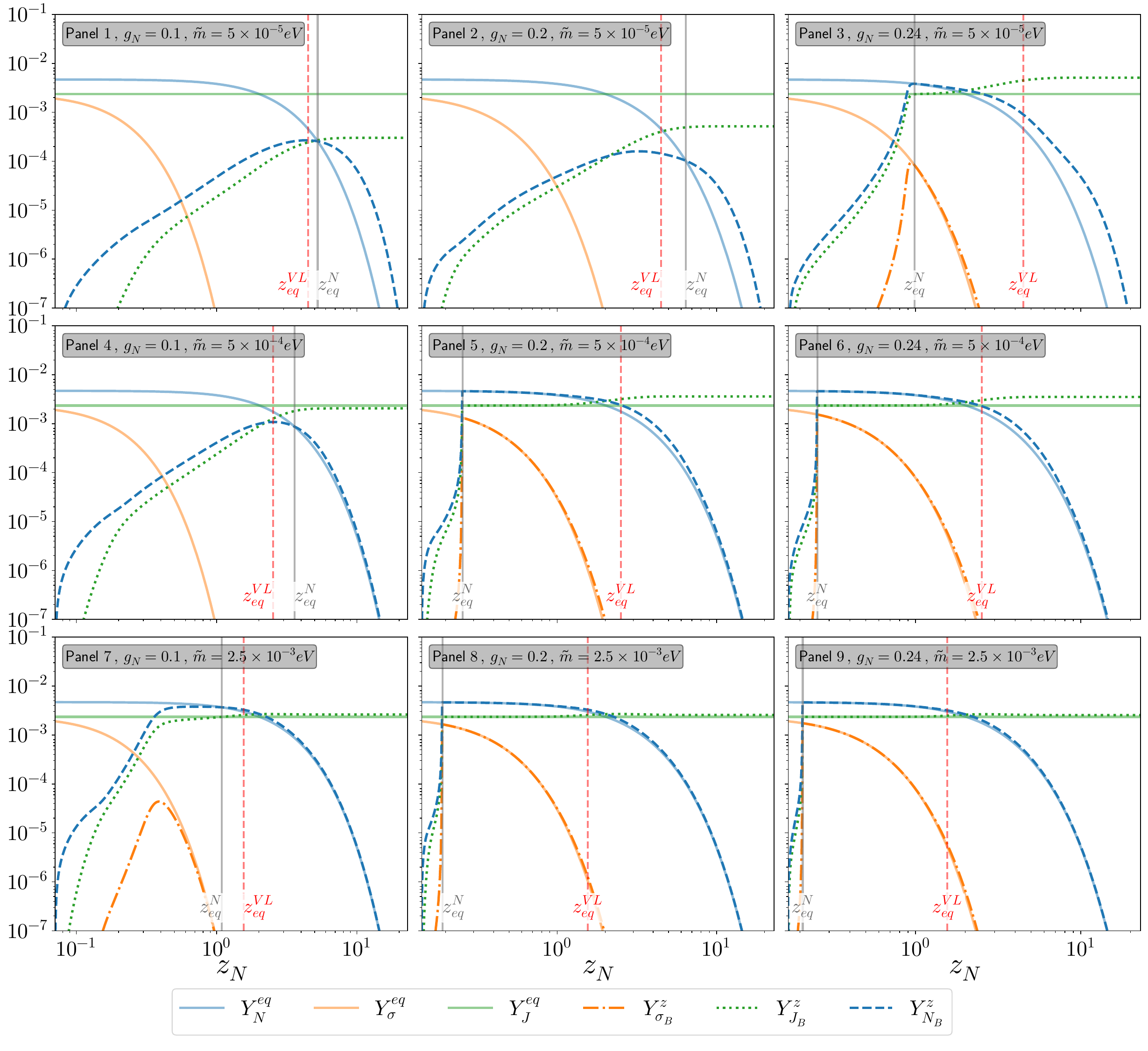}
  \caption{
  Evolution of $Y_{N, \sigma, J}(\zn)$, compared to the respective equilibrium abundances, for $\gn = [0.1,0.2,0.24]$ and $\mtl = [\num{5e-5}, \num{5e-4}, \num{2.5 e-3}]\si{\electronvolt}$. 
  The vertical solid line indicates $\zneq$ while the vertical dashed line indicates $\zvleq$. 
  We do not show $Y_\latexchi$ as it is always close to thermal equilibrium. 
  For the parameter choices in panels $1$, $2$ and $4$, no sizable $Y_\sigma$ abundance is produced and therefore it does not appear in the respective figures. In the same panels, the majoron is never thermalized and hence freezes in once the interactions with $N$ and $\sigma$ decouple, resulting in a relic abundance smaller than the thermal one. Comparing e.g.\ panels $1$ and $2$, it is apparent how slightly increasing $\gn$ slows the neutrino thermalization down due to an enhanced scattering rate of neutrinos to majorons until $\gn$ reaches a value that results in sufficiently strong interactions between $N, \sigma$ and $J$ that briefly thermalized them before these interactions decouple (panel $3$). Increasing $\mtl$ enhances the neutrino production and therefore indirectly the production of $\sigma$ and $J$, thermalizing them faster (see e.g.\ panels $2$, $4$ and $8$). 
  }
  \label{fig:yPlots:11:1:49:1:57abundancies2}
\end{figure}

\begin{figure}[h!]
  \centering
    \includegraphics[width=\textwidth]{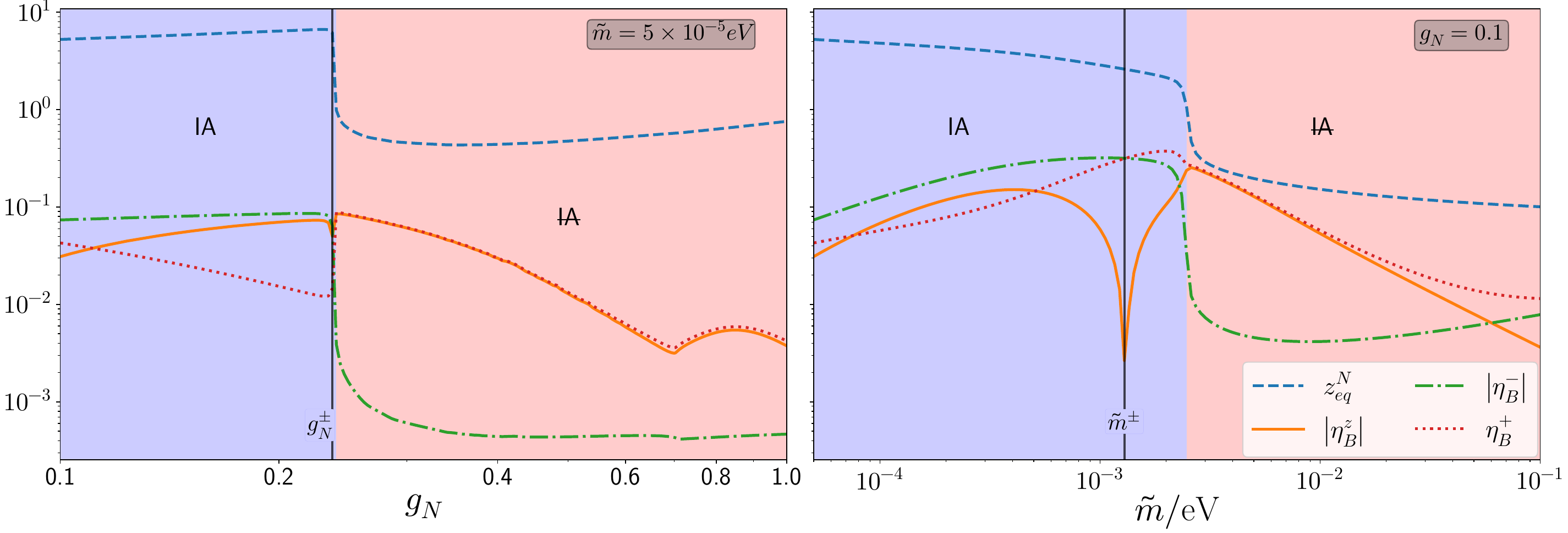}
  \caption{
  \textit{Left:} $\etaBzabs$, $|\etaBzm|$, $\etaBzp$ and $\zneq$ as functions of $\gn$ with $\mtl = \SI{5e-5}{\electronvolt}$. In the blue region, the final efficiency depends on the initial abundances while in the red region, the final efficiency is independent from the initial abundance. The vertical line denotes $\gn^{\pm}$. \\ 
  \textit{Right:} $\etaBzabs$, $|\etaBzm|$, $\etaBzp$ and $\zneq$ as functions of $\mtl$ with $\gn =  0.1$. In the blue region, the final efficiency depends on the initial abundances while in the red region, the final efficiency is independent from the initial abundance. The vertical line denotes $\mtl^{\pm}$. Note that as $\mtl \gtrsim \SI{e-2}{\electronvolt}$, $|\etaBzm|$ begins to increase again. This is due to washout processes reducing the efficiency and independent from the scattering processes in the majoron+triplet model. 
  }
  \label{fig:IaEqPlots:3:11}
\end{figure}

\paragraph{Abundances}
Due to the strong gauge interactions, $\latexchi$ is thermalized instantly at $\zi$ irrespective of $\gn$ and $\mtl$ and as in case $B_t$, its evolution is independent from $N, \sigma$ and $J$. On the other hand,  the thermalization of $N,\sigma$ and $J$ depends strongly on $\gn$ and $\mtl$. 
This is highlighted in Fig. \ref{fig:heatmaps:zeq:11} where we show density plots of $\zalleq(\mtl, \gn)$ where $\zalleq$ are defined via
\begin{align}
  Y_{i_B}^z(\zieq) = Y_i^{eq}(\zieq)
\end{align}
with 
\begin{align}
  Y_{i_B}^z(\zn) 
  \begin{cases}
   < Y_i^{eq}(\zn)\,,\quad &\zn < \zieq\,,\\
   > Y_i^{eq}(\zn)\,,\quad &\zn > \zieq\,.
  \end{cases}
\end{align}
Further, we define $(\gic,\mic)$ which we already introduced in Sec. \ref{sec:comparison} so that
\begin{align}
  \zneq(\mic, \gic) = 1\,, \label{eq:def:ic} 
\end{align} 
with
\begin{align}
  \zneq(\mtl <\mic, \gn < \gic) > 1\,, \\
  \zneq(\mtl >\mic, \gn > \gic) < 1\,.
\end{align}
\begin{figure}[h!]
  \centering
    \includegraphics[width=\textwidth]{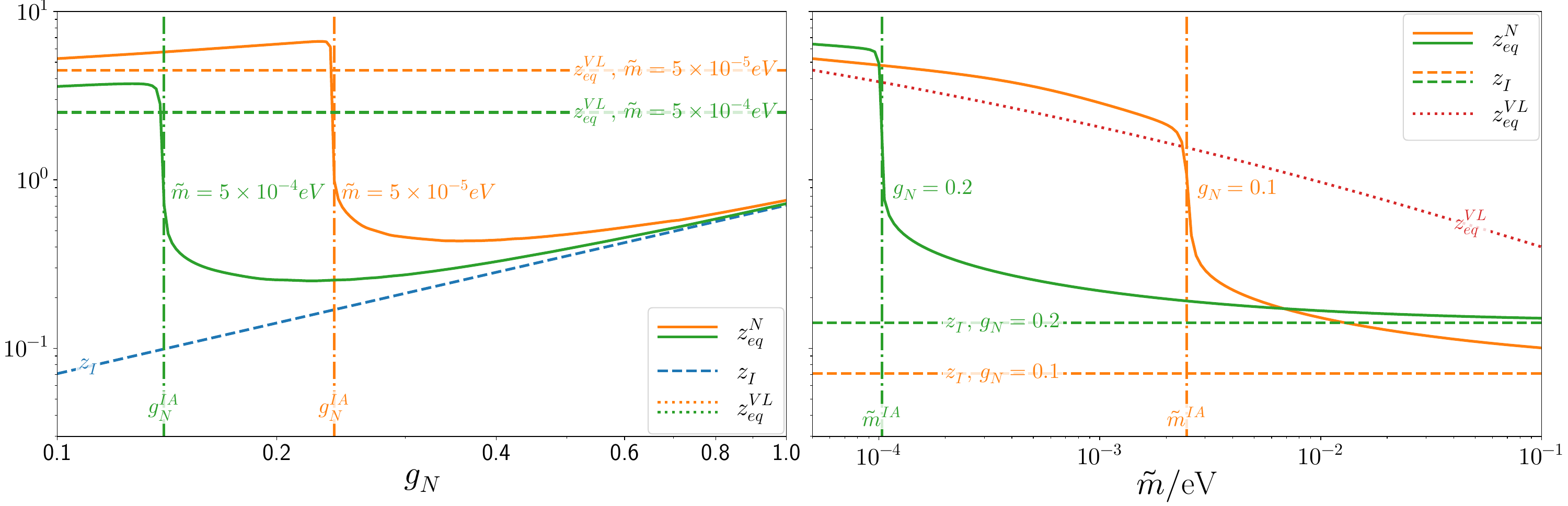}
  \caption{
  \textit{Left:} $\zneq$ as a function of $\gn$ for $\mtl = [\num{5e-5}, \num{5e-4}]\si{\electronvolt}$, compared to $\zvleq$ and $\zi$. 
  \textit{Right:} $\zneq$ as a function of $\mtl$ for $\gn = [0.1,0.2]$, compared to $\zvleq$ and $\zi$. \\
  We note that $\zvleq$ ($\zi$) is constant in the left(right) panel as it does not depend on $\gn$ ($\mtl$). The dash-dotted line denotes $\gic$ (left) and $\mic$ (right), respectively. Seet text for details.
  }
  \label{fig:IaEqPlots:11}
\end{figure}
\noindent
In Fig. \ref{fig:heatmaps:zeq:11}, we show $(\mic, \gic)$ as a white line while in the green regions, we have $\zieq < 1$. 
We stress that there is no fundamental reason to explicitly choose $\zneq(\mic, \gic) = 1$ as the implicit definition of $\mic, \gic$. As Boltzmann suppression of neutrinos sets in at $\zn \approx 1$, it is reasonable to assume that once $\zneq \lesssim 1$, the effect of the initial abundance on the neutrino evolution and the efficiency is small which motivates our choice of \eqref{eq:def:ic}.
Moreover, comparing \ref{fig:densitycomp2} and \ref{fig:heatmaps:zeq:11} , we can easily see that this definition of $(\gic,\mic)$ reasonably well seperates the IA and the \st{IA} regime. As we will see later, the explicit value used in \eqref{eq:def:ic} is irrelevant as long as it is close to $1$. Additionally, we used a cut-off at $\zn = 10$ to simplify the visualization of $\zseq$ and $\zjeq$.\footnote{Due to Boltzmann suppression, we expect that around $\zn \approx 10$, the effects of $\sigma$ and $J$ on the neutrino evolution are no longer significant which motivates the choice of $\zn = 10$. The exact value is however of no relevance for the following discussion. }\\
From Fig. \ref{fig:heatmaps:zeq:11}, we easily see that $\sigma$ and $J$ do not reach thermal equilibrium until $\zn = 10$ over a significant parameter space where $(\gn,\mtl)<(\gic,\mic)$. On the other hand, similarly to the neutrinos, we find that $\sigma$ and $J$ are quickly thermalized once $(\gn,\mtl)>(\gic,\mic)$, indicating that there is a close correlation between their thermalizations. In the following, we will discuss this in greater detail. \\
As the initial abundances of $N, \sigma$ and $J$ vanish, the Boltzmann equation for the neutrino evolution at $\zn \approx \zi$ is the same as in VL, 
\begin{align}
sH\zn \ddd{Y_{N_{B,VL}}^z} \approx \gamma_D + 2\gamma_Q\,, \label{eq:nproduction:vl}
\end{align}
and the plasma is populated with neutrinos via inverse decays and quark scatterings. 
As the interactions of neutrinos with $\sigma$ and $J$ are significantly stronger than $\gamma_{D,Q}$, neutrinos soon begin to populate the plasma with $\sigma$ and $J$ and the Boltzmann equation for case $B_z$ becomes
\begin{align}
  sH\zn \ddd{Y_{N_{B}}^z}  \approx& (\gamma_D + 2\gamma_Q) + \kappa \,,\label{eq:nproduction}
\end{align}
where we defined
\begin{align}
  \kappa \equiv&-2 (\overline{\delta_N}^2-\overline{\delta_J}^2)\gat{NNJJ}-2 (\overline{\delta_N}^2-\overline{\delta_\sigma}^2)\gat{NN\sigma\sigma}-2 (\overline{\delta_N}^2-\overline{\delta_J}\overline{\delta_\sigma})\gat{NN\sigma J} + 2\overline{\delta_{sub}}\gat{\sigma,NN}\,,
\end{align} 
with 
\begin{align}
  \overline{\delta_i} \equiv \delta_{i_B}^z\,,
\end{align}
and \eqref{eq:nproduction} holds as long as $\overline{\delta_N} \ll 1$. 
Here, $\kappa$ plays an essential role in the neutrino evolution as depending on the sign, it can either enhance the neutrino abundance via scatterings and $\sigma$ decays to neutrinos or diminish it via the inverse processes. In the first case, neutrinos are thermalized faster than in VL while in the second case, neutrino thermalization is slower compared to VL. \\
In order to make the effect of the scattering processes appearing in $\kappa$ more clear, let us discuss the effects of $\mtl$ and $\gn$ on $\zneq$ individually. To that end, we show $\zneq(\gn)$ and $\zneq(\mtl)$ in Fig. \ref{fig:IaEqPlots:11} for $\gn = [0.1,0.2,0.24]$ and $\mtl = [\num{5e-5}, \num{5e-4}, \num{2.5e-3}]\si{\electronvolt}$, respectively, while in Fig. \ref{fig:yPlots:11:1:49:1:57abundancies2}, we show $Y_{(N,\sigma,J)_B}^z(\zn)$ and compare them to the respective thermal abundances. \\
First, let us discuss the effect of $\mtl$ on $\zneq$ and for definiteness, we set $\gn = 0.1$.  
For $\mtl = \SI{5e-5}{\electronvolt}$(panel $1$ in Fig. \ref{fig:yPlots:11:1:49:1:57abundancies2}), the plasma is populated with neutrinos very slowly as $\gamma_D$ and $\gamma_Q$ are initially small. As a result, the scatterings in $\kappa$ become Boltzmann suppressed before thermal $\sigma, J$ abundances can be created. As the scatterings to $\sigma$ and $J$ reduce the neutrino abundance, we have $\ynzB(\zn) < \ynzVL(\zn)$ and consequently, neutrinos are thermalized later than in VL, i.e.\ $\zneq > \zvleq$.\\
As $\mtl$ increases to e.g.\ $\mtl = \SI{5e-4}{\electronvolt}$ (panel $4$ plot in Fig. \ref{fig:yPlots:11:1:49:1:57abundancies2}), inverse decays and quark scatterings populate the plasma with neutrinos faster and a larger abundance of $\sigma, J$ can be produced before the corresponding scattering processes become Boltzmann suppressed. As in VL, this implies that $\zneq$ and $\zvleq$ decrease with $\mtl$ (right plot in Fig. \ref{fig:IaEqPlots:11}) while  $\zneq > \zvleq$ still holds. \\
Eventually, as $\mtl \approx \SI{2.5e-3}{\electronvolt}$(panel $7$ in Fig. \ref{fig:yPlots:11:1:49:1:57abundancies2}), a sufficiently large neutrino  abundance is created for small $\zn$, allowing efficient interactions between $N, \sigma, J$ before they become Boltzmann suppressed. Thus, $N, \sigma, J$ are thermalized significantly faster than before and $\zneq(\mtl)$ drops rapidly and approaches $\zi$ (right plot in Fig. \ref{fig:IaEqPlots:11}). 
On the other hand, the neutrino thermalization in VL depends only on $\gamma_{D,Q} \sim\mtl$ and consequently $\zvleq(\mtl)$ smoothly decreases, eventually resulting in $\zneq < \zvleq$.\\
Next, let us discuss the effect of $\gn$ on $\zneq$. As discussed previously, the Boltzmann suppression of the scattering rates involving $N, \sigma, J$ is shifted to larger $\zn$ and the magnitude of the overall thermal rates increases as we move to larger $\gn$. Considering for example $\gn = 0.2$ and $\mtl = \SI{5e-5}{\electronvolt}$(panel $2$ in Fig. \ref{fig:yPlots:11:1:49:1:57abundancies2}), we find that this initially slows the neutrino thermalization down compared to $\gn=0.1$. 
This can be traced back to a larger number of neutrinos scattering to $\sigma$ and $J$ while the interaction rates are not yet fast enough to thermalize them, consequently reducing the neutrino abundance. 
Thus, $\zneq(\gn)$ initially increases with $\gn$, while $\zvleq$ is constant with $\zneq(\gn) > \zvleq$. Eventually though, at $\gn \approx 0.24$(panel $3$ in Fig. \ref{fig:yPlots:11:1:49:1:57abundancies2}), the scattering processes between $N, \sigma, J$ are fast enough to thermalize them efficiently and $\zneq(\gn)$ drops rapidly and approaches $\zi$, resulting in $\zneq(\gn) < \zvleq$ (left plot in Fig. \ref{fig:IaEqPlots:11}). \\
To summarize, we find that while $(\mtl, \gn) < (\mic, \gic)$, the scattering processes in $\kappa$ suppress the neutrino abundance as neutrinos scatter to $\sigma$ and $J$, resulting in a slower thermalization compared to VL. On the other hand, for $(\mtl, \gn) > (\mic, \gic)$, we find that the scattering processes enhance the thermalization of $N, \sigma, J$. We further note that since $\zneq(\mtl, \gn)$ drops rapidly around $(\mic, \gic)$, they are very insensitive to the exact value of $\zneq$ used in the definition \eqref{eq:def:ic}. 
\begin{figure}[h!]
  \centering
    \includegraphics[width=\textwidth]{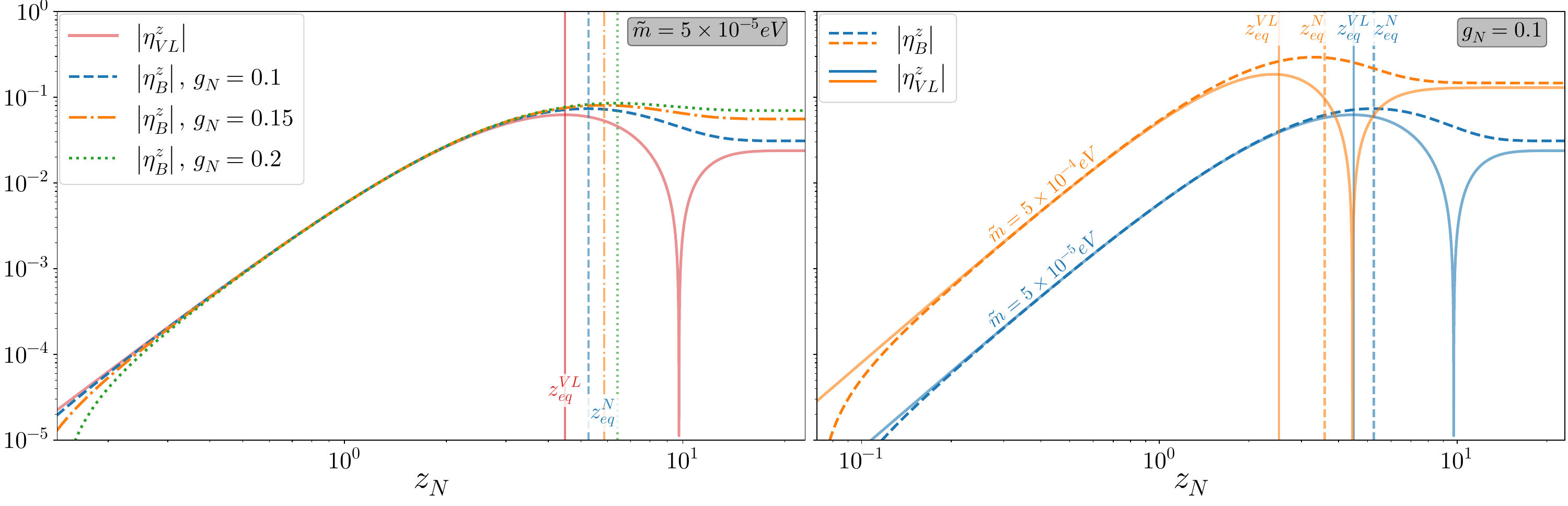}
  \caption{\textit{Left:} Evolution of $\etaBzabs(\zn)$ for $\gn = [0.1,0.15,0.2]$(dashed, dash-dotted, dotted) and $\mtl = \SI{5e-5}{\electronvolt}$, compared to $\etaVLzabs$ (solid). For $\zn \ll \zneq, \zvleq$, the evolution of the efficiency is dominated by (inverse) neutrino decays and therefore independent from $\gn$. The efficiencies $\etaBzabs(\zn)$ and $\etaVLzabs(\zn)$ increase until they peak at $\zneq$ and $\zvleq$, respectively. In VL, $\etaVLz(\zn)$ eventually changes sign while $\etaBz(\zn)$ stays negative. As $\zneq$ increases with $\gn$, we have $\etaBz(\zneq)|_{\gn=0.1} <\etaBz(\zneq)|_{\gn=0.15} <\etaBz(\zneq)|_{\gn=0.2}$. 
  \\
  \textit{Right:} Evolution of $\etaBzabs(\zn)$ (solid) for $\gn = 0.1$ and $\mtl = [\num{5e-5}, \num{5e-4}]\si{\electronvolt}$ compared to $\etaVLzabs$ (dashed). For $\zn \ll \zneq, \zvleq$, the evolution of the efficiency is determined by $\gamma_D\sim \mtl$, i.e.\ a larger $\mtl$ enhances the negative part of $\etaBz$ and $\etaVLz$ produced in $\zn \ll \zneq$ and $\zn \ll \zvleq$, respectively. As neutrinos are thermalized faster, $\zneq$ and $\zvleq$ decrease with $\mtl$.}
  \label{fig:yPlotsIA:11}
\end{figure}
\paragraph{Efficiency} 
In the IA regime, $\mtl$ is small so that we can safely neglect WO effects and the Boltzmann equation for $\etaBz$ becomes
\begin{align}
  sH\zn \ddd\etaBz \approx \fnorm (\overline{\delta_N}-1)\gamma_D \,. \label{eq:etaz}
\end{align}
Clearly, before neutrinos are thermalized at $\zneq$, we have $\overline{\delta_N}-1 < 0$ and a negative efficiency is created. We therefore define
\begin{align}
  \etaBzm \equiv \eta(\zneq) < 0\,. 
\end{align}
For $\zn \ll \zneq$, we can simplify \eqref{eq:etaz} to 
\begin{align}
  sH\zn \ddd\etaBz \stackrel{\zn \ll \zneq}{\approx}-\fnorm \gamma_D\, \label{eq:etaminus}
\end{align}
and as can be seen in Fig. \ref{fig:yPlotsIA:11}, this implies that $\etaBz(\zn\ll\zneq)$ closely follows $\etaVLz(\zn\ll\zneq)$ and is mostly independent from $\gn$. 
For $\zn > \zneq$ on the other hand, a positive efficiency is created with 
\begin{align}
  \etaBzp \equiv \int_{\zneq}^\infty \mathrm{d}\zn\,\fnorm (\overline{\delta_N}-1)\gamma_D\frac{1}{sH\zn}  >0  \,. \label{eq:etaplus}
\end{align}
Assuming that the scattering processes involving $N, \sigma, J$ are already Boltzmann suppressed and (inverse) decays $N\leftrightarrow LH$ are the only relevant processes for $\zn >\zneq$, we can simplify this to 
\begin{align}
  \etaBzp \eqsim \fnorm \ynzB(\zneq)\,.
\end{align}
Thus, the final efficiency is given by 
\begin{align}
  \etaBz(\mtl, \gn) = \etaBzp(\mtl, \gn) + \etaBzm(\mtl, \gn)\,
\end{align}
where $\etaBzp(\mtl,\gn)$ and $\etaBzm(\mtl,\gn)$ clearly depend significantly on $\zneq(\mtl,\gn)$ and in particular, $\etaBzp(\mtl, \gn) < |\etaBzm(\mtl, \gn)|$ results in a negative final efficiency. From Fig. \ref{fig:density:A:B}, recall that $\etaBz$ is indeed negative if $\mtl$ and $\gn$ are small and for convenience, we define that $\etaBz(\mtl, \gn)$ changes sign at $(\mtl^\pm, \gn^\pm)$, i.e.\ 
\begin{align}
  \etaBzp(\mtl^\pm, \gn^\pm) = |\etaBzm(\mtl^\pm, \gn^\pm)|
\end{align}
so that 
\begin{align}
  \etaBz(\mtl, \gn) <& 0 \,,\quad (\mtl,\gn)< (\mtl^\pm, \gn^\pm)\,,\\
  \etaBz(\mtl, \gn) >& 0 \,,\quad (\mtl,\gn)> (\mtl^\pm, \gn^\pm)\,.
\end{align}
In Fig. \ref{fig:IaEqPlots:3:11}, we show $|\etaBzm|, \etaBzp$ and $\etaBz$ as functions of $\gn$ (left) and $\mtl$ (right) with $\mtl = \SI{5e-5}{\electronvolt}$ and $\gn =0.1$, respectively. In the blue regions, the initial conditions are relevant while in the red regions, the initial conditions are irrelevant for the final efficiency. \\
Let us first focus on the blue region in the left plot. 
We immediatly see that $\etaBz$ changes sign at $\gn^\pm \approx \gic$ and for simplicity, we will assume equality in the following discussion so that the blue region also corresponds to $\etaB < 0$. 
From \eqref{eq:etaminus} it is clear that the dependence of $\etaBzm$ on $\gn$ stems only from $\zneq(\gn)$. More precisely, as $\zneq$ increases with $\gn$, the $\zn$-range in which inverse decays produce a sizable negative efficiency increases, resulting in $|\etaBzm(\gn)| \sim \zneq(\gn)$. This is further highlighted in Fig. \ref{fig:yPlotsIA:11}(left). On the other hand, the neutrino abundance is Boltzmann suppressed with 
\begin{align}
  \ynzB(\zneq) \sim \mathrm{e}^{-\zneq}\,,
\end{align}
and consequently, an increasing $\zneq(\gn)$ diminishes $\etaBzp(\gn)$. 
Thus, in the blue region, $|\etaBzm(\gn)|$ increases while $\etaBzp(\gn)$ decreased and as $|\etaBzm(\gn)|>\etaBzp(\gn)$, this clearly results in $\etaBz$ decreasing (or $|\etaBz|$ increasing since $\etaBz < 0$ ) with $\gn$. 
Once neutrinos are quickly thermalized for $\gn \gtrsim 0.24$ (red region in Fig. \ref{fig:IaEqPlots:3:11}), the initial conditions are irrelevant and the contribution of $\etaBzm$ can be neglected, resulting in $\etaBz(\gn) \sim \etaBzp(\gn) >0$. \\
Next, let us consider the blue region in the right plot in Fig. \ref{fig:IaEqPlots:3:11}. 
First, we note that $\mtl^{\pm} < \mic$, i.e.\ $\etaBz$ changes sign in the blue region before the initial conditions are irrelevant. 
From \eqref{eq:etaminus}, it is clear that a larger $\mtl$ enhances the neutrino production via inverse decays already at $\zn \ll \zneq$ (see also Fig. \ref{fig:yPlotsIA:11}(right)) and consequently, $|\etaBzm|$ increases with $\mtl$.
On the other hand, $\etaBzp$ depends strongly on $\zneq(\mtl)$ due to the Boltzmann suppression of neutrinos. 
More precisely, recalling that $\zneq(\mtl)$ decreases, this implies that the Boltzmann suppression at $\zneq$ decreases as well and consequently,  $\etaBzp(\mtl)$ increases. 
In contrast to the scenario shown in the left hand plot, the sign change of the efficiency is not due to the rapid drop of $|\etaBzm|$ at $\mic$ but due to $\etaBzp$ approaching and eventually overtaking $|\etaBzm|$ at $\mtl^\pm < \mic$. Clearly, this results in a severly diminished final efficiency around $\mtl^\pm \approx \SI{1.3e-3}{\electronvolt}$.
Only once neutrino thermalization becomes fast and $\zneq$ drops at $\mic \sim\SI{2.5e-3}{\electronvolt}$, we find that $|\etaBzm|$ drops as well, resulting in $\eta(\mtl)\sim\etaBzp(\mtl)$ (red region) and the initial conditions are rendered irrelevant. 
Cleary, the $\mtl$-range around $\mtl^{\pm}$ where $|\etaBz|$ almost vanishes depends crucially on the difference between $\mtl^\pm$ and $\mic$. As $\gn$ increases, $\mtl^\pm$ approaches $\mic$ and the $\mtl$-range where the efficiency is diminished shrinks until at $\gn \approx 0.15$, we have  $\mtl^\pm\approx\mic$. \\
To summarize, we find that $\etaBz$ significantly depends on how fast neutrinos are thermalized. While the evolution of the efficiency for small $\zn$ is not significantly changed compared to VL, the suppression of the neutrino abundance for $(\mtl, \gn) < (\mic, \gic)$ 
results in a later thermalization compared to VL and thus reduces the number of neutrinos that can contribute to $\etaBzp$. Moreover, for $(\mtl, \gn) \approx (\mtl^{\pm}, \gn^{\pm})$, the negative and the positive contribution almost cancel each other, resulting in an almost vanishing final efficiency $\etaBz$. 

\subsection{General Remarks}
\label{sec:results:final} 
\begin{figure}[h!]
    \centering
    \includegraphics[width=0.8\textwidth]{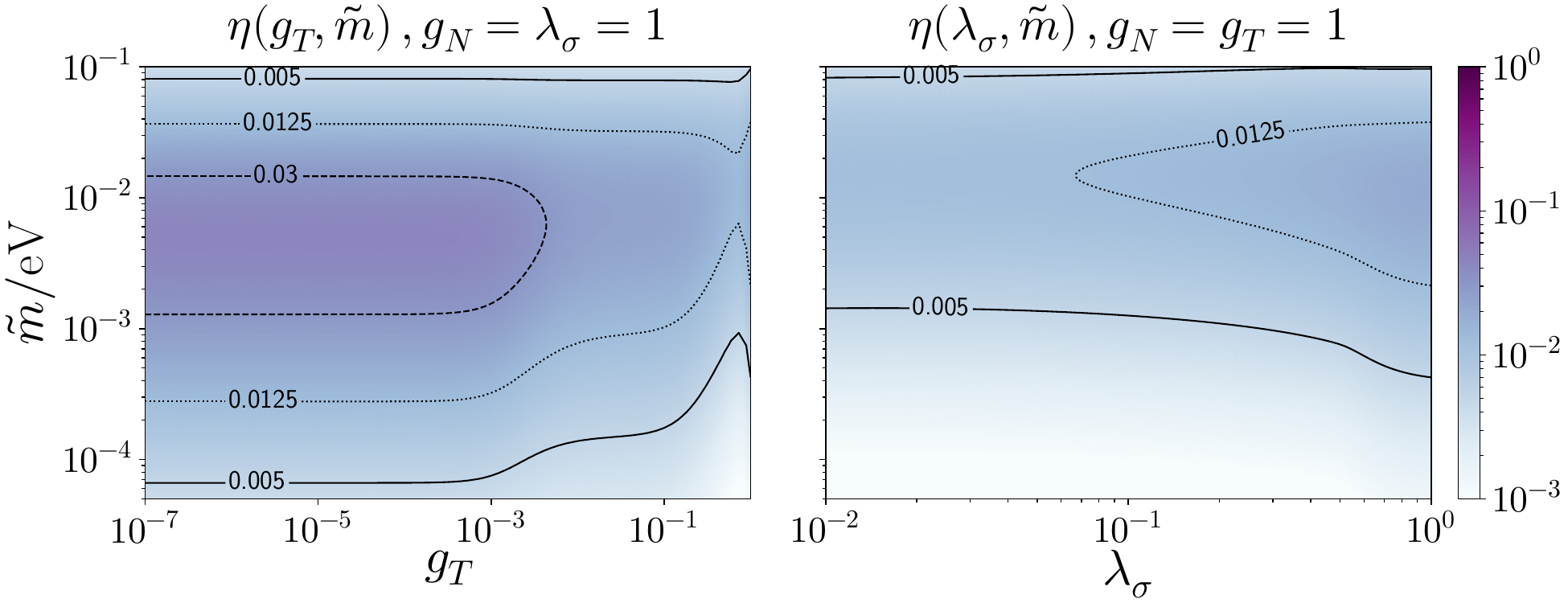}
    \caption{Efficiency $\eta(\geta, \mtl)$ (left) and $\eta(\ls,\mtl)$ (right) for $\gn=0.1$. Note that we used the same color scaling as in Fig. \ref{fig:density:A:B}. We can easily see that the efficiencies do not change significatly with $\geta$ and $\ls$, respectively. }
    \label{fig:etacompplots:case23:ylt}
\end{figure}

Before discussing other implications of the model, let us adress some additional points. \\
So far, we have kept $\ls$ constant and discussed the effect of $\geta = \num{e-7}$ compared to $\geta = 1$ on the efficiency. 
Before we proceed, let us comment in greater detail on the effect these couplings have on the efficiency. To that end, we show $\eta(\geta, \mtl)$ (left) and $\eta(\ls, \mtl)$ (right) for $\gn = 1$ in the scenario where all particles are in thermal equilibrium at $\zi$ in Fig. \ref{fig:etacompplots:case23:ylt}. As expected, the sizes of $\geta$ and $\ls$ have significatly smaller effects on $\eta$ compared to $\gn$. What is more, below $\geta \sim \num{e-3}$, $\eta(\geta)$ becomes constant, i.e.\ the Yukawa coupling of the triplet is irrelevant for the value of the final efficiency. In fact, we checked explicitly that the final efficiency in case $B$ agrees with the final efficiency in scenario where $\geta = 0$. Thus, the results in case $B$ are applicable to a broader class of models, in particular to the conventional singlet majoron model, and hold even for larger values of $\geta$ than considered in case $B$. Similarly, $\eta(\ls)$ is barely affected by the changes in $\lambda$ and becomes almost constant for $\ls \lesssim 0.1$.\\
Next, recall that we neglected the doublet-singlet mixing term $\sim\lambda_{mix}$ in \eqref{eq:potential}. We expect that depending on the size of the $\lambda_{mix}$, including this term could drive the efficiencies in cases $A,B$ closer to $\hat A, \hat B$ as the additional interaction with $H$ could result in a more efficient thermalization of $\sigma$ and $J$. \\
After these general considerations, we will focus on a more realistic version of the model and examine dark matter constraints on the new particles.

\section{Dark Matter}
\label{sec:dm}
The most precise measurements of the DM energy density today come the CMB anisotropies with an observed value of \cite{Planck:2018vyg}
\begin{align}
	\Omega_{DM}h_0^2 &= 0.120 \pm 0.0001 \,, 
\end{align}
where $h_0$ is the scaling factor for the Hubble parameter today, $H_0$,  
\begin{align}  
	h_0 &= 0.674\pm 0.0005\,
\end{align}
with 
\begin{align}
    H_0 \equiv h_0 \SI{100}{\kilo\meter\per\second\per\mega\parsec}\,.
\end{align}
A DM candidate needs to be non-relativistic at the onset of galaxy formation, non-baryonic, have at most weak interactions with the SM and it has to be stable on cosmological timescales. As we will discuss in the following, both the majoron $J$ and the triplet fermion $\latexchi$ in the majoron+triplet model can in principle fulfull these requirements.
\subsection{Majoron}
The majoron as a DM candidate has already been extensively discussed in various scenarios \cite{Brune:2018sab,Frigerio:2011in,Hall:2009bx,Rothstein:1992rh,Berezinsky:1993fm}. In this work, we do not specify a mechanism that generates the necessary majoron mass $\mj$ and merely focus on the implications in light of the majoron+triplet model. Moreover, we note that the majoron can decay to a pair of light neutrinos if $\mj$ exceeds the light neutrino masses. This implies that in order to be a viable DM candidate, the majoron decay width needs to be sufficiently small in order to render the majoron stable on cosmological timescales. For simplicity, we assume for the remainder of this section that the majoron is indeed longlived enough to constitute DM and explore the limits the majoron+triplet model places on the majoron mass.\\
During the leptogenesis era, the majoron interacts via scatterings and inverse decays with $N, \sigma$ and $\latexchi$. As discussed in Sec. \ref{sec:results}, these interactions are very effective at populating the plasma with majorons in cases $A$ and $B_t$, for some parameter sets even reaching an abundance larger than the thermal one. Those interactions decouple at $\zn^d \sim \mathcal{O}(10)$ and the majoron abundance reaches a constant value $Y_J(\zn^d)\gtrsim Y_J^{eq}(\zn^d)$. In case $B_z$ with $(\mtl,\gn)< (\mtl^{IA}, \gn^{IA})$ on the other hand, the majoron abundance tends to be smaller than the corresponding thermal abundance when the interactions with $N, \sigma$ and $\latexchi$ decouple, resulting in $Y_J(\zn^d)\lesssim Y_J^{eq}(\zn^d)$. In both cases, a majoron relic density 
\begin{align}
	\Omega_{J} = \frac{Y_J(\zn \to\infty) m_J s_0}{\rho_c}
\end{align}
arises. In the equation above, $s_0$ is the entropy density today, defined via the photon density today $n_{\gamma,0}$ as
\begin{align}
	s_0 \approx 7.04 n_{\gamma,0}\,
\end{align}
while $\rho_c$ is the critical density, 
\begin{align}
	\rho_c = \num{1.87834(4)e-29}h^2\si{\gram\per\centi\metre\cubed}\,.
\end{align}
From the condition $\Omega_{J}(\mj^{DM})\stackrel{!}{=}\Omega_{DM}$, we determine the maximal acceptable majoron mass $\mj^{DM}$ that does not result in a majoron relic density that exceeds the DM relic density. Thus, if $\mj = \mj^{DM}$, the majoron constitues all DM while for $\mj < \mj^{DM}$, the majoron accounts only for a fraction of the DM relic density, requiring contributions from additional DM particles. In Fig. \ref{fig:densityDM}, we show density plots of $\mj^{DM}$ in the $\gn-\mtl$ plane for case $B_t$(left) and case $B_z$(middle).\footnote{Note that the constraints from case $A$ and $B_t$ result in almost identical constraints which is why we present only case $B_t$.}
We can immediatly see that in the case of thermal initial abundances, the large majoron abundance constrains the majoron mass to be of order $\mj \sim \SI{e2}{\electronvolt}$. In case $B_z$ with $\mtl \sim \SI{5e-5}{\electronvolt}, \gn \sim 0.1$, i.e.\ in the parameter range where the majoron is not thermalized, values up to $\mj \sim \SI{e3}{\electronvolt}$ are allowed. In the context of DM, this mass range is however problematic as the majoron with such a low mass would be relativistic at the onset of galaxy formation and constitute hot DM. This implies that in this setup, the majoron mass needs to be significantly smaller than $\mj^{DM}$ in order to result in a neglectable contribution to the DM relic density. We stress however that we implicitly assumed that the majoron is in kinetic equilibrium when the relic density freezes out and that majoron decays to neutrinos are slow. If the former assumption does not hold, even a small majoron mass could result in experimentally viable cold majoron DM (similarly to the axion) while in the case that the latter assumption does not hold, the majoron would be unstable and not a DM candidate. Moreover, depending on the mechanism that is employed in order to generate the majoron mass, the limits derived in this section can drastically change. Thus, whether the majoron in the majoron+triplet model is a viable DM candidate or not depends significantly on the origin of the majoron mass.

\subsection{Triplet}
\begin{figure}[H]
	\centering
		\includegraphics[width=\textwidth]{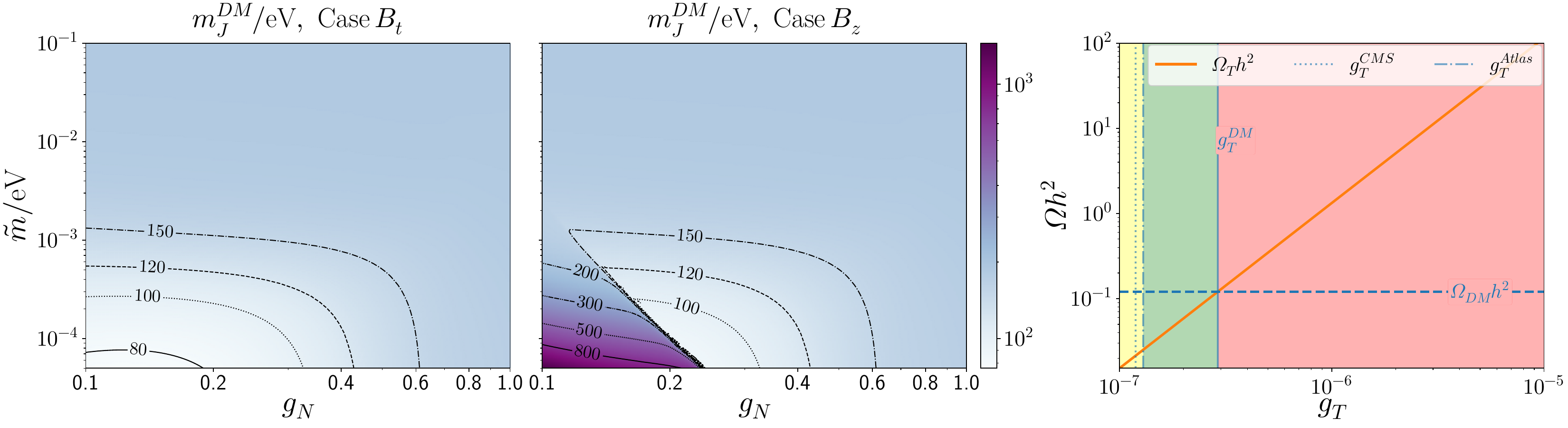}
	\caption{
	\textit{Left, middle:} Upper limits on the majoron mass that does not overproduce DM for case $A, B_t$ (left) and $B_z$ (middle) in the $\gn-\mtl$ plane. Note that these constraints are only valid if majoron decays to light neutrinos are sufficiently slow. \\
	\textit{Right:} Triplet relic density $\Omega_\latexchi h^2$ as a function of $\geta$, compared to the DM relic density $\Omega_{DM}h^2$. In the green region, the triplet relic density does not exceed the DM relic density and is in agreement with bounds from Atlas and CMS placed on the triplet mass. In the red region, DM is overproduced while in the yellow region, bounds from Atlas are violated. 
	}
	\label{fig:densityDM}
\end{figure}

As discussed previously, the triplet $T$ is effectively thermalized via gauge interactions while the scattering processes with $N, \sigma, J$ tend to be subdominant. Thus, once gauge interactions decouple at $z_\latexchi \sim 10$, the triplet freezes out and $Y_\latexchi$ reaches a constant value. This implies that the relic density is determined mostly by $\geta$ and rather independent from $\gn$ and $\ls$.  
For simplicity, we therefore neglect all scattering processes except these involving gauge bosons and accordingly solve a simple Boltzmann equation for the triplet evolution,
\begin{align}
	  s H \zn  \dy{\latexchi} &= -2 \left(\delta_\latexchi^2  -1\right) \gat{\gaugeww}\,,
\end{align}
for $\num{e-7}\leq \geta \leq \num{e-5}$. The corresponding relic density given by 
\begin{align}
	\Omega_{\latexchi} = \frac{Y_\latexchi(\zn \to\infty) \meta s_0}{\rho_c}
\end{align}
is shown in Fig. \ref{fig:densityDM}(left). We can immediatly see that the triplet relic density exceeds the observed DM relic density already for $\geta \gtrsim \num{2.9e-7}$ or equivalently $\meta \gtrsim \SI{2}{\tera\electronvolt}$.
Searches for triplet fermions in the framework of the type III seesaw by Atlas\cite{ATLAS:2022yhd} and CMS\cite{CMS:2017ybg} place lower bounds on the triplet mass of 
\begin{align}
	\meta^{Atlas} \approx \SI{910}{\giga\electronvolt}\,,\qquad\meta^{CMS} \approx \SI{840}{\giga\electronvolt}\,,
\end{align}
corresponding to   
\begin{align}
	\geta^{Atlas} \approx \num{1.19e-7}\,,\qquad\geta^{CMS} \approx \num{1.29e-7}\,.
\end{align}
In terms of the discussed scenarios of majoron+triplet leptogenesis, this implies that case $A$ is excluded while case $B$ is tightly constrained but remains viable. \footnote{Recall from Sec. \ref{sec:results:final} that the exact value of $\geta$ is not relevant in case $B$ as the triplet does not affect the efficiency.}
We stress nontheless that the discussion of the effects on the efficiency for different values of $\geta$ gives valuable insights on the dynamics of the leptogenesis scenario despite being largely experimentally forbidden.
We also stress that these constraints are only valid in the absence of the $Z_2$ forbidden mixing term with SM leptons and the Higgs doublet. Reintroducing this term allows lepton number violating decays of $\latexchi$ and thus renders the triplet instable. We note however that this the inclusion of the mixing term is non-trivial and beyond the scope of this work. 

\section{Baryon Asymmetry}
\label{sec:basym}
In this section, we finally discuss the baryon asymmetry, 
\begin{align}
    Y_\basym \equiv Y_{\bnumber}-Y_{\overline{\bnumber}}\,,
\end{align}
that can be created in the majoron+triplet model. As we found in the previous section that DM constraints exclude case $A$, we will focus on case $B$. 
    
\subsection{Sphaleron Rate}
As discussed in \cite{Brune:2022vzd}, the presence of the triplet changes the $\left[SU(2)_L\right]^2\times U(1)_\lnumber$ anomaly factor $\mathcal A_\lnumber$ from $\mathcal A_\lnumber = 3$ in the SM to $\mathcal A_\lnumber = -1$ in the majoron+triplet model. As a consequence, sphaleron transitions no longer conserve $Y_\basym-Y_\lasym$ as in the SM but instead $Y_\basym+3Y_\lasym$, thereby affecting the rate with which sphaleron processes convert the lepton asymmetry to a baryon asymmetry. \\
For simplicity, we calculate the conversion rate at a temperature $T \sim \si{\tera\electronvolt}$ where all gauge and Yukawa interactions are fast. This induces relations between the chemical potentials of the SM particles, 
\begin{align}
	\mu_{W^-} &= \mu_{\phi^0} + \mu_{\phi^-} = \mu_{d_L} - \mu_{u_L} = \mu_{e_L} - \mu_{\nu_L}\,, \\
	\mu_{\phi^0} &= \mu_{u_R} - \mu_{u_L} = \mu_{d_L} - \mu_{u_R}  = \mu_{e_L} - \mu_{e_R}\,.
\end{align}
With $\geta$ of order $\num{e-7}$, gauge interactions of the triplet are still fast as well, resulting in the additional relation 
\begin{align}
	\mu_{W^\pm} = \mu_{\latexchi^\pm}\,
\end{align}
while the chemical potentials of $N, \latexchi^0, \sigma, J$ vanish. 
Moreover, fast Sphaleron transitions in the majoron+triplet model imply 
\begin{align}
	\mu_{d_R} -6\mu_{e_L} + 11 \mu_{e_R} - 6\mu_{\nu_L} + 8\mu_{u_R} = 0
\end{align}
while in the SM, the corresponding relation reads
\begin{align}
	2\mu_{d_L} + \mu_{u_L} + \mu_{\nu_L} = 0 \,. 	
\end{align}
In the relativistic limit, the asymmetry in a charge $q$ can be expressed as
\begin{align}
	Y_q \equiv \frac{n_q-n_{\overline q}}{s} = \frac{T^2}{6s}\left[ \sum_{i\in f} q_i g_i \mu_i + 2 \sum_{j\in f} q_j g_j \mu_j \right]  \label{eq:charge:asym}
\end{align}
where the sum over $i (j)$ is for fermions (bosons), $q_{i,j}$ is the charge and $g_{i,i}$ are the numbers of degrees of freedom of the corresponding particle. Before EWSB, the plasma is charge and hypercharge neutral which imposes additional conditions between the chemical potentials by means of \eqref{eq:charge:asym}, finally resulting in 
\cite{Brune:2022vzd}
\begin{align} 
	Y_\basym = \frac{76}{679}Y_{(\basym + 3\lasym)}\,,
\end{align}
while the respective relation in the SM reads
\begin{align}
	Y_\basym = \frac{28}{79}Y_{(\basym - \lasym)}\,.
\end{align}
Assuming that no other sources of baryon number violation are present, we find that the baryon asymmetry is given by
\begin{align}
	|Y_\basym| = 
	\begin{cases}
		 3\times\frac{76}{679}Y_\lasym^B &\approx  0.35443 |Y_\lasym^B|\,, \\
		 \frac{28}{79}Y_\lasym^{VL} &\approx  0.335788 |Y_\lasym^{VL}|\,, \label{eq:basym1}
	\end{cases}
\end{align}
i.e.\ apart from the different signs, we find that the baryon asymmetries that can be generated from a given lepton asymmetries in each model do not differ significantly. 

\subsection{Comparison with experimental data}
\begin{figure}[h!]
    \centering
    \begin{subfigure}{\textwidth}
    	\includegraphics[width=\textwidth]{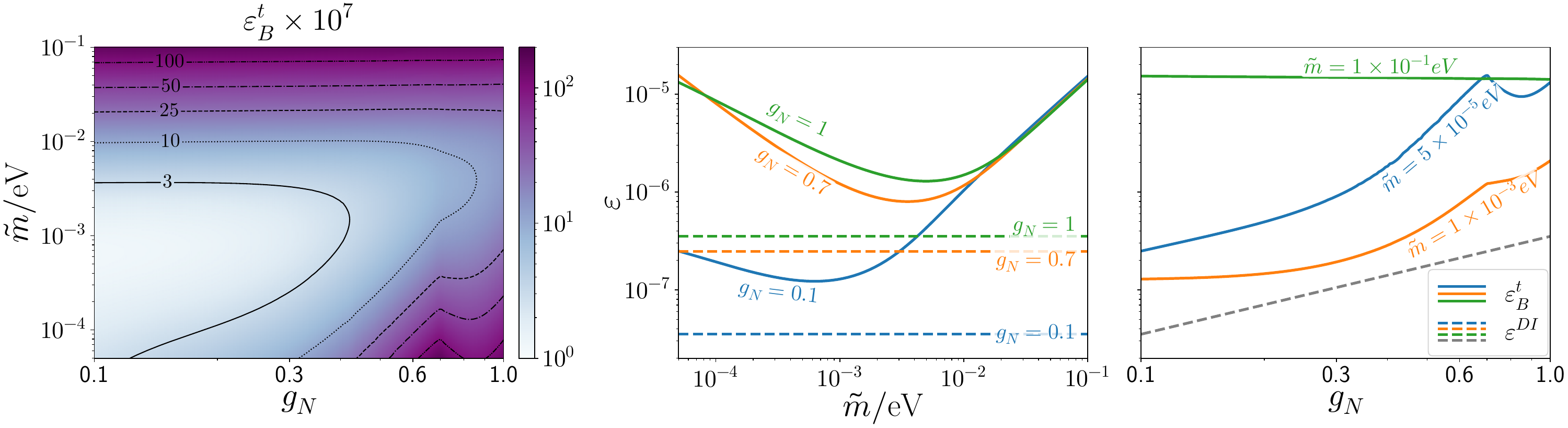}
    \end{subfigure}\\
    \centering
    \begin{subfigure}{\textwidth}
    \includegraphics[width=\textwidth]{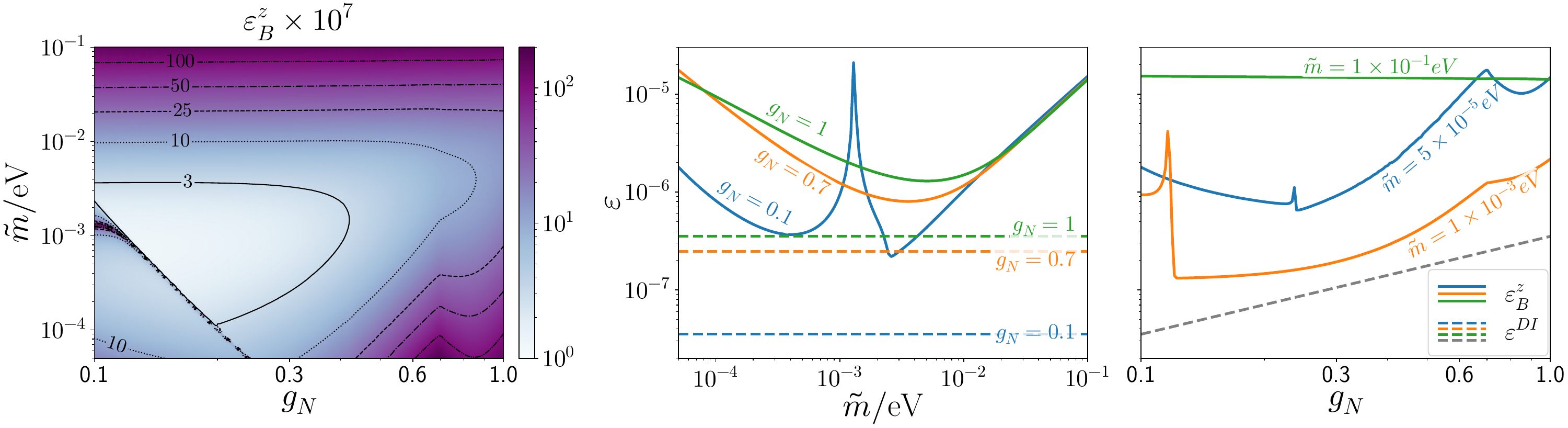}
    \end{subfigure}
    \caption{
    \textit{Left:} Density plots of the CP violation $\left|\varepsilon_B^t\right|$ (top) and $\left|\varepsilon_B^z\right|$ (bottom) required to explain the experimentally observed baryon asymmetry in case $B$ in the majoron+triplet model.\\
    \textit{Middle:} $\left|\varepsilon_B^t\right|$ (top) and $\left|\varepsilon_B^z\right|$ (bottom) as functions of $\mtl$ (solid lines) for $\gn = [0.1,0.7,1]$. The dashed lines corresponds to $\varepsilon^{DI}$. \\
    \textit{Right:} $\left|\varepsilon_B^t\right|$ (top) and $\left|\varepsilon_B^z\right|$ (bottom) as functions of $\gn$ (solid lines) for $\mtl = [\num{5e-5}, \num{e-3}, \num{e-1}]\si{\electronvolt}$. The dashed line corresponds to $\varepsilon^{DI}$. 
    }
    \label{fig:cp}
\end{figure}

\begin{figure}[h!]
    \centering
    \includegraphics[width=0.5\textwidth]{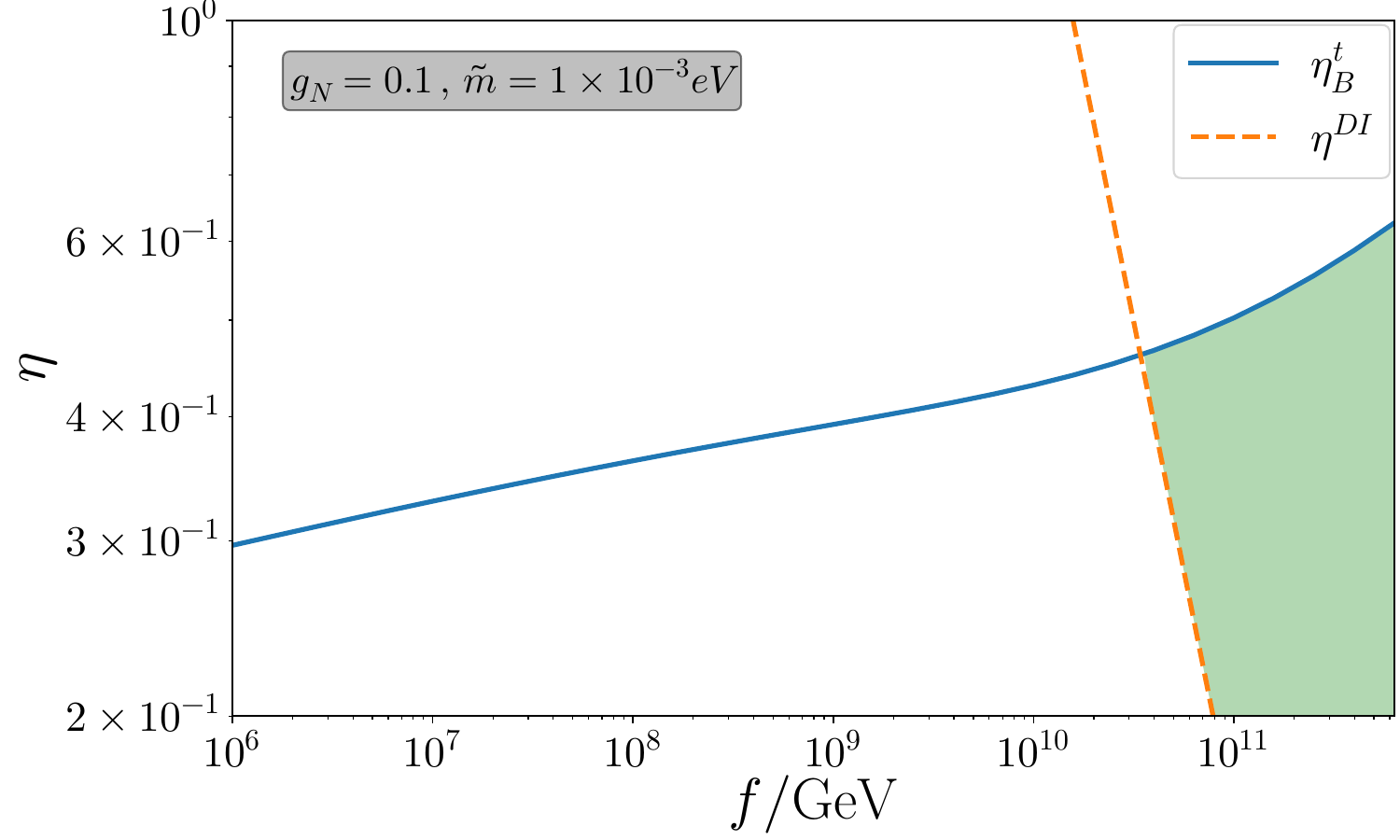}
    \caption{Efficiency $\etaBt$ for $\gn=0.1$ and $\mtl = \SI{e-3}{\electronvolt}$ as a function of the VEV $f$, compared to $\eta^{DI}$ as defined in \eqref{eq:eta:di}. In the green region, $\etaBt$ is sizable enough to reproduce the experimentally observed baryon asymmetry with a CP violation $\varepsilon$ that fulfills the DI bound.   }
    \label{fig:etaVEV}
\end{figure}

From observations, the baryon asymmetry of the universe is given by 
\begin{align}
	Y_\basym^{exp} &\equiv \frac{n_\bnumber-n_{\overline{\bnumber}}}{s}|_0 \approx  \frac{\num{274 e-10}}{7.04}\Omega_\bnumber h^2  \approx \num{8.718 e-11} \label{eq:basym:exp}
\end{align}
where 
\begin{align}
	\Omega_\bnumber h^2 &= 0.0224 \pm 0.0001 
\end{align}
is the baryon density obtained from CMB data \cite{Planck:2018vyg}.\\
Combining \eqref{eq:basym1} and \eqref{eq:basym:exp} we can determine the CP violation that is necessary in the majoron+triplet model in order to reproduce the experimentally observed baryon asymmetry as 
\begin{align}
	\varepsilon = \frac{679}{76\times3}\frac{Y_\basym^{exp}}{\norm \eta}\,.
\end{align} 
The results for case $B$ are shown in Fig. \ref{fig:cp} where we show density plots of $\varepsilon$ in the $\gn-\mtl$ plane (left), $\varepsilon(\mtl)$ (middle) and $\varepsilon(\gn)$ (right) for $\gn = [0.1,0.7,1]$ and $\mtl = [\num{5e-5},\num{e-3},\num{e-1}]\si{\electronvolt}$, respectively. For comparison, we also show $\varepsilon^{DI}(\mtl)$ and $\varepsilon^{DI}(\gn)$. 
We can immediatly see that even in the best case scenario of $\gn = 0.1$ and $\mtl \approx \SI{e-3}{\electronvolt}$, the CP violation necessary to explain experimental data is about one order of magnitude larger than $\varepsilon^{DI}$. The reason for this large difference is quite simple: The efficiency $\etaB$ is maximal for small $\gn$ while the DI bound $\varepsilon^{DI}$ decreases with $\mn = \sfrac{\gn}{\sqrt 2}f$. This is in contrast to VL where the efficiency is independent from $\mn$ and thus in principle, fulfilling the DI bound only requires a sufficiently large neutrino mass. 
Nevertheless, we stress that this does not per se exclude the viability of majoron+triplet leptogenesis. \\
First, recall that $\eta$ depends on the VEV $f$. 
In Fig. \ref{fig:etaVEV}, we show $\eta_{B}(f)$ with $f$ ranging from $\SI{e6}{\giga\electronvolt}$ to $\SI{e12}{\giga\electronvolt}$ for $\gn = 0.1$ and $\mtl = \SI{e-3}{\electronvolt}$. 
We also translate the DI bound into a lower limit on the efficiency necessary to reproduce experimental data as
\begin{align}
	\eta^{DI}(\mn)\equiv \frac{679}{3\times76}\frac{Y_B^{exp}}{\norm \varepsilon^{DI}(\mn)} \,. \label{eq:eta:di}
\end{align}
From Fig. \ref{fig:etaVEV} it is apparent that a higher breaking scale $f$ results in a larger efficiency while a smaller VEV diminishes the efficiency. Additionally, 
we have $\eta^{DI} \sim \sfrac{1}{f}$ and we can therefore conclude that a larger VEV can close the gap between $\varepsilon$ and $\varepsilon^{DI}$ from both ends while in the same manner, a smaller breaking scale would worsen the situation.  \\
Additionally, it has been shown that the DI bound can be violated if the hierarchy between $\mn$ and $M_{N_{2,3}}$ is not too strong \cite{Hambye:2003rt}. In fact, as the masses of $N_{2,3}$ arise via the same VEV as the mass of $N_1$ and taking into account perturbativity of $g_{N_{2,3}}$, it is even highly plausible that the hierarchy between $\mn$ and $M_{N_{2,3}}$ is rather weak which could easily enhance $\varepsilon$ past the DI bound. Finally, we neglected flavour effects in the Boltzmann equations. In the type III leptogenesis scenario, flavour effects have been shown to enhance the efficiency in the regime where WO processes dominate the efficiency \cite{AristizabalSierra:2010mv}. We expect a similar enhancement in the WO regime while in the \st{WO} regime, the efficiency is dominated by flavor-independent scattering processes and we thus expect only a weak effect on the efficiency.   

\section{Summary}
\label{sec:summary}
We have studied the implications of leptogenesis in the majoron model extended by an additional right-handed triplet with a $Z_2$ symmetry. 
The existence of the triplet changes the $[SU(2)_L]^2\times U(1)_\lnumber$ anomaly factor and hence affects how Sphaleron processes convert a lepton asymmetry to a baryon asymmetry. \\
In particular, we solved the Boltzmann equations relevant for the particle evolutions under two different assumptions: First, we considered a simplified scenario where only neutrinos deviate from thermal equilibirium before considering a scenario where all additional particles in the majoron+triplet model can deviate from thermal equilibrium.
Moreover, we have considered various different assignments for the parameters of the model and explored the relevance of the initial particle abundances. 
For certain parameter sets, solving only the Boltzmann equation for the neutrino evolution drastically underestimates the efficiency compared to the scenario where all Boltzmann equations are considered. For $\mtl \to \SI{e-1}{\electronvolt}$, i.e.\ the strong WO regime in VL, we generally find that inverse neutrino decays dominate so that the same efficiencies as in VL are reached while for small $\mtl$, the scattering processes with the new particles dominate the neutrino evolution. In the small $\mtl$ regime, we find that sizable efficiencies can be generated if the Yukawa coupling of the heavy neutrino is small, $\gn \sim 0.1$.\\
Moreover, we found that the initial abundances are mostly irrelevant for the final efficiency and only play a role in a very restricted subset of the considered parameters. Additionally, we found that the triplet does not effect the efficiency if the corresponding Yukawa coupling is small, $\geta \lesssim \num{e-3}$, thus restricting the relevance of the triplet to the effects on the conversion of the lepton asymmetry to the baryon asymmetry. \\
Due to the $Z_2$ symmetry, the triplet $\latexchi$ does not decay and obtains a relic density, qualifying it as a DM candidate. For a VEV of $f = \SI{e10}{\electronvolt}$, the Yukawa coupling of the triplet is limited from above with $\geta \lesssim \num{2.29e-7}$ in order to avoid overproduction of DM. 
We also considered the possibility of majoron DM, assuming that the majoron obtains a mass by an unspecified mechanism. We found that the upper limit on the experimentally observed DM relic density requires a majoron mass too small to render the majoron cold at structure formation, in contrast to observations. We note however that those bounds may drastically change, depending on the details of the mechanism that generates the majoron mass. \\
We compared the results for the efficiency in the scenario where the triplet Yukawa coupling is small, $\geta \sim \num{e-7}$, with the experimentally observed baryon asymmetry, taking into account that the Sphaleron conversion rate is different from the SM. In order to reproduce experimental data, in many cases a CP violation $\varepsilon$ that exceeds the DI bound $\varepsilon^{DI}$ is necessary. We note however that this bound can easily be avoided if resonance effects are taken into account. Moreover, we briefly examined the effects of the VEV $f$ on the efficiency. While we performed our analysis with $f = \SI{e10}{\giga\electronvolt}$, a larger VEV could enhance the efficiency that can be reached while additionally enhancing the DI bound. In this work, we neglected flavour effects in the Boltzmann equations. It should be stressed that while we expect that flavour effects are subdominant in the regimes where scattering processes dominate, they can have significant effects on the efficiencies if $\mtl$ is large. \\
In summary, we find that succesfull leptogenesis in the majoron+triplet model is in principle viable. For future works, a dedicated analysis of the leptogenesis scenario taking e.g.\ flavour effects or the resonant enhancement of the CP asymmetry into account seems interesting. Moreover, we expect that the inclusion of the mixing term between the triplet and the Higgs which was omitted in this work can have interesting effects on the creation of the lepton asymmetry which are worth examining in future works. 

\section*{Acknowledgements}
I would like to thank Heinrich Päs, Dominik Hellmann, Sara Krieg, Mustafa Tabet and Max Berbig for useful discussions. This work was supported by the \textit{Studienstiftung des deutschen Volkes}.
\appendix
\section{Appendix}
\label{sec:appendix}

\subsection{General Formulae}
\label{sec:interactions}
For scattering processes, we compute the thermal rates as 
\begin{align}
\begin{split}
   \gat{abij} \equiv \gamma(ab \leftrightarrow ij)  &= \frac{T}{64\pi^4}\int_{s_{min}}^{\infty} \!\! \mathrm d s \sqrt{s} \,\hat\sigma(s)\,\besselk{1}{\frac{\sqrt{s}}{T}} \\
   &= \frac{\mn M_X^3}{64\pi^4}\int_{x_{min}}^{\infty} \!\! \mathrm d x_X \frac{\sqrt{x_X}}{\zn} \,\hat\sigma(x_X)\,\besselk{1}{\frac{M_X}{\mn }\sqrt{x_X}\zn} \label{eq:gamma}
   \end{split}
\end{align}
where 
\begin{align}
  \zn &= \frac{\mn}{T}\,,\quad x_X = \frac{s}{M_X^2}\,,\label{eq:zn} \\
  s_{min} &= \mathrm{max}\left[ (M_a+M_b)^2,(M_i+M_j)^2  \right]\,, \label{eq:smin} \\
  x_{min} &= \mathrm{max}\left[ \left(\frac{M_a+M_b}{M_X}\right)^2,\left(\frac{M_i+M_j}{M_X}\right)^2 \right]\label{eq:xmin}\,.
\end{align}
Moreover, $\hat\sigma(s)$ is the reduced crossection defined as 
\begin{align}
  \hat\sigma(s) = 2s\lambda(1, \frac{M_a^2}{s}, \frac{M_b^2}{s})\sigma(s)\,,\quad \sigma(s) = \int_{-1}^1 \mathrm{d}\cos\theta \frac{1}{32\pi s}|\overline M|^2\,, \label{eq:sigma}
\end{align}
where $\sigma$ is the total cross section summed over initial and final spins and
\begin{align}
  \lambda(a,b,c) &= (a-b-c)^2\,. \label{eq:lambda}
\end{align}
For decays, we can write the thermal rate as 
\begin{align}
  \gat{a,ij}\equiv\gamma(a\leftrightarrow ij) &= n_a^{eq} \frac{\besselk{1}{z}}{\besselk{2}{z}}\Gamma_a \,,\label{eq:gammadecay}
\end{align}
where $\Gamma_a$ is the decay rate, $\besselk{n}{z}$ is the modified Bessel function of the second kind and 
\begin{align}
  n_a^{eq} &= g_a \int \frac{\mathrm d^3 p}{(2\pi)^3}f = \frac{g_a M_a T}{2\pi^2} \besselk{1}{\frac{M_a}{T}} = \begin{cases}  \frac{g_a T^3}{\pi^2}\,, &T\gg M_a\,,  \\ g_a\left(\frac{M_a T}{2\pi}\right)^{\sfrac{3}{2}}\e^{-\sfrac{M_a}{T}}\,, &T\ll M_a\,,   \end{cases} \label{eq:numberdensity}
\end{align}
is the number density of particle $a$ in thermal equilibrium at a temperature $T$ with $g_a$ degrees of freedom and $f$ is approximated as a Maxwell-Boltzmann distribution, $f\approx \e^{-\frac{E_a}{T}}$.
Note that unless CP-violating effects are present, the thermal rates for direct and inverse processes are the same.\\
The evolution of the number density $n_a$ of a particle $a$ in an expanding universe can be written as \cite{Giudice:2003jh}
\begin{align}
  \dot{n_a} +3H n_a &= -\sum_{b,i,j} \be{ab\leftrightarrow ij}
\end{align}
where $H$ is the Hubble parameter, 
\begin{align}
  H &= \sqrt{\frac{8}{\pi m_{pl}^2} g_s \frac{1}{T^4}}\,, 
\end{align}
$g_s= 106.75$ is the effective number of degrees of freedom, $m_{pl}$ is the Planck mass and 
\begin{align}
  \left[ab\leftrightarrow ij\right] &= \left(\delta_a\delta_b - \delta_i\delta_j\right)\gamma(ab \leftrightarrow ij)\,,
\end{align}
with
\begin{align}
  \delta_a = \frac{n_a}{n_a^{eq}}\,.
\end{align}
It is convenient to normalize the number density $n_a$ to the entropy density $s$, given by
\begin{align}
  s(z,m) &= \frac{4 m^3}{z^3 \pi^2}g_s \,, \label{eq:entropy}
\end{align}
yielding the abundance of $a$ as 
\begin{align}
  Y_a = \frac{n_a}{s}\,.
\end{align}
Using
\begin{align}
  3 H = -\frac{\dot{s}}{s}\,,
\end{align}
casting the Boltzmann equation in terms of $Y_a$ allows us to reabsorb the $3H$ term, i.e.
\begin{align}
  s H \zn\dy{a} &= -\sum_{b,i,j} \be{ab\leftrightarrow ij}\,. \label{eq:gen:boltzmann}
\end{align}
One subtlety that needs to be addressed is the substraction of on-shell processes. For example, let us consider the Boltzmann equation for a particle $X$ that interacts via a scattering process $XX \leftrightarrow YY$ with s-channel exchange of a particle $Z$. Considering the scattering process alone, the Boltzmann equation reads
\begin{align}
  sH z \dyz{X} = -2\left(\delta_X^2 -\delta_Y^2\right)\gamma(XX \leftrightarrow YY)\,. \label{eq:on:shell:example}
\end{align}
Moreover, if decays of $Z$ to $XX, YY$ are kinematically allowed, this would account for a term 
\begin{align}
   -2\left(\delta_X^2 -\delta_Z\right)\gamma(Z\leftrightarrow XX)\,.
\end{align}
However, the on-shell contribution is also accounted for in \eqref{eq:on:shell:example} and we therefore need to replace $\gamma(XX \leftrightarrow YY)$ with the off-shell contribution, i.e.\ 
\begin{align}
\begin{split}
  \gamma^{off}(XX \leftrightarrow YY) &= \gamma(XX \leftrightarrow YY)- \br{Z \leftrightarrow YY}\gamma(Z\leftrightarrow XX) \\
  &= \gamma(XX \leftrightarrow YY)- \br{Z \leftrightarrow XX}\gamma(Z\leftrightarrow YY)\,,
\end{split}
\end{align}
where 
\begin{align}
  \br{Z \leftrightarrow YY} = \frac{\Gamma\left(Z \to YY\right)}{\Gamma\left(Z \to \mathrm{everything}\right)}
\end{align}
is the branching ratio and 
\begin{align}
  \br{Z \leftrightarrow YY}\gamma(Z\leftrightarrow XX) =  \br{Z \leftrightarrow XX}\gamma(Z\leftrightarrow YY)\,.
\end{align}
Consquently, the resulting Boltzmann equation is given by 
\begin{align}
  sH z \dyz{X} = -2\left(\delta_X^2 -\delta_Y^2\right)\gamma^{off}(XX \leftrightarrow YY) - 2\left(\delta_X^2 -\delta_Z\right)\gamma(Z\leftrightarrow XX)\,.
\end{align}

\subsection{Feynman Diagrams}
\label{sec:app:feynman}
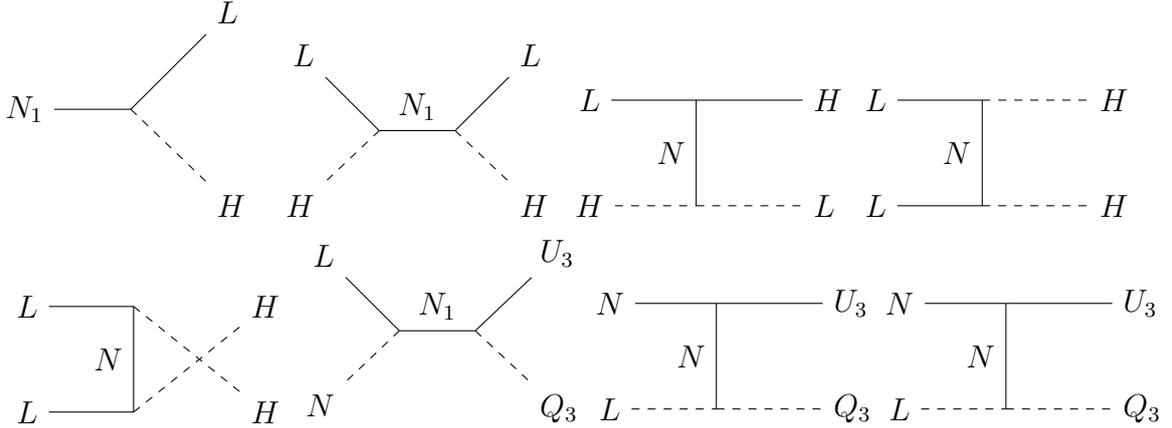
\begin{figure}[H]
\begin{subfigure}{0.22\textwidth}
  \centering
  \begin{tikzpicture}
  \begin{feynman}
  \vertex (a1) {\(N_1\)};
  \vertex[right=1.4cm of a1] (a2);
  \vertex[above right=1.4cm of a2] (b1){\(L\)};
  \vertex[below right=1.4cm of a2](b2){\(H\)};
  \diagram*{
    (a1) --(a2) , 
    (a2) -- (b1),
     (a2) -- [scalar](b2),
  };
  \end{feynman}
  \end{tikzpicture}
  \end{subfigure}
\begin{subfigure}{0.22\textwidth}
  \centering
  \begin{tikzpicture}
  \begin{feynman}
  \vertex (a1);
  \vertex[right=1cm of a1] (a2);
  \vertex[above right=1cm of a2] (b1){\(L\)};
  \vertex[below right=1cm of a2](b2){\(H\)};
  \vertex[above left=1cm of a1] (c1){\(L\)};
  \vertex[below left=1cm of a1](c2){\(H\)};
  \diagram*{
    (a1) --[edge label=\(N_1\)](a2) , 
    (a2) -- (b1),
    (a2) -- [scalar](b2),
    (a1) -- (c1),
    (a1) -- [scalar](c2)
  };
  \end{feynman}
  \end{tikzpicture}
  \end{subfigure}
  \begin{subfigure}{0.22\textwidth}
  \centering
  \begin{tikzpicture}
  \begin{feynman}
  \vertex (a1){\(L\)};
  \vertex[right=1.4cm of a1] (a2);
  \vertex[right=1.4cm of a2] (a3){\(H\)};
  \vertex[below=1.4cm of a1] (b1){\(H\)};
  \vertex[right=1.4cm of b1] (b2);
  \vertex[right=1.4cm of b2] (b3){\(L\)};
  \diagram*{
    (a1) --(a2) --(a3),
    (b1) --[scalar](b2) --[scalar](b3),
    (a2) -- [edge label'=\(N\)](b2),
  };
  \end{feynman}
  \end{tikzpicture}
  \end{subfigure}
  \begin{subfigure}{0.22\textwidth}
  \centering
  \begin{tikzpicture}
  \begin{feynman}
  \vertex (a1){\(L\)};
  \vertex[right=1.4cm of a1] (a2);
  \vertex[right=1.4cm of a2] (a3){\(H\)};
  \vertex[below=1.4cm of a1] (b1){\(L\)};
  \vertex[right=1.4cm of b1] (b2);
  \vertex[right=1.4cm of b2] (b3){\(H\)};
  \diagram*{
    (a1) --(a2) --[scalar](a3),
    (b1) --(b2) --[scalar](b3),
    (a2) -- [edge label'=\(N\)](b2),
  };
  \end{feynman}
  \end{tikzpicture}
  \end{subfigure}\\
  \begin{subfigure}{0.22\textwidth}
  \centering
  \begin{tikzpicture}
  \begin{feynman}
  \vertex (a1){\(L\)};
  \vertex[right=1.4cm of a1] (a2);
  \vertex[right=1.4cm of a2] (a3){\(H\)};
  \vertex[below=1.4cm of a1] (b1){\(L\)};
  \vertex[right=1.4cm of b1] (b2);
  \vertex[right=1.4cm of b2] (b3){\(H\)};
  \diagram*{
    (a1) --(a2) --[scalar](b3),
    (b1) --(b2) --[scalar](a3),
    (a2) -- [edge label'=\(N\)](b2),
  };
  \end{feynman}
  \end{tikzpicture}
  \end{subfigure}
  \centering
  \begin{subfigure}{0.22\textwidth}
  \centering
  \begin{tikzpicture}
  \begin{feynman}
  \vertex (a1);
  \vertex[right=1cm of a1] (a2);
  \vertex[above right=1cm of a2] (b1){\(U_3\)};
  \vertex[below right=1cm of a2](b2){\(Q_3\)};
  \vertex[above left=1cm of a1] (c1){\(L\)};
  \vertex[below left=1cm of a1](c2){\(N\)};
  \diagram*{
    (a1) --[edge label=\(N_1\)](a2) , 
    (a2) -- (b1),
    (a2) -- [scalar](b2),
    (a1) -- (c1),
    (a1) -- [scalar](c2)
  };
  \end{feynman}
  \end{tikzpicture}
  \end{subfigure}
  \begin{subfigure}{0.22\textwidth}
  \centering
  \begin{tikzpicture}
  \begin{feynman}
  \vertex (a1){\(N\)};
  \vertex[right=1.4cm of a1] (a2);
  \vertex[right=1.4cm of a2] (a3){\(U_3\)};
  \vertex[below=1.4cm of a1] (b1){\(L\)};
  \vertex[right=1.4cm of b1] (b2);
  \vertex[right=1.4cm of b2] (b3){\(Q_3\)};
  \diagram*{
    (a1) --(a2) --(a3),
    (b1) --[scalar](b2) --[scalar](b3),
    (a2) -- [edge label'=\(N\)](b2),
  };
  \end{feynman}
  \end{tikzpicture}
  \end{subfigure}
    \begin{subfigure}{0.22\textwidth}
  \centering
  \begin{tikzpicture}
  \begin{feynman}
  \vertex (a1){\(N\)};
  \vertex[right=1.4cm of a1] (a2);
  \vertex[right=1.4cm of a2] (a3){\(U_3\)};
  \vertex[below=1.4cm of a1] (b1){\(L\)};
  \vertex[right=1.4cm of b1] (b2);
  \vertex[right=1.4cm of b2] (b3){\(Q_3\)};
  \diagram*{
    (a1) --(a2) --(a3),
    (b1) --[scalar](b2) --[scalar](b3),
    (a2) -- [edge label'=\(N\)](b2),
  };
  \end{feynman}
  \end{tikzpicture}
  \end{subfigure}
  \caption{Feynman diagrams contributing both to VL and to leptogenesis in the majoron+triplet model.}
  \label{fig:feynman:vl}
\end{figure}

\begin{figure}[H]
  \centering
    \begin{subfigure}{0.22\textwidth}
  \centering
  \begin{tikzpicture}
  \begin{feynman}
  \vertex (a1){\(\sigma\)};
  \vertex[right=1.2cm of a1] (a2);
  \vertex[above right=1.2cm of a2] (b1){\(N\)};
  \vertex[below right=1.2cm of a2](b2){\(N\)};
  \diagram*{
    (a1) --[scalar](a2) , 
    (a2) -- (b1),
    (a2) -- (b2),
  };
  \end{feynman}
  \end{tikzpicture}
  \end{subfigure}
  \begin{subfigure}{0.22\textwidth}
  \centering
  \begin{tikzpicture}
  \begin{feynman}
  \vertex (a1);
  \vertex[right=1cm of a1] (a2);
  \vertex[above right=1cm of a2] (b1){\(\latexchi\)};
  \vertex[below right=1cm of a2](b2){\(\latexchi\)};
  \vertex[above left=1cm of a1] (c1){\(N\)};
  \vertex[below left=1cm of a1](c2){\(N\)};
  \diagram*{
    (a1) --[scalar, edge label=\(J\)](a2) , 
    (a2) -- (b1),
    (a2) -- (b2),
    (a1) -- (c1),
    (a1) -- (c2)
  };
  \end{feynman}
  \end{tikzpicture}
  \end{subfigure}
  \begin{subfigure}{0.22\textwidth}
  \centering
  \begin{tikzpicture}
  \begin{feynman}
  \vertex (a1);
  \vertex[right=1cm of a1] (a2);
  \vertex[above right=1cm of a2] (b1){\(\latexchi\)};
  \vertex[below right=1cm of a2](b2){\(\latexchi\)};
  \vertex[above left=1cm of a1] (c1){\(N\)};
  \vertex[below left=1cm of a1](c2){\(N\)};
  \diagram*{
    (a1) --[scalar, edge label=\(\sigma\)](a2) , 
    (a2) --(b1),
    (a2) -- (b2),
    (a1) -- (c1),
    (a1) -- (c2)
  };
  \end{feynman}
  \end{tikzpicture}
  \end{subfigure}
  \begin{subfigure}{0.22\textwidth}
  \centering
  \begin{tikzpicture}
  \begin{feynman}
  \vertex (a1);
  \vertex[right=1cm of a1] (a2);
  \vertex[above right=1cm of a2] (b1){\(\sigma\)};
  \vertex[below right=1cm of a2](b2){\(J\)};
  \vertex[above left=1cm of a1] (c1){\(N\)};
  \vertex[below left=1cm of a1](c2){\(N\)};
  \diagram*{
    (a1) --[scalar, edge label=\(J\)](a2) , 
    (a2) -- [scalar](b1),
    (a2) -- [scalar](b2),
    (a1) -- (c1),
    (a1) -- (c2)
  };
  \end{feynman}
  \end{tikzpicture}
  \end{subfigure}\\
  \begin{subfigure}{0.3\textwidth}
  \centering
  \begin{tikzpicture}
  \begin{feynman}
  \vertex (a1){\(N\)};
  \vertex[right=1.4cm of a1] (a2);
  \vertex[right=1.4cm of a2] (a3){\(\sigma\)};
  \vertex[below=1.4cm of a1] (b1){\(N\)};
  \vertex[right=1.4cm of b1] (b2);
  \vertex[right=1.4cm of b2] (b3){\(J\)};
  \diagram*{
    (a1) --(a2) --[scalar](a3),
    (b1) --(b2) --[scalar](b3),
    (a2) -- [edge label'=\(N\)](b2),
  };
  \end{feynman}
  \end{tikzpicture}
  \end{subfigure}
  \begin{subfigure}{0.22\textwidth}
  \centering
  \begin{tikzpicture}
  \begin{feynman}
  \vertex (a1);
  \vertex[right=1cm of a1] (a2);
  \vertex[above right=1cm of a2] (b1){\(N\)};
  \vertex[below right=1cm of a2](b2){\(J\)};
  \vertex[above left=1cm of a1] (c1){\(\sigma\)};
  \vertex[below left=1cm of a1](c2){\(N\)};
  \diagram*{
    (a1) --[ edge label=\(N\)](a2) , 
    (a2) -- (b1),
    (a2) -- [scalar](b2),
    (a1) --[scalar] (c1),
    (a1) -- (c2)
  };
  \end{feynman}
  \end{tikzpicture}
  \end{subfigure}
  \begin{subfigure}{0.22\textwidth}
  \centering
  \begin{tikzpicture}
  \begin{feynman}
  \vertex (a1){\(\sigma\)};
  \vertex[right=1.4cm of a1] (a2);
  \vertex[right=1.4cm of a2] (a3){\(N\)};
  \vertex[below=1.4cm of a1] (b1){\(N\)};
  \vertex[right=1.4cm of b1] (b2);
  \vertex[right=1.4cm of b2] (b3){\(J\)};
  \diagram*{
    (a1) --[scalar](a2) --(a3),
    (b1) --(b2) --[scalar](b3),
    (a2) -- [edge label'=\(N\)](b2),
  };
  \end{feynman}
  \end{tikzpicture}
  \end{subfigure}
  \begin{subfigure}{0.22\textwidth}
  \centering
  \begin{tikzpicture}
  \begin{feynman}
  \vertex (a1);
  \vertex[right=1cm of a1] (a2);
  \vertex[above right=1cm of a2] (b1){\(J\)};
  \vertex[below right=1cm of a2](b2){\(J\)};
  \vertex[above left=1cm of a1] (c1){\(N\)};
  \vertex[below left=1cm of a1](c2){\(N\)};
  \diagram*{
    (a1) --[scalar, edge label=\(\sigma\)](a2) , 
    (a2) -- [scalar](b1),
    (a2) -- [scalar](b2),
    (a1) -- (c1),
    (a1) -- (c2)
  };
  \end{feynman}
  \end{tikzpicture}
  \end{subfigure}\\
  \begin{subfigure}{0.22\textwidth}
  \centering
  \begin{tikzpicture}
  \begin{feynman}
  \vertex (a1){\(N\)};
  \vertex[right=1.4cm of a1] (a2);
  \vertex[right=1.4cm of a2] (a3){\(J\)};
  \vertex[below=1.4cm of a1] (b1){\(N\)};
  \vertex[right=1.4cm of b1] (b2);
  \vertex[right=1.4cm of b2] (b3){\(J\)};
  \diagram*{
    (a1) --(a2) --[scalar](a3),
    (b1) --(b2) --[scalar](b3),
    (a2) -- [edge label'=\(N\)](b2),
  };
  \end{feynman}
  \end{tikzpicture}
  \end{subfigure}
  \begin{subfigure}{0.3\textwidth}
  \centering
  \begin{tikzpicture}
  \begin{feynman}
  \vertex (a1){\(N\)};
  \vertex[right=1.4cm of a1] (a2);
  \vertex[right=1.4cm of a2] (a3){\(J\)};
  \vertex[below=1.4cm of a1] (b1){\(N\)};
  \vertex[right=1.4cm of b1] (b2);
  \vertex[right=1.4cm of b2] (b3){\(J\)};
  \diagram*{
    (a1) --(a2) --[scalar](b3),
    (b1) --(b2) --[scalar](a3),
    (a2) -- [edge label'=\(N\)](b2),
  };
  \end{feynman}
  \end{tikzpicture}
  \end{subfigure} 
   \begin{subfigure}{0.25\textwidth}
  \centering
  \begin{tikzpicture}
  \begin{feynman}
  \vertex (a1);
  \vertex[right=1cm of a1] (a2);
  \vertex[above right=1cm of a2] (b1){\(\sigma\)};
  \vertex[below right=1cm of a2](b2){\(\sigma\)};
  \vertex[above left=1cm of a1] (c1){\(N\)};
  \vertex[below left=1cm of a1](c2){\(N\)};
  \diagram*{
    (a1) --[scalar, edge label=\(J\)](a2) , 
    (a2) -- [scalar](b1),
    (a2) -- [scalar](b2),
    (a1) -- (c1),
    (a1) -- (c2)
  };
  \end{feynman}
  \end{tikzpicture}
  \end{subfigure}
  \begin{subfigure}{0.25\textwidth}
  \centering
  \begin{tikzpicture}
  \begin{feynman}
  \vertex (a1);
  \vertex[right=1cm of a1] (a2);
  \vertex[above right=1cm of a2] (b1){\(\sigma\)};
  \vertex[below right=1cm of a2](b2){\(\sigma\)};
  \vertex[above left=1cm of a1] (c1){\(N\)};
  \vertex[below left=1cm of a1](c2){\(N\)};
  \diagram*{
    (a1) --[scalar, edge label=\(\sigma\)](a2) , 
    (a2) -- [scalar](b1),
    (a2) -- [scalar](b2),
    (a1) -- (c1),
    (a1) -- (c2)
  };
  \end{feynman}
  \end{tikzpicture}
  \end{subfigure}
  \begin{subfigure}{0.23\textwidth}
  \centering
  \begin{tikzpicture}
  \begin{feynman}
  \vertex (a1){\(N\)};
  \vertex[right=1.4cm of a1] (a2);
  \vertex[right=1.4cm of a2] (a3){\(\sigma\)};
  \vertex[below=1.4cm of a1] (b1){\(N\)};
  \vertex[right=1.4cm of b1] (b2);
  \vertex[right=1.4cm of b2] (b3){\(\sigma\)};
  \diagram*{
    (a1) --(a2) --[scalar](a3),
    (b1) --(b2) --[scalar](b3),
    (a2) -- [edge label'=\(N\)](b2),
  };
  \end{feynman}
  \end{tikzpicture}
  \end{subfigure}
  \begin{subfigure}{0.23\textwidth}
  \centering
  \begin{tikzpicture}
  \begin{feynman}
  \vertex (a1){\(N\)};
  \vertex[right=1.4cm of a1] (a2);
  \vertex[right=1.4cm of a2] (a3){\(\sigma\)};
  \vertex[below=1.4cm of a1] (b1){\(N\)};
  \vertex[right=1.4cm of b1] (b2);
  \vertex[right=1.4cm of b2] (b3){\(\sigma\)};
  \diagram*{
    (a1) --(a2) --[scalar](b3),
    (b1) --(b2) --[scalar](a3),
    (a2) -- [edge label'=\(N\)](b2),
  };
  \end{feynman}
  \end{tikzpicture}
  \end{subfigure}
  \caption{Feynman diagrams for interactions that change the neutrino abundance in the majoron+triplet model.}
  \label{fig:feynman:N}
\end{figure}
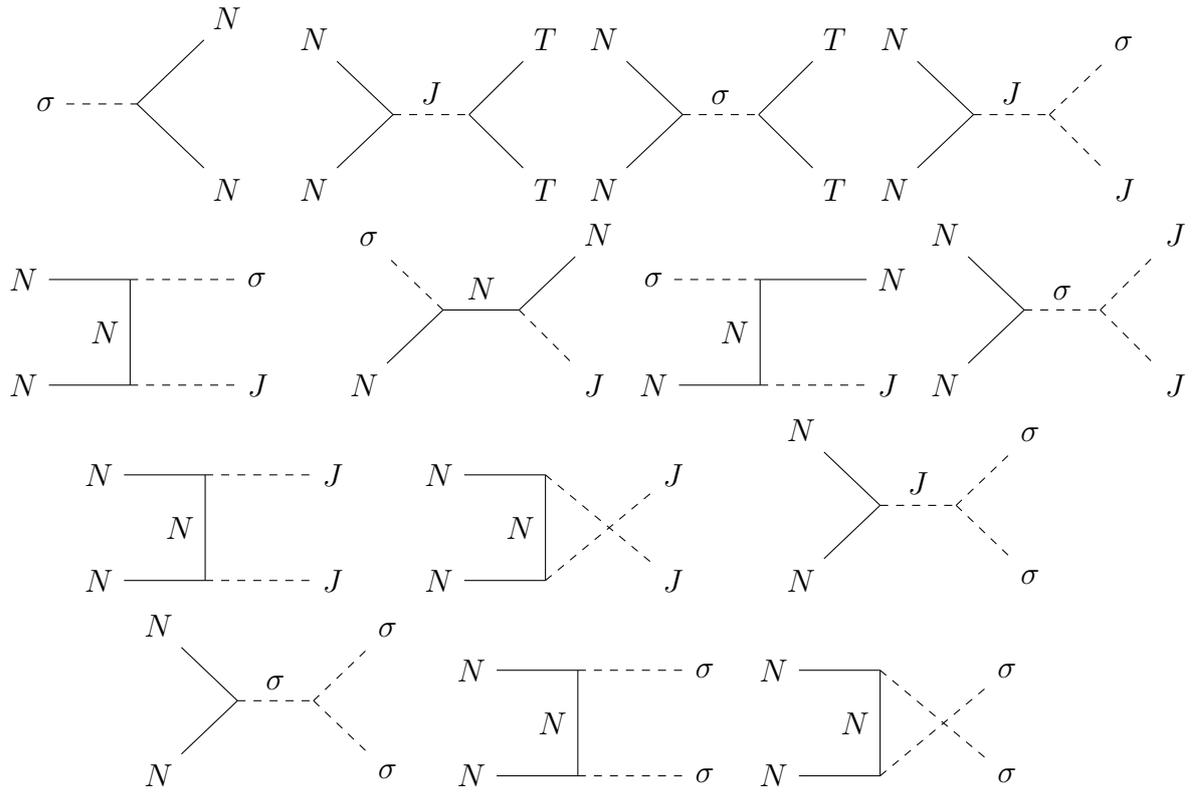

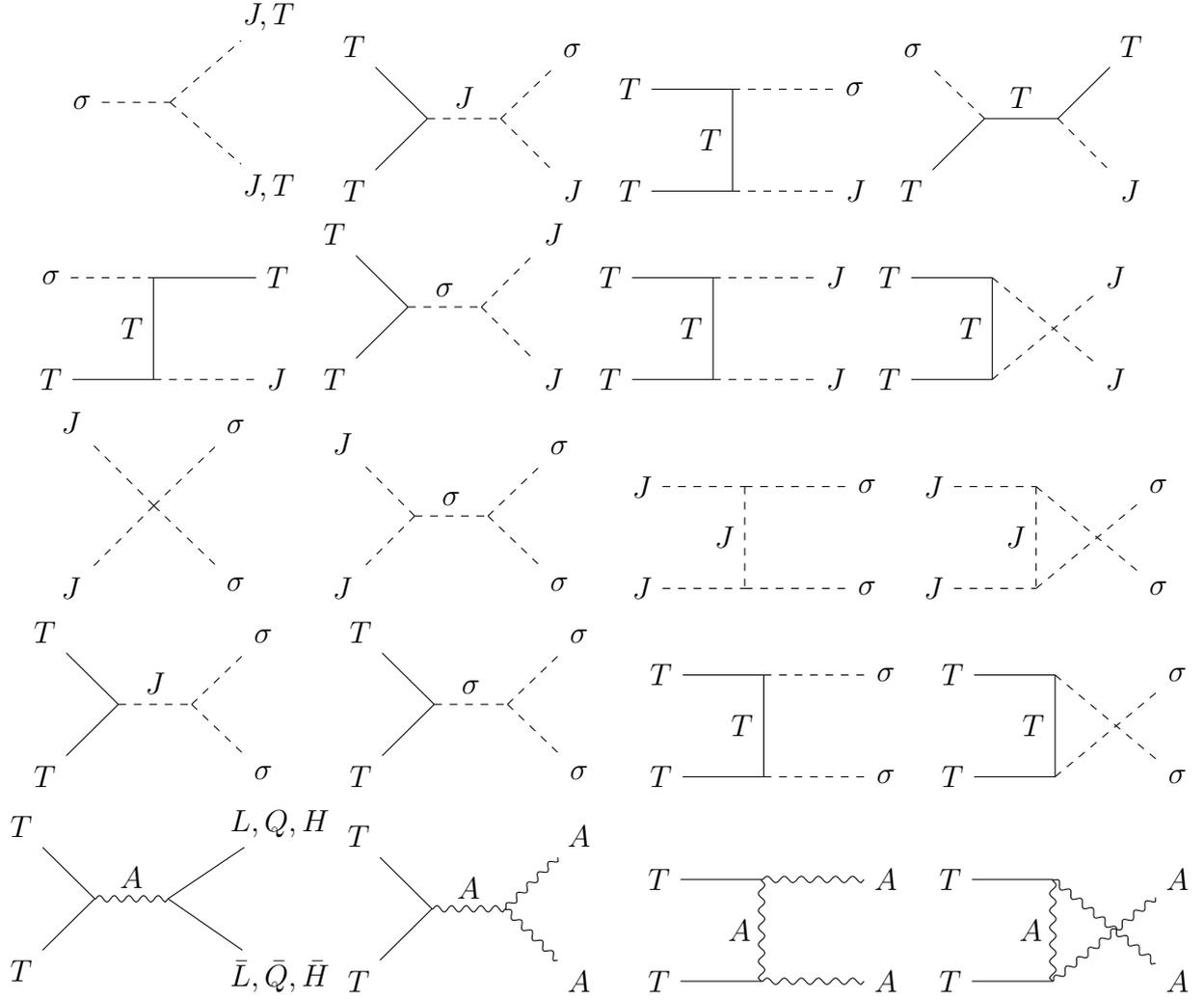
\begin{figure}[H]
  \centering
  \begin{subfigure}{0.22\textwidth}
  \centering
  \begin{tikzpicture}
  \begin{feynman}
  \vertex (a1){\(\sigma\)};
  \vertex[right=1.2cm of a1] (a2);
  \vertex[above right=1.2cm of a2] (b1){\(J,\latexchi\)};
  \vertex[below right=1.2cm of a2](b2){\(J,\latexchi\)};
  \diagram*{
    (a1) --[scalar](a2) , 
    (a2) -- [scalar](b1),
    (a2) -- [scalar](b2),

  };
  \end{feynman}
  \end{tikzpicture}
  \end{subfigure}
  \begin{subfigure}{0.22\textwidth}
  \centering
  \begin{tikzpicture}
  \begin{feynman}
  \vertex (a1);
  \vertex[right=1cm of a1] (a2);
  \vertex[above right=1cm of a2] (b1){\(\sigma\)};
  \vertex[below right=1cm of a2](b2){\(J\)};
  \vertex[above left=1cm of a1] (c1){\(\latexchi\)};
  \vertex[below left=1cm of a1](c2){\(\latexchi\)};
  \diagram*{
    (a1) --[scalar, edge label=\(J\)](a2) , 
    (a2) -- [scalar](b1),
    (a2) -- [scalar](b2),
    (a1) -- (c1),
    (a1) -- (c2)
  };
  \end{feynman}
  \end{tikzpicture}
  \end{subfigure}
  \begin{subfigure}{0.22\textwidth}
  \centering
  \begin{tikzpicture}
  \begin{feynman}
  \vertex (a1){\(\latexchi\)};
  \vertex[right=1.4cm of a1] (a2);
  \vertex[right=1.4cm of a2] (a3){\(\sigma\)};
  \vertex[below=1.4cm of a1] (b1){\(\latexchi\)};
  \vertex[right=1.4cm of b1] (b2);
  \vertex[right=1.4cm of b2] (b3){\(J\)};
  \diagram*{
    (a1) --(a2) --[scalar](a3),
    (b1) --(b2) --[scalar](b3),
    (a2) -- [edge label'=\(\latexchi\)](b2),
  };
  \end{feynman}
  \end{tikzpicture}
  \end{subfigure}
  \begin{subfigure}{0.22\textwidth}
  \centering
  \begin{tikzpicture}
  \begin{feynman}
  \vertex (a1);
  \vertex[right=1cm of a1] (a2);
  \vertex[above right=1cm of a2] (b1){\(\latexchi\)};
  \vertex[below right=1cm of a2](b2){\(J\)};
  \vertex[above left=1cm of a1] (c1){\(\sigma\)};
  \vertex[below left=1cm of a1](c2){\(\latexchi\)};
  \diagram*{
    (a1) --[ edge label=\(\latexchi\)](a2) , 
    (a2) -- (b1),
    (a2) -- [scalar](b2),
    (a1) --[scalar] (c1),
    (a1) -- (c2)
  };
  \end{feynman}
  \end{tikzpicture}
  \end{subfigure} 
  \begin{subfigure}{0.22\textwidth}
  \centering
  \begin{tikzpicture}
  \begin{feynman}
  \vertex (a1){\(\sigma\)};
  \vertex[right=1.4cm of a1] (a2);
  \vertex[right=1.4cm of a2] (a3){\(\latexchi\)};
  \vertex[below=1.4cm of a1] (b1){\(\latexchi\)};
  \vertex[right=1.4cm of b1] (b2);
  \vertex[right=1.4cm of b2] (b3){\(J\)};
  \diagram*{
    (a1) --[scalar](a2) --(a3),
    (b1) --(b2) --[scalar](b3),
    (a2) -- [edge label'=\(\latexchi\)](b2),
  };
  \end{feynman}
  \end{tikzpicture}
  \end{subfigure}
  \begin{subfigure}{0.22\textwidth}
  \centering
  \begin{tikzpicture}
  \begin{feynman}
  \vertex (a1);
  \vertex[right=1cm of a1] (a2);
  \vertex[above right=1cm of a2] (b1){\(J\)};
  \vertex[below right=1cm of a2](b2){\(J\)};
  \vertex[above left=1cm of a1] (c1){\(\latexchi\)};
  \vertex[below left=1cm of a1](c2){\(\latexchi\)};
  \diagram*{
    (a1) --[scalar, edge label=\(\sigma\)](a2) , 
    (a2) -- [scalar](b1),
    (a2) -- [scalar](b2),
    (a1) -- (c1),
    (a1) -- (c2)
  };
  \end{feynman}
  \end{tikzpicture}
  \end{subfigure}
  \begin{subfigure}{0.22\textwidth}
  \centering
  \begin{tikzpicture}
  \begin{feynman}
  \vertex (a1){\(\latexchi\)};
  \vertex[right=1.4cm of a1] (a2);
  \vertex[right=1.4cm of a2] (a3){\(J\)};
  \vertex[below=1.4cm of a1] (b1){\(\latexchi\)};
  \vertex[right=1.4cm of b1] (b2);
  \vertex[right=1.4cm of b2] (b3){\(J\)};
  \diagram*{
    (a1) --(a2) --[scalar](a3),
    (b1) --(b2) --[scalar](b3),
    (a2) -- [edge label'=\(\latexchi\)](b2),
  };
  \end{feynman}
  \end{tikzpicture}
  \end{subfigure}
  \begin{subfigure}{0.22\textwidth}
  \centering
  \begin{tikzpicture}
  \begin{feynman}
  \vertex (a1){\(\latexchi\)};
  \vertex[right=1.4cm of a1] (a2);
  \vertex[right=1.4cm of a2] (a3){\(J\)};
  \vertex[below=1.4cm of a1] (b1){\(\latexchi\)};
  \vertex[right=1.4cm of b1] (b2);
  \vertex[right=1.4cm of b2] (b3){\(J\)};
  \diagram*{
    (a1) --(a2) --[scalar](b3),
    (b1) --(b2) --[scalar](a3),
    (a2) -- [edge label'=\(\latexchi\)](b2),
  };
  \end{feynman}
  \end{tikzpicture}
  \end{subfigure} \newline
  \begin{subfigure}{0.22\textwidth}
  \centering
  \begin{tikzpicture}
  \begin{feynman}
  \vertex (a1);
  \vertex[above right=1.2cm of a1] (b1){\(\sigma\)};
  \vertex[below right=1.2cm of a1](b2){\(\sigma\)};
  \vertex[above left=1.2cm of a1] (c1){\(J\)};
  \vertex[below left=1.2cm of a1](c2){\(J\)};
  \diagram*{
    (a1) -- [scalar](b1),
    (a1) -- [scalar](b2),
    (a1) -- [scalar](c1),
    (a1) -- [scalar](c2)
  };
  \end{feynman}
  \end{tikzpicture}
  \end{subfigure}
  \begin{subfigure}{0.25\textwidth}
  \centering
  \begin{tikzpicture}
  \begin{feynman}
  \vertex (a1);
  \vertex[right=1cm of a1] (a2);
  \vertex[above right=1cm of a2] (b1){\(\sigma\)};
  \vertex[below right=1cm of a2](b2){\(\sigma\)};
  \vertex[above left=1cm of a1] (c1){\(J\)};
  \vertex[below left=1cm of a1](c2){\(J\)};
  \diagram*{
    (a1) --[scalar, edge label=\(\sigma\)](a2) , 
    (a2) -- [scalar](b1),
    (a2) -- [scalar](b2),
    (a1) -- [scalar](c1),
    (a1) -- [scalar](c2)
  };
  \end{feynman}
  \end{tikzpicture}
  \end{subfigure}
  \begin{subfigure}{0.23\textwidth}
  \centering
  \begin{tikzpicture}
  \begin{feynman}
  \vertex (a1){\(J\)};
  \vertex[right=1.4cm of a1] (a2);
  \vertex[right=1.4cm of a2] (a3){\(\sigma\)};
  \vertex[below=1.4cm of a1] (b1){\(J\)};
  \vertex[right=1.4cm of b1] (b2);
  \vertex[right=1.4cm of b2] (b3){\(\sigma\)};
  \diagram*{
    (a1) --[scalar](a2) --[scalar](a3),
    (b1) --[scalar](b2) --[scalar](b3),
    (a2) -- [scalar,edge label'=\(J\)](b2),
  };
  \end{feynman}
  \end{tikzpicture}
  \end{subfigure}
  \begin{subfigure}{0.23\textwidth}
  \centering
  \begin{tikzpicture}
  \begin{feynman}
  \vertex (a1){\(J\)};
  \vertex[right=1.4cm of a1] (a2);
  \vertex[right=1.4cm of a2] (a3){\(\sigma\)};
  \vertex[below=1.4cm of a1] (b1){\(J\)};
  \vertex[right=1.4cm of b1] (b2);
  \vertex[right=1.4cm of b2] (b3){\(\sigma\)};
  \diagram*{
    (a1) --[scalar](a2) --[scalar](b3),
    (b1) --[scalar](b2) --[scalar](a3),
    (a2) -- [scalar,edge label'=\(J\)](b2),
  };
  \end{feynman}
  \end{tikzpicture}
  \end{subfigure}\\
   \begin{subfigure}{0.25\textwidth}
  \centering
  \begin{tikzpicture}
  \begin{feynman}
  \vertex (a1);
  \vertex[right=1cm of a1] (a2);
  \vertex[above right=1cm of a2] (b1){\(\sigma\)};
  \vertex[below right=1cm of a2](b2){\(\sigma\)};
  \vertex[above left=1cm of a1] (c1){\(\latexchi\)};
  \vertex[below left=1cm of a1](c2){\(\latexchi\)};
  \diagram*{
    (a1) --[scalar, edge label=\(J\)](a2) , 
    (a2) -- [scalar](b1),
    (a2) -- [scalar](b2),
    (a1) -- (c1),
    (a1) -- (c2)
  };
  \end{feynman}
  \end{tikzpicture}
  \end{subfigure}
  \begin{subfigure}{0.25\textwidth}
  \centering
  \begin{tikzpicture}
  \begin{feynman}
  \vertex (a1);
  \vertex[right=1cm of a1] (a2);
  \vertex[above right=1cm of a2] (b1){\(\sigma\)};
  \vertex[below right=1cm of a2](b2){\(\sigma\)};
  \vertex[above left=1cm of a1] (c1){\(\latexchi\)};
  \vertex[below left=1cm of a1](c2){\(\latexchi\)};
  \diagram*{
    (a1) --[scalar, edge label=\(\sigma\)](a2) , 
    (a2) -- [scalar](b1),
    (a2) -- [scalar](b2),
    (a1) -- (c1),
    (a1) -- (c2)
  };
  \end{feynman}
  \end{tikzpicture}
  \end{subfigure}
  \begin{subfigure}{0.23\textwidth}
  \centering
  \begin{tikzpicture}
  \begin{feynman}
  \vertex (a1){\(\latexchi\)};
  \vertex[right=1.4cm of a1] (a2);
  \vertex[right=1.4cm of a2] (a3){\(\sigma\)};
  \vertex[below=1.4cm of a1] (b1){\(\latexchi\)};
  \vertex[right=1.4cm of b1] (b2);
  \vertex[right=1.4cm of b2] (b3){\(\sigma\)};
  \diagram*{
    (a1) --(a2) --[scalar](a3),
    (b1) --(b2) --[scalar](b3),
    (a2) -- [edge label'=\(\latexchi\)](b2),
  };
  \end{feynman}
  \end{tikzpicture}
  \end{subfigure}
  \begin{subfigure}{0.23\textwidth}
  \centering
  \begin{tikzpicture}
  \begin{feynman}
  \vertex (a1){\(\latexchi\)};
  \vertex[right=1.4cm of a1] (a2);
  \vertex[right=1.4cm of a2] (a3){\(\sigma\)};
  \vertex[below=1.4cm of a1] (b1){\(\latexchi\)};
  \vertex[right=1.4cm of b1] (b2);
  \vertex[right=1.4cm of b2] (b3){\(\sigma\)};
  \diagram*{
    (a1) --(a2) --[scalar](b3),
    (b1) --(b2) --[scalar](a3),
    (a2) -- [edge label'=\(\latexchi\)](b2),
  };
  \end{feynman}
  \end{tikzpicture}
  \end{subfigure}\\
  \begin{subfigure}{0.25\textwidth}
  \centering
  \begin{tikzpicture}
  \begin{feynman}
  \vertex (a1) ;
  \vertex[right=1cm of a1] (a2);
  \vertex[above right=1cm of a2] (b1){\(L,Q,H\)};
  \vertex[below right=1cm of a2](b2){\(\bar L,\bar Q, \bar H\)};
  \vertex[above left=1cm of a1] (c1){\(\latexchi\)};
  \vertex[below left=1cm of a1](c2){\(\latexchi\)};
  \diagram*{
    (a1) --[boson,edge label=\(A\)](a2) , 
    (a2) -- (b1),
    (a2) -- (b2),
    (a1) -- (c1),
    (a1) -- (c2)
  };
  \end{feynman}
  \end{tikzpicture}
  \end{subfigure}
  \begin{subfigure}{0.25\textwidth}
  \centering
  \begin{tikzpicture}
  \begin{feynman}
  \vertex (a1) ;
  \vertex[right=1cm of a1] (a2);
  \vertex[above right=1cm of a2] (b1){\(A\)};
  \vertex[below right=1cm of a2](b2){\(A\)};
  \vertex[above left=1cm of a1] (c1){\(\latexchi\)};
  \vertex[below left=1cm of a1](c2){\(\latexchi\)};
  \diagram*{
    (a1) --[boson,edge label=\(A\)](a2) , 
    (a2) -- [boson](b1),
    (a2) -- [boson](b2),
    (a1) -- (c1),
    (a1) -- (c2)
  };
  \end{feynman}
  \end{tikzpicture}
  \end{subfigure}
  \begin{subfigure}{0.23\textwidth}
  \centering
  \begin{tikzpicture}
  \begin{feynman}
  \vertex (a1){\(\latexchi\)};
  \vertex[right=1.4cm of a1] (a2);
  \vertex[right=1.4cm of a2] (a3){\(A\)};
  \vertex[below=1.4cm of a1] (b1){\(\latexchi\)};
  \vertex[right=1.4cm of b1] (b2);
  \vertex[right=1.4cm of b2] (b3){\(A\)};
  \diagram*{
    (a1) --(a2) --[boson](a3),
    (b1) --(b2) --[boson](b3),
    (a2) -- [boson,edge label'=\(A\)](b2),
  };
  \end{feynman}
  \end{tikzpicture}
  \end{subfigure}
  \begin{subfigure}{0.23\textwidth}
  \centering
  \begin{tikzpicture}
  \begin{feynman}
  \vertex (a1){\(\latexchi\)};
  \vertex[right=1.4cm of a1] (a2);
  \vertex[right=1.4cm of a2] (a3){\(A\)};
  \vertex[below=1.4cm of a1] (b1){\(\latexchi\)};
  \vertex[right=1.4cm of b1] (b2);
  \vertex[right=1.4cm of b2] (b3){\(A\)};
  \diagram*{
    (a1) --(a2) --[boson](b3),
    (b1) --(b2) --[boson](a3),
    (a2) -- [boson,edge label'=\(A\)](b2),
  };
  \end{feynman}
  \end{tikzpicture}
  \end{subfigure}
  \caption{Feynman diagrams for interactions that change the abundances of $\latexchi, \sigma$ and $J$ in the majoron+triplet model. Here, $A = \{B, W_{1,2,3}\}$ corresponds to the gauge bosons. }
  \label{fig:feynman:diagrams:rest}
\end{figure}

In this section, we present the feynman diagrams relevant for the discussed leptogenesis scenarios. \\
In Fig. \eqref{fig:feynman:vl}, we show the feynman diagrams for interactions that appear already in the VL scenario and are hence also relevant for leptogenesis in the majoron+triplet model. Note that we neglect scatterings involving gauge bosons. \\
In Fig. \eqref{fig:feynman:N}, we present feynman diagrams for interactions with neutrinos in the intial or final state that appear in the majoron+triplet model. Those interactions are particularly relevant for leptogenesis as they change the neutrino abundance. \\
In Fig. \eqref{fig:feynman:diagrams:rest}, we show feynman diagrams for the remaining scattering processes in the majoron+triplet model that change the abundances of $\latexchi, \sigma$ and $J$. Note that these interactions are only relevant in cases $A, B$ and do not affect leptogenesis in the simplified scenarios $\hat A, \hat B$. 


\subsection{Cross Sections and Matrix Elements}
\label{sec:app:rates}
The reduced cross sections for neutrino interactions involvolving third generation quarks can e.g.\ be found in \cite{Buchmuller:2004nz} and are given by 
\begin{align}
  \sigma_{Q_3U_3 NL} &= \frac{3 y_t^2}{4 \pi} \frac{\tilde m_1 \mn}{v^2}\left(\frac{x_N-1}{x_N}\right)^2 \\
  \sigma_{L\overline{Q_3} N\overline{U_3}} &=\sigma_{L\overline{U_3} N\overline{Q_3}} = \frac{3 yt^2}{4 \pi} \frac{\mn \tilde m_1 }{v^2} \frac{x-1}{x} \left( \frac{x-2+2a_r}{x-1+a_r} + \frac{1-2a_r}{x-1}\log\left( \frac{x_N-1+a_r }{a_r}\right) \right)\,,
\end{align}
where $a_r \equiv \frac{m_H}{\mn}$. \\
The reduced cross section for scatterings of the triplet $\latexchi$ involvolving gauge bosons, $\latexchi\latexchi\leftrightarrow AA, L\overline{L}, Q\overline{Q}, H\overline{H}$, is the same as in type III leptogenesis and given by \cite{Cirelli:2007xd,Hambye:2012fh,Hambye:2003rt,Strumia:2008cf}
\begin{align}
\hat\sigma_{\latexchi,gauge} &= \frac{6 g_2^4}{72\pi}\left( \frac{45}{2}\beta -\frac{27}{2} \beta^3 - \left( 9 \left( \beta^2 -2 \right)^2+18 \left( \beta^2 -1\right)^2   \right)\log\left[ \frac{1+\beta}{1-\beta} \right]   \right)\,, 
\end{align}
where $\beta = \sqrt{1-\frac{4}{x_\latexchi}}$.\\
Next, we present the cross sections and matrix elements that appear in the majoron+triplet model. For convenience, we introduce the following abbreviations: 
\begin{align}
\overline{x} &= x-1\,,\qquad \hat{x} = x+1\,,\qquad\tilde x = \left(x^2-1\right)\,,\\
A &= \sqrt{\frac{\left(\mn^2 \overline{x}+\ms^2\right)^2}{\mn^2 x}} \,,&\quad
B &= 4 \mj^2-\mn^2 x \,,\\
C &= 4 \ms^2-\mn^2 x\,,&\quad
D &= 4 \meta^2-\mn^2 x \,,\\
E &= \mj^2-\mn^2 x \,,&\quad
F &= \ms^2-\mn^2 x \,,\\
G &= \meta^2-\mn^2 x\,,&\quad
H &= \sqrt{\frac{\mn^4 \overline{x}^2-2 \mn^2 \ms^2 \hat{x}+\ms^4}{\mn^2 x}} \,,\\
O &= \sqrt{\frac{\mj^4-2 \mj^2 \mn^2 \hat{x}+\mn^4 \overline{x}^2}{\mn^2 x}}\,,&\quad
Q &= \left(-\mj^2+\mn^2 x+\ms^2\right)^2\,,\\ 
R &= \mn^2 (x-4) \,,&\quad
S &= \left(\mn^2 x+\ms^2\right)\,,\\
T &= \left(x-c_\vartheta^2 (x-4)\right)\,, &\quad U &= \left(F^2+\Gamma_\sigma^2 \ms^2\right)\,.
\end{align}  
The reduced cross sections for $NN\to\latexchi\latexchi$ can then be written as
\begin{align} 
 \hat\sigma(NN\leftrightarrow \latexchi\latexchi) &=  \frac{3 \sqrt{-D} \geta^2 \gn^2 \sqrt{R} \left(\frac{\mn^4 x^2}{E^2}-\frac{D R}{F^2+\Gamma_\sigma^2 \ms^2}\right)}{8 \pi  \mn^2 x}\,.
\end{align} 
For brevity, we only present the squablack matrix elements (summed over initial and final spins) for the remaining interactions, 
\begingroup
\allowdisplaybreaks

\begin{align}
\begin{split}
  |\overline M(NN\leftrightarrow JJ)|^2 &= \frac{\gn^2 R}{2} \times M_{NN}^{JJ}\,,
\end{split}\\
|\overline M(N\latexchi\latexchi\leftrightarrow JJ)|^2 &= 3|\overline M(NN\leftrightarrow JJ)|^2\big|_{\gn\to\geta, \mn\to\meta}\\
  |\overline M(NN\leftrightarrow \sigma J)|^2 &= \frac{\gn^2}{E^2 \mn^2 x \left(c_\vartheta \sqrt{R} \sqrt{\frac{F^2+\mj^4-2 \mj^2 S}{\mn^2 x}}+\sqrt{Q}-2 \ms^2\right)^2} \times M_{NN}^{\sigma J} \\
   |\overline M(\latexchi\latexchi\leftrightarrow \sigma J)|^2 &=  3|\overline M(NN\leftrightarrow \sigma J)|^2\big|_{\gn\to\geta, \mn\to\meta}\,,\\
  |\overline M(\sigma N\leftrightarrow JN)|^2&= \frac{\gn^4 x\left(\mn^2 x\right)^{-7/2} \times (A x \mn^2M_{\sigma N,1}^{JN} + \sqrt{\mn^2 x}M_{\sigma N,2}^{JN})}{ \left(A (\mn^2 \hat{x}-\mj^2)+\left(c_\vartheta H O-2 \ms^2\right) \sqrt{\mn^2 x}\right)^2 \left(\overline{x}^2 \mn^2+\Gamma_N^2\right)} \\
  |\overline M(\sigma \latexchi\leftrightarrow J\latexchi)|^2&=3|\overline M(\sigma N\leftrightarrow JN)|^2|_{\gn\to\geta, \mn\to\meta}\,.\\
 |\overline M(NN\leftrightarrow \sigma\sigma)|^2 &= \frac{\gn^2 R}{2 U \left(C c_\vartheta^2 R+\left(\mn^2 x-2 \ms^2\right)^2\right)^2}\times M_{NN}^{\sigma\sigma} \\
 |\overline M(\latexchi\latexchi\leftrightarrow \sigma\sigma)|^2 &=  3|\overline M(NN\leftrightarrow \sigma\sigma)|^2\big|_{\gn\to\geta, \mn\to\meta}\,,\\
  |\overline M(\sigma\sigma\leftrightarrow JJ)|^2 &= \frac{k_\sigma^4}{\ms^4 U \left(\left(\mn^2 x-2 \ms^2\right)^2-B C c_\vartheta^2\right)^2}\times M_{\sigma\sigma}^{JJ}\,,
\end{align}
where 
\begin{align}
\begin{split}
M_{NN}^{JJ}&\equiv\frac{k_\sigma^2}{U}+\frac{4 B c_\vartheta^2 \gn \mn \left(B \gn \mn T+\frac{2 F k_\sigma \left(4 \mj^2 \mn^2 T-\mn^4 x T-4 \mj^4\right)}{U}\right)}{\left(-4 \mj^2 \mn^2T+\mn^4 x T+4 \mj^4\right)^2}\,,\\
M_{NN}^{\sigma J}&\equiv-c_\vartheta^2 R \left(E^2 \gn^2-k_\sigma^2 \mn^2 x\right) \left(F^2+\mj^4-2 \mj^2 S\right) \\
  &-2 c_\vartheta k_\sigma \mn^3 \sqrt{R} x \sqrt{\frac{F^2+\mj^4-2 \mj^2 S}{\mn^2 x}} \left(2 E \gn \left(\sqrt{Q}-2 \mn^2 x\right)+k_\sigma \mn x \left(2 \ms^2-\sqrt{Q}\right)\right)\\
  &+\mn^2 x \color{black}\left\{\color{black}E^2 \gn^2 \left(\mj^4-2 \mj^2 S+\mn^4 x (x+16)-2 \mn^2 \left(8 \sqrt{Q}+\ms^2 (x-8)\right)+\ms^4\right) \right.\\
  &\left.-4 E \gn k_\sigma \mn \left\{\mj^4-2 \mj^2 S+\mn^4 x^2-2 \ms^2 \sqrt{Q}+6 \mn^2 \ms^2 x-2 \left(\mn^2 x\right)^{3/2} \sqrt{\frac{Q}{\mn^2 x}}\right.\right.\\
  &\left.\left.+\ms^4\right\}+k_\sigma^2 \mn^2 x \left(\mj^4-2 \mj^2 S+\mn^4 x^2+\ms^2 \left(2 \mn^2 x-4 \sqrt{Q}\right)+5 \ms^4\right)\color{black}\right\}\,,
\end{split} \\
\begin{split}
  M_{NN}^{\sigma\sigma}&\equiv2 C c_\vartheta^2 \left(-2 \gn^2 \left(\mn^4 (x-16) x-4 \mn^2 \ms^2 (x-8)\right) U\right. \\
  &\left.+4 \gn k_\sigma \mn \left(\mn^4 x^2-3 \mn^2 \ms^2 x+2 \ms^4\right) \left(\mn^2 (8-3 x)+2 \ms^2\right)+k_\sigma^2 R \left(\mn^2 x-2 \ms^2\right)^2\right)\\
  &+\left(\mn^2 x-2 \ms^2\right)^2 \left(64 \Gamma_\sigma^2 \gn^2 \mn^2 \ms^2+\left(\mn^2 x (k_\sigma-8 \gn \mn)-2 \ms^2 (k_\sigma-4 \gn \mn)\right)^2\right)\\
  &-\left(c_\vartheta^4 (x-4) \left(\mn^3 x-4 \mn \ms^2\right)^2 \left(-8 F \gn k_\sigma \mn+4 \gn^2 U-k_\sigma^2 R\right)\right)\,,
  \end{split}\\
\begin{split}
  M_{\sigma\sigma}^{JJ} &\equiv8 \ms^6 \left(\Gamma_\sigma^2 \left(2 B^2 c_\vartheta^4+11 B c_\vartheta^2 \mn^2 x+11 \mn^4 x^2\right)-2 \mn^4 x^2 \left(9 B c_\vartheta^2+17 \mn^2 x\right)\right)\\
  &+8 \mn^2 \ms^4 x \left(2 \mn^2 x \left(B^2 c_\vartheta^4+6 B c_\vartheta^2 \mn^2 x+6 \mn^4 x^2\right)-\Gamma_\sigma^2 \left(B c_\vartheta^2+\mn^2 x\right) \left(B c_\vartheta^2+2 \mn^2 x\right)\right)\\
  &+\mn^4 \ms^2 x^2 \left(\left(c_\vartheta^2-1\right) \mn^2 x-4 c_\vartheta^2 \mj^2\right) \left(8 \mn^2 x \left(B c_\vartheta^2+2 \mn^2 x\right)-\Gamma_\sigma^2 \left(B c_\vartheta^2+\mn^2 x\right)\right)\\
  &+32 \ms^8 \left(2 \mn^2 x \left(B c_\vartheta^2+6 \mn^2 x\right)-3 \Gamma_\sigma^2 \left(B c_\vartheta^2+2 \mn^2 x\right)\right)\\
  &+\mn^8 x^4 \left(\left(c_\vartheta^2-1\right) \mn^2 x-4 c_\vartheta^2 \mj^2\right)^2+16 \ms^{10} \left(9 \Gamma_\sigma^2-16 \mn^2 x\right)+64 \ms^{12}\,,
\end{split}
\end{align}
\endgroup

\begin{landscape}
\begin{align}
\begin{split}
	M_{\sigma N, 1}^{JN}&\equiv 
	\left(-\left(\left(3 c_\vartheta^2+1\right) \overline{x}^3 \mn^6\right)+\ms^2 \overline{x} \left((7 x+5) c_\vartheta^2-x+1\right) \mn^4+\ms^4 \left((1-5 x) c_\vartheta^2+x-1\right) \mn^2+\left(c_\vartheta^2+1\right) \ms^6\right)\mj^6 \\
	&+\mn^2 \left(3 \left(3 c_\vartheta^2+1\right) \overline{x}^3 \hat{x} \mn^6-\ms^2 \tilde{x} \left(3 (7 x+5) c_\vartheta^2+x+3\right) \mn^4+\ms^2 \left(\ms^2 \left(3 \hat{x} (5 x-1) c_\vartheta^2+5 x^2+3\right)-2 c_\vartheta H O x \hat{x}\right)\mn^2 \right.\\
	&\left.	+2 c_\vartheta H \ms^4 O x-\ms^6 \left(3 \hat{x} c_\vartheta^2+7 x+3\right)\right) \mj^4+\mn^4 \left(-\overline{x}^2 \overline{x} \left(\left(9 x^2+6 x+9\right) c_\vartheta^2+x (7 x+6)+3\right)
	\right.\\
	&\left. \mn^6+\left(\overline{x} \left((7 x+5) (x (3 x+2)+3) c_\vartheta^2+x (17 x \hat{x}+11)+3\right) \ms^2-4 x^2 (x+3) \Gamma_N^2-8 c_\vartheta H O x^2 \tilde{x}\right)\mn^4
\right.\\
	&\left.
	 -\ms^2 \left(\left((5 x-1) (x (3 x+2)+3) c_\vartheta^2+x (x (25 x+17)+3)+3\right) \ms^2-4 x^2 \Gamma_N^2-4 c_\vartheta H O x \left(4 x^2+x+1\right)\right) \mn^2
\right.\\
	&\left.
	 -4 c_\vartheta H \ms^4 O x (2 x+1)+\ms^6 \left((x (3 x+2)+3) c_\vartheta^2+(3 x+1) (5 x+3)\right)\right)\\
	& \mj^2+\mn^6 \hat{x} \left(\overline{x}^3 \left(3 \overline{x}^2 c_\vartheta^2+x (5 x+2)+1\right) \mn^6+\left(4 x^2 \left((x+3) \Gamma_N^2+2 c_\vartheta H O \tilde{x}\right)
\right.\right.\\
	&\left.\left.	
	-\ms^2 \overline{x}	\left(\overline{x}^2 (7 x+5) c_\vartheta^2+(3 x+1) (x (5 x+2)+1)\right)\right) \mn^4+\ms^2 \left(\ms^2 \left(\overline{x}^2 (5 x-1) c_\vartheta^2+x+x^2 (19 x+11)+1\right)-
\right.\right.\\
	&\left.\left.	
	2 x \left(2 x \Gamma_N^2+c_\vartheta H O \left(7 x^2+1\right)\right)\right) \mn^2+2 c_\vartheta H \ms^4 O x (3 x+1)-\ms^6 \left(\overline{x}^2 c_\vartheta^2+(3 x+1)^2\right)\right)\,,
	\end{split}\nonumber \\
\begin{split}
	M_{\sigma N}^{JN, 2}&\equiv\ms^2 \left(\mn^2 \hat{x}-\ms^2\right) \left(\left(c_\vartheta^2+1\right) \overline{x}^2 \mn^4-2 \ms^2 \left(\hat{x} c_\vartheta^2-x+1\right) \mn^2+\left(c_\vartheta^2+1\right) \ms^4\right)\mj^6\\
	& +\mn^2 \left(4 \left(c_\vartheta^2+1\right) \overline{x}^3 x \hat{x} \mn^8-\overline{x} \left(\ms^2 \left(3 \overline{x}^3+c_\vartheta^2 (5 x (3 x \hat{x}+1)-3)\right)-c_\vartheta \left(c_\vartheta^2+3\right) H O \overline{x}^2 x\right)\mn^6
\right.\\
	&\left.
	+\ms^2 \left(\ms^2 \left(\hat{x} (x (21 x+2)+9) c_\vartheta^2-5 x+x^2 (7-11 x)+9\right)-2 c_\vartheta H O \overline{x} x \left(\hat{x} c_\vartheta^2-2 x+2\right)\right) \mn^4
\right.\\
	&\left.
	-\ms^4 \left(\ms^2 \left((x (13 x+18)+9) c_\vartheta^2+x (10-3 x)+9\right)-c_\vartheta \left(c_\vartheta^2-1\right) H O \overline{x} x\right) \mn^2-2 c_\vartheta H \ms^6 O x+\ms^8 \left(3 \hat{x} c_\vartheta^2+7 x+3\right)\right) \mj^4\\
	&-\mn^4 \left(8 \overline{x}^2 x \hat{x} \left(\tilde{x} c_\vartheta^2+x (x+2)-1\right) \mn^8+\left(2 x \hat{x} \left(c_\vartheta \left(c_\vartheta^2+3\right) H O \overline{x}^3+8 x \Gamma_N^2\right)-\ms^2 \overline{x} \left(\hat{x} (x (x (27 x+23)+17)-3) c_\vartheta^2
\right.\right.\right.\\
	&\left.\left.\left.	
	+\overline{x}^2 (x (7 x+12)-3)\right)\right) \mn^6+\ms^2 \left(\left((33 x^4+52 x^3+30 x^2+4 x+9) c_\vartheta^2-4 x+x^2 (6-x (7 x+4))+9\right) \ms^2-4 x^2 (x+5) \Gamma_N^2
\right.\right.\\
&\left.\left.
	-4 c_\vartheta H O x \tilde{x} \left(\hat{x} c_\vartheta^2-x+2\right)\right)	
	 \mn^4-\ms^4 \left(\left(\hat{x} (x (17 x+6)+9) c_\vartheta^2+x (x (9 x+7)+23)+9\right) \ms^2-4 x^2 \Gamma_N^2
\right.\right.\\
&\left.\left.
	 -2 c_\vartheta H O x \left(\tilde{x} c_\vartheta^2+3 x^2+1\right)\right) \mn^2-4 c_\vartheta H \ms^6 O x (2 x+1)+\ms^8 \left((x (3 x+2)+3) c_\vartheta^2+(3 x+1) (5 x+3)\right)\right)\\
	& \mj^2+\mn^6 \left(4 \left(\overline{x}^2 c_\vartheta^2+\overline{x}^2\right) \overline{x}^3 x \hat{x} \mn^8-\overline{x} \left(\left(\overline{x}^2 (x (13 x \hat{x}+7)-1) c_\vartheta^2+\overline{x} (x (x (x (5 x-32)+2)-8)+1)\right) \ms^2\right.\right.\\
	&\left.\left.+x \left(16 x \hat{x} \Gamma_N^2-c_\vartheta H O \overline{x} \left(c_\vartheta^2 \overline{x}^3+x (x (7 x-17)-3)-3\right)\right)\right) \mn^6+\left(\hat{x} \left(\overline{x}^2 (x (15 x-2)+3) c_\vartheta^2+\overline{x}^2 (x (3 x+2)+3)\right) \ms^4
\right.\right.\\
	&\left.\left.
	+2 x \left(-2 x (x (x+2)+5) \Gamma_N^2-c_\vartheta H O \tilde{x} \left(c_\vartheta^2 \overline{x}^2+2 \left(2 x^2+x+1\right)\right)\right) \ms^2\right.\right.\\
	&\left.\left.
	+4 c_\vartheta H O \overline{x} x^3 \Gamma_N^2\right) \mn^4
	+\ms^4 \left(x \left(4 x \hat{x} \Gamma_N^2+c_\vartheta H O \left(c_\vartheta^2 \overline{x}^3+x+x^2 (11 x+19)+1\right)\right)\right.\right.\\
	&\left.\left.-\ms^2 \left(\overline{x}^2 (x (7 x+6)+3) c_\vartheta^2+11 x^4+28 x^3+10 x^2+12 x+3\right)\right) \mn^2+\ms^6 \hat{x} \left(\ms^2 \left(\overline{x}^2 c_\vartheta^2+(3 x+1)^2\right)-2 c_\vartheta H O x (3 x+1)\right)\right) \,.
	\end{split}\nonumber
\end{align}
\end{landscape}  
   
\subsection{Boltzmann Equations}
\label{sec:app_boltzmann}
In order to obtain the correct Boltzmann equations for the evolution of the particle abundances, we need to substract on-shell contributions from $2\leftrightarrow2$ processes as discussed in App. \ref{sec:interactions}. In the majoron+triplet model, the on-shell exchange of $\sigma$ needs to be substracted in the thermal rates for $\gat{NN\latexchi\latexchi}$, $\gat{NNJJ}$ and $\gat{\latexchi\latexchi JJ}$, resulting in 
\begin{align}
	\gat{NN\latexchi\latexchi}^{off} &= \gat{\latexchi\latexchi NN}^{off} =\gat{NN\latexchi\latexchi}-\br{\sigma,\latexchi\latexchi}\gat{\sigma,NN}\,, \\
	\gat{NNJJ}^{off} &= \gat{JJNN}^{off} =\gat{NNJJ}-\br{\sigma, JJ}\gat{\sigma,NN} \,, \\
	\gat{\latexchi\latexchi JJ}^{off} &= \gat{JJ \latexchi\latexchi}^{off} =\gat{\latexchi\latexchi JJ}-\br{\sigma, JJ}\gat{\sigma,\latexchi\latexchi}\,.
\end{align}
With this in mind, the relevant processes for the neutrino evolution are given by 
\begin{align}
	\be{N \leftrightarrow LH } &= \left(\delta_N-1\right)\gamma_D\,, &\quad 
	\be{NL\leftrightarrow Q_3U_3} &= 2\left(\delta_N-1\right)\gat{NLQ_3U_3}\,, \\
	\be{N\overline{U_3}\leftrightarrow L\overline{Q_3}} &= 2\left(\delta_N-1\right)\gat{N\overline{U_3}L\overline{2_3}}\,, &\quad
	\be{N\overline{Q_3}\leftrightarrow L\overline{U_3}} &= 2\left(\delta_N-1\right)\gat{N\overline{Q_3}L\overline{U_3}}\,, \\
	\be{NN \leftrightarrow \sigma } &= 2\left(\delta_N^2-\delta_\sigma\right)\gat{\sigma,NN}\,, &\quad 
	\be{NN\leftrightarrow\latexchi\latexchi} &= 2\left(\delta_N^2-\delta_\latexchi^2\right)\gat{NN\latexchi\latexchi}^{off}\,, \\
	\be{NN\leftrightarrow\sigma\sigma} &= 2\left(\delta_N^2-\delta_\sigma^2\right)\gat{NN\sigma\sigma}\,, &\quad 
	\be{NN\leftrightarrow JJ} &= 2\left(\delta_N^2-\delta_J^2\right)\gat{NNJJ}^{off}\,, \\
	\be{NN\leftrightarrow\sigma J} &= 2\left(\delta_N^2-\delta_\sigma\delta_J\right)\gat{NN\sigma J} \,,
\end{align}
where the scattering processes in the first two lines also appear in VL and we assumed that $H$, $L$, $Q_3$ and $U_3$ are in thermal equilibrium, $Y_{H, L, Q_3, U_3} = Y_{H, L, Q_3, U_3}^{eq}$.
Next, the relevant scattering processes for the evolution of $\latexchi$ are given by 
\begin{align}
	\be{\latexchi\latexchi \leftrightarrow \sigma } &= 2\left(\delta_\latexchi^2-\delta_\sigma\right)\gat{\sigma,\latexchi\latexchi}\,, &\quad 
	\be{\latexchi\latexchi\leftrightarrow NN} &= 2\left(\delta_\latexchi^2-\delta_\latexchi^2\right)\gat{\latexchi\latexchi NN}^{off}\,, \\
	\be{\latexchi\latexchi\leftrightarrow\sigma\sigma} &= 2\left(\delta_\latexchi^2-\delta_\sigma^2\right)\gat{\latexchi\latexchi\sigma\sigma}\,, &\quad 
	\be{\latexchi\latexchi\leftrightarrow JJ} &= 2\left(\delta_\latexchi^2-\delta_J^2\right)\gat{\latexchi\latexchi JJ}^{off}\,, \\
	\be{\latexchi\latexchi\leftrightarrow\sigma J} &= 2\left(\delta_\latexchi^2-\delta_\sigma\delta_J\right)\gat{\latexchi\latexchi\sigma J}\,, &\quad 
	\be{\latexchi\latexchi\leftrightarrow AA} &= 2\left(\delta_\latexchi^2-\delta_AA^2\right)\gat{\latexchi\latexchi AA} 
\end{align}
while for $J$, we have
\begin{align}
	\be{JJ \leftrightarrow \sigma\sigma} &= 2\left(\delta_J^2-\delta_\sigma^2\right) \gat{JJ\sigma\sigma}\,, &\quad
	\be{JJ \leftrightarrow \latexchi\latexchi} &= 2\left(\delta_J^2 -\delta_\sigma\right)\gat{\sigma,JJ}\,, \\
	\be{JJ \leftrightarrow NN} &= 2\left(\delta_J^2-\delta_N^2\right) \gat{JJNN}^{off}\,, &\quad
	\be{JJ \leftrightarrow \latexchi\latexchi} &= 2\left(\delta_J^2-\delta_\latexchi^2\right) \gat{JJ\latexchi\latexchi}^{off}\,, \\
	\be{\sigma J \leftrightarrow NN} &= \left(\delta_J\delta_\sigma-\delta_N^2\right) \gat{\sigma JNN}\,, &\quad
	\be{\sigma J \leftrightarrow \latexchi\latexchi} &= \left(\delta_J\delta_\sigma-\delta_\latexchi^2\right) \gat{\sigma J\latexchi\latexchi}\,, \\
	\be{JN \leftrightarrow \sigma N} &= \left(\delta_J\delta_N-\delta_\sigma\delta_N\right)\gat{JN\sigma N}\,, &\quad
	\be{J\latexchi \leftrightarrow \sigma \latexchi} &= \left(\delta_J\delta_\latexchi-\delta_\sigma\delta_\latexchi\right)\gat{J\latexchi\sigma \latexchi} 
\end{align}
Finally, the relevant scattering processes for the evolution of $\sigma$ are
\begin{align}
	\be{\sigma \leftrightarrow JJ} &= \left(\delta_\sigma -\delta_J^2\right)\gat{\sigma,JJ}\,, &\quad
	\be{\sigma \leftrightarrow NN} &= \left(\delta_\sigma -\delta_N^2\right)\gat{\sigma,NN}\,, \\
	\be{\sigma \leftrightarrow \latexchi\latexchi} &= \left(\delta_\sigma -\delta_\latexchi^2\right)\gat{\sigma,\latexchi\latexchi}\,, &\quad
	\be{\sigma\sigma \leftrightarrow JJ} &= 2\left(\delta_\sigma^2 -\delta_J^2\right)\gat{\sigma\sigma JJ}\,, \\
	\be{\sigma\sigma \leftrightarrow NN} &= 2\left(\delta_\sigma^2 -\delta_N^2\right)\gat{\sigma\sigma NN}\,, &\quad
	\be{\sigma\sigma \leftrightarrow \latexchi\latexchi} &= 2\left(\delta_\sigma^2 -\delta_\latexchi^2\right)\gat{\sigma\sigma \latexchi\latexchi}\,, \\
	\be{\sigma J \leftrightarrow NN} &= \left(\delta_\sigma\delta_J -\delta_N^2\right)\gat{\sigma J NN}\,, &\quad
	\be{\sigma J \leftrightarrow \latexchi\latexchi} &= \left(\delta_\sigma\delta_J -\delta_\latexchi^2\right)\gat{\sigma J \latexchi\latexchi}\,, \\
	\be{\sigma N \leftrightarrow JN }&= \left(\delta_\sigma\delta_N-\delta_J\delta_N\right) \gat{\sigma N JN}\,, &\quad
	\be{\sigma \latexchi \leftrightarrow J\latexchi }&= \left(\delta_\sigma\delta_\latexchi-\delta_J\delta_\latexchi\right) \gat{\sigma \latexchi J\latexchi}\,.
\end{align}
Summing the respective processes according to \eqref{eq:gen:boltzmann} and using $\br{\sigma, NN}+\br{\sigma, \latexchi\latexchi}+\br{\sigma, JJ} = 1$, we obtain \eqref{eq:ben}, \eqref{be:chi}, \eqref{be:s}, \eqref{be:j} and \eqref{eq:dyn_ad}.


\begin{thebibliography}{44}


\providecommand{\url}[1]{\texttt{#1}}
\expandafter\ifx\csname urlstyle\endcsname\relax
  \providecommand{\doi}[1]{doi: #1}\else
  \providecommand{\doi}{doi: \begingroup \urlstyle{rm}\Url}\fi

\bibitem[1]{Planck:2018vyg}
\textsc{Aghanim}, N. u.\,a.:
\newblock {Planck 2018 results. VI. Cosmological parameters}.
\newblock {In: }\emph{Astron. Astrophys.} 641 (2020), S.~A6.
\newblock \url{http://dx.doi.org/10.1051/0004-6361/201833910}. --
\newblock DOI 10.1051/0004--6361/201833910. --
\newblock [Erratum: Astron.Astrophys. 652, C4 (2021)]

\bibitem[2]{KamLAND:2002uet}
\textsc{Eguchi}, K. u.\,a.:
\newblock {First results from KamLAND: Evidence for reactor anti-neutrino
  disappearance}.
\newblock {In: }\emph{Phys. Rev. Lett.} 90 (2003), S. 021802.
\newblock \url{http://dx.doi.org/10.1103/PhysRevLett.90.021802}. --
\newblock DOI 10.1103/PhysRevLett.90.021802

\bibitem[3]{SNO:2002tuh}
\textsc{Ahmad}, Q.~R. u.\,a.:
\newblock {Direct evidence for neutrino flavor transformation from neutral
  current interactions in the Sudbury Neutrino Observatory}.
\newblock {In: }\emph{Phys. Rev. Lett.} 89 (2002), S. 011301.
\newblock \url{http://dx.doi.org/10.1103/PhysRevLett.89.011301}. --
\newblock DOI 10.1103/PhysRevLett.89.011301

\bibitem[4]{Super-Kamiokande:1998kpq}
\textsc{Fukuda}, Y. u.\,a.:
\newblock {Evidence for oscillation of atmospheric neutrinos}.
\newblock {In: }\emph{Phys. Rev. Lett.} 81 (1998), S. 1562--1567.
\newblock \url{http://dx.doi.org/10.1103/PhysRevLett.81.1562}. --
\newblock DOI 10.1103/PhysRevLett.81.1562

\bibitem[5]{Minkowski:1977sc}
\textsc{Minkowski}, Peter:
\newblock {$\mu \to e\gamma$ at a Rate of One Out of $10^{9}$ Muon Decays?}
\newblock {In: }\emph{Phys. Lett. B} 67 (1977), S. 421--428.
\newblock \url{http://dx.doi.org/10.1016/0370-2693(77)90435-X}. --
\newblock DOI 10.1016/0370--2693(77)90435--X

\bibitem[6]{Fukugita:1986hr}
\textsc{Fukugita}, M. ; \textsc{Yanagida}, T.:
\newblock {Baryogenesis Without Grand Unification}.
\newblock {In: }\emph{Phys. Lett. B} 174 (1986), S. 45--47.
\newblock \url{http://dx.doi.org/10.1016/0370-2693(86)91126-3}. --
\newblock DOI 10.1016/0370--2693(86)91126--3

\bibitem[7]{Khlebnikov:1988sr}
\textsc{Khlebnikov}, S.~Y. ; \textsc{Shaposhnikov}, M.~E.:
\newblock {The Statistical Theory of Anomalous Fermion Number Nonconservation}.
\newblock {In: }\emph{Nucl. Phys. B} 308 (1988), S. 885--912.
\newblock \url{http://dx.doi.org/10.1016/0550-3213(88)90133-2}. --
\newblock DOI 10.1016/0550--3213(88)90133--2

\bibitem[8]{Chikashige:1980ui}
\textsc{Chikashige}, Y. ; \textsc{Mohapatra}, Rabindra~N.  ; \textsc{Peccei},
  R.~D.:
\newblock {Are There Real Goldstone Bosons Associated with Broken Lepton
  Number?}
\newblock {In: }\emph{Phys. Lett. B} 98 (1981), S. 265--268.
\newblock \url{http://dx.doi.org/10.1016/0370-2693(81)90011-3}. --
\newblock DOI 10.1016/0370--2693(81)90011--3

\bibitem[9]{Schechter:1981cv}
\textsc{Schechter}, J. ; \textsc{Valle}, J. W.~F.:
\newblock {Neutrino Decay and Spontaneous Violation of Lepton Number}.
\newblock {In: }\emph{Phys. Rev. D} 25 (1982), S. 774.
\newblock \url{http://dx.doi.org/10.1103/PhysRevD.25.774}. --
\newblock DOI 10.1103/PhysRevD.25.774

\bibitem[10]{Georgi:1981pg}
\textsc{Georgi}, Howard~M. ; \textsc{Glashow}, Sheldon~L.  ; \textsc{Nussinov},
  Shmuel:
\newblock {Unconventional Model of Neutrino Masses}.
\newblock {In: }\emph{Nucl. Phys. B} 193 (1981), S. 297--316.
\newblock \url{http://dx.doi.org/10.1016/0550-3213(81)90336-9}. --
\newblock DOI 10.1016/0550--3213(81)90336--9

\bibitem[11]{Brune:2018sab}
\textsc{Brune}, Tim ; \textsc{P\"as}, Heinrich:
\newblock {Massive majorons and constraints on the majoron-neutrino coupling}.
\newblock {In: }\emph{Phys. Rev. D} 99 (2019), Nr. 9, S. 096005.
\newblock \url{http://dx.doi.org/10.1103/PhysRevD.99.096005}. --
\newblock DOI 10.1103/PhysRevD.99.096005

\bibitem[12]{Frigerio:2011in}
\textsc{Frigerio}, Michele ; \textsc{Hambye}, Thomas  ; \textsc{Masso}, Eduard:
\newblock {Sub-GeV dark matter as pseudo-Goldstone from the seesaw scale}.
\newblock {In: }\emph{Phys. Rev. X} 1 (2011), S. 021026.
\newblock \url{http://dx.doi.org/10.1103/PhysRevX.1.021026}. --
\newblock DOI 10.1103/PhysRevX.1.021026

\bibitem[13]{Hall:2009bx}
\textsc{Hall}, Lawrence~J. ; \textsc{Jedamzik}, Karsten ;
  \textsc{March-Russell}, John  ; \textsc{West}, Stephen~M.:
\newblock {Freeze-In Production of FIMP Dark Matter}.
\newblock {In: }\emph{JHEP} 03 (2010), S. 080.
\newblock \url{http://dx.doi.org/10.1007/JHEP03(2010)080}. --
\newblock DOI 10.1007/JHEP03(2010)080

\bibitem[14]{Rothstein:1992rh}
\textsc{Rothstein}, I.~Z. ; \textsc{Babu}, K.~S.  ; \textsc{Seckel}, D.:
\newblock {Planck scale symmetry breaking and majoron physics}.
\newblock {In: }\emph{Nucl. Phys. B} 403 (1993), S. 725--748.
\newblock \url{http://dx.doi.org/10.1016/0550-3213(93)90368-Y}. --
\newblock DOI 10.1016/0550--3213(93)90368--Y

\bibitem[15]{Berezinsky:1993fm}
\textsc{Berezinsky}, V. ; \textsc{Valle}, J. W.~F.:
\newblock {The KeV majoron as a dark matter particle}.
\newblock {In: }\emph{Phys. Lett. B} 318 (1993), S. 360--366.
\newblock \url{http://dx.doi.org/10.1016/0370-2693(93)90140-D}. --
\newblock DOI 10.1016/0370--2693(93)90140--D

\bibitem[16]{Pilaftsis:2008qt}
\textsc{Pilaftsis}, Apostolos:
\newblock {Electroweak Resonant leptogenesis in the Singlet majoron Model}.
\newblock {In: }\emph{Phys. Rev. D} 78 (2008), S. 013008.
\newblock \url{http://dx.doi.org/10.1103/PhysRevD.78.013008}. --
\newblock DOI 10.1103/PhysRevD.78.013008

\bibitem[17]{Gu:2009hn}
\textsc{Gu}, Pei-Hong ; \textsc{Sarkar}, Utpal:
\newblock {Leptogenesis Bound on Spontaneous Symmetry Breaking of Global Lepton
  Number}.
\newblock {In: }\emph{Eur. Phys. J. C} 71 (2011), S. 1560.
\newblock \url{http://dx.doi.org/10.1140/epjc/s10052-011-1560-2}. --
\newblock DOI 10.1140/epjc/s10052--011--1560--2

\bibitem[18]{AristizabalSierra:2014uzi}
\textsc{Aristizabal~Sierra}, Diego ; \textsc{Tortola}, M. ; \textsc{Valle}, J.
  W.~F.  ; \textsc{Vicente}, A.:
\newblock {Leptogenesis with a dynamical seesaw scale}.
\newblock {In: }\emph{JCAP} 07 (2014), S. 052.
\newblock \url{http://dx.doi.org/10.1088/1475-7516/2014/07/052}. --
\newblock DOI 10.1088/1475--7516/2014/07/052

\bibitem[19]{Lazarides:2018aev}
\textsc{Lazarides}, George ; \textsc{Reig}, Mario ; \textsc{Shafi}, Qaisar ;
  \textsc{Srivastava}, Rahul  ; \textsc{Valle}, Jos\'e W.~F.:
\newblock {Spontaneous Breaking of Lepton Number and the Cosmological Domain
  Wall Problem}.
\newblock {In: }\emph{Phys. Rev. Lett.} 122 (2019), Nr. 15, S. 151301.
\newblock \url{http://dx.doi.org/10.1103/PhysRevLett.122.151301}. --
\newblock DOI 10.1103/PhysRevLett.122.151301

\bibitem[20]{Heeck:2019guh}
\textsc{Heeck}, Julian ; \textsc{Patel}, Hiren~H.:
\newblock {Majoron at two loops}.
\newblock {In: }\emph{Phys. Rev. D} 100 (2019), Nr. 9, S. 095015.
\newblock \url{http://dx.doi.org/10.1103/PhysRevD.100.095015}. --
\newblock DOI 10.1103/PhysRevD.100.095015

\bibitem[21]{FileviezPerez:2014xju}
\textsc{Fileviez~Perez}, Pavel ; \textsc{Patel}, Hiren~H.:
\newblock {The Electroweak Vacuum Angle}.
\newblock {In: }\emph{Phys. Lett. B} 732 (2014), S. 241--243.
\newblock \url{http://dx.doi.org/10.1016/j.physletb.2014.03.064}. --
\newblock DOI 10.1016/j.physletb.2014.03.064

\bibitem[22]{Anselm:1992yz}
\textsc{Anselm}, A.~A. ; \textsc{Johansen}, A.~A.:
\newblock {Baryon nonconservation in standard model and Yukawa interaction}.
\newblock {In: }\emph{Nucl. Phys. B} 407 (1993), S. 313--330.
\newblock \url{http://dx.doi.org/10.1016/0550-3213(93)90060-3}. --
\newblock DOI 10.1016/0550--3213(93)90060--3

\bibitem[23]{Anselm:1993uj}
\textsc{Anselm}, A.~A. ; \textsc{Johansen}, A.~A.:
\newblock {Can electroweak theta term be observable?}
\newblock {In: }\emph{Nucl. Phys. B} 412 (1994), S. 553--573.
\newblock \url{http://dx.doi.org/10.1016/0550-3213(94)90392-1}. --
\newblock DOI 10.1016/0550--3213(94)90392--1

\bibitem[24]{Csaki:2023ziz}
\textsc{Cs\'aki}, Csaba ; \textsc{D'Agnolo}, Raffaele~T. ; \textsc{Kuflik},
  Eric  ; \textsc{Ruhdorfer}, Maximilian:
\newblock {Instanton NDA and applications to axion models}.
\newblock {In: }\emph{JHEP} 04 (2024), S. 074.
\newblock \url{http://dx.doi.org/10.1007/JHEP04(2024)074}. --
\newblock DOI 10.1007/JHEP04(2024)074

\bibitem[25]{Brune:2022vzd}
\textsc{Brune}, Tim:
\newblock {Leptogenesis in majoron models without domain walls}.
\newblock {In: }\emph{Phys. Rev. D} 107 (2023), Nr. 9, S. 096023.
\newblock \url{http://dx.doi.org/10.1103/PhysRevD.107.096023}. --
\newblock DOI 10.1103/PhysRevD.107.096023

\bibitem[26]{Plumacher:1996kc}
\textsc{Plümacher}, Michael:
\newblock {Baryogenesis and lepton number violation}.
\newblock {In: }\emph{Z. Phys. C} 74 (1997), S. 549--559.
\newblock \url{http://dx.doi.org/10.1007/s002880050418}. --
\newblock DOI 10.1007/s002880050418

\bibitem[27]{Kolb:1979qa}
\textsc{Kolb}, Edward~W. ; \textsc{Wolfram}, Stephen:
\newblock {Baryon Number Generation in the Early Universe}.
\newblock {In: }\emph{Nucl. Phys. B} 172 (1980), S. 224.
\newblock \url{http://dx.doi.org/10.1016/0550-3213(82)90012-8}. --
\newblock DOI 10.1016/0550--3213(82)90012--8. --
\newblock [Erratum: Nucl.Phys.B 195, 542 (1982)]

\bibitem[28]{Covi:1996wh}
\textsc{Covi}, Laura ; \textsc{Roulet}, Esteban  ; \textsc{Vissani}, Francesco:
\newblock {CP violating decays in leptogenesis scenarios}.
\newblock {In: }\emph{Phys. Lett. B} 384 (1996), S. 169--174.
\newblock \url{http://dx.doi.org/10.1016/0370-2693(96)00817-9}. --
\newblock DOI 10.1016/0370--2693(96)00817--9

\bibitem[29]{Davidson:2002qv}
\textsc{Davidson}, Sacha ; \textsc{Ibarra}, Alejandro:
\newblock {A Lower bound on the right-handed neutrino mass from leptogenesis}.
\newblock {In: }\emph{Phys. Lett. B} 535 (2002), S. 25--32.
\newblock \url{http://dx.doi.org/10.1016/S0370-2693(02)01735-5}. --
\newblock DOI 10.1016/S0370--2693(02)01735--5

\bibitem[30]{Salas}
\textsc{Salas}, P. ; \textsc{Forero}, Danna ; \textsc{Gariazzo}, S. ;
  \textsc{Martínez-Miravé}, P. ; \textsc{Mena}, O. ; \textsc{Ternes}, C. ;
  \textsc{Tórtola}, Mariam  ; \textsc{Valle}, Jose:
\newblock 2020 global reassessment of the neutrino oscillation picture.
\newblock {In: }\emph{Journal of High Energy Physics} 2021 (2021), 02.
\newblock \url{http://dx.doi.org/10.1007/JHEP02(2021)071}. --
\newblock DOI 10.1007/JHEP02(2021)071

\bibitem[31]{Barbieri:1999ma}
\textsc{Barbieri}, Riccardo ; \textsc{Creminelli}, Paolo ; \textsc{Strumia},
  Alessandro  ; \textsc{Tetradis}, Nikolaos:
\newblock {Baryogenesis through leptogenesis}.
\newblock {In: }\emph{Nucl. Phys. B} 575 (2000), S. 61--77.
\newblock \url{http://dx.doi.org/10.1016/S0550-3213(00)00011-0}. --
\newblock DOI 10.1016/S0550--3213(00)00011--0

\bibitem[32]{AristizabalSierra:2010mv}
\textsc{Aristizabal~Sierra}, D. ; \textsc{Kamenik}, Jernej~F.  ;
  \textsc{Nemevsek}, Miha:
\newblock {Implications of Flavor Dynamics for Fermion Triplet leptogenesis}.
\newblock {In: }\emph{JHEP} 10 (2010), S. 036.
\newblock \url{http://dx.doi.org/10.1007/JHEP10(2010)036}. --
\newblock DOI 10.1007/JHEP10(2010)036

\bibitem[33]{Hambye:2012fh}
\textsc{Hambye}, Thomas:
\newblock {Leptogenesis: beyond the minimal type I seesaw scenario}.
\newblock {In: }\emph{New J. Phys.} 14 (2012), S. 125014.
\newblock \url{http://dx.doi.org/10.1088/1367-2630/14/12/125014}. --
\newblock DOI 10.1088/1367--2630/14/12/125014

\bibitem[34]{Hambye:2003rt}
\textsc{Hambye}, Thomas ; \textsc{Lin}, Yin ; \textsc{Notari}, Alessio ;
  \textsc{Papucci}, Michele  ; \textsc{Strumia}, Alessandro:
\newblock {Constraints on neutrino masses from leptogenesis models}.
\newblock {In: }\emph{Nucl. Phys. B} 695 (2004), S. 169--191.
\newblock \url{http://dx.doi.org/10.1016/j.nuclphysb.2004.06.027}. --
\newblock DOI 10.1016/j.nuclphysb.2004.06.027

\bibitem[35]{Strumia:2008cf}
\textsc{Strumia}, Alessandro:
\newblock {Sommerfeld corrections to type-II and III leptogenesis}.
\newblock {In: }\emph{Nucl. Phys. B} 809 (2009), S. 308--317.
\newblock \url{http://dx.doi.org/10.1016/j.nuclphysb.2008.10.007}. --
\newblock DOI 10.1016/j.nuclphysb.2008.10.007

\bibitem[36]{Zhuridov:2012hb}
\textsc{Zhuridov}, Dmitry~V.:
\newblock {Neutrino Masses and leptogenesis from Extra Fermions}.
\newblock {In: }\emph{Int. J. Mod. Phys. A} 28 (2013), S. 1350104.
\newblock \url{http://dx.doi.org/10.1142/S0217751X13501042}. --
\newblock DOI 10.1142/S0217751X13501042

\bibitem[37]{AristizabalSierra:2012js}
\textsc{Aristizabal~Sierra}, D. ; \textsc{Medeiros~Varzielas}, I. de:
\newblock {The role of lepton flavor symmetries in leptogenesis}.
\newblock {In: }\emph{Fortsch. Phys.} 61 (2013), S. 645--665.
\newblock \url{http://dx.doi.org/10.1002/prop.201200122}. --
\newblock DOI 10.1002/prop.201200122

\bibitem[38]{Buchmuller:2003gz}
\textsc{Buchmüller}, W. ; \textsc{Di~Bari}, P.  ; \textsc{Plümacher}, M.:
\newblock {The Neutrino mass window for baryogenesis}.
\newblock {In: }\emph{Nucl. Phys. B} 665 (2003), S. 445--468.
\newblock \url{http://dx.doi.org/10.1016/S0550-3213(03)00449-8}. --
\newblock DOI 10.1016/S0550--3213(03)00449--8

\bibitem[39]{Buchmuller:2004nz}
\textsc{Buchmüller}, W. ; \textsc{Di~Bari}, P.  ; \textsc{Plümacher}, M.:
\newblock {Leptogenesis for pedestrians}.
\newblock {In: }\emph{Annals Phys.} 315 (2005), S. 305--351.
\newblock \url{http://dx.doi.org/10.1016/j.aop.2004.02.003}. --
\newblock DOI 10.1016/j.aop.2004.02.003

\bibitem[40]{ATLAS:2022yhd}
\textsc{Aad}, Georges u.\,a.:
\newblock {Search for type-III seesaw heavy leptons in leptonic final states in
  pp collisions at $\sqrt{s} = 13~\text {TeV}$ with the ATLAS detector}.
\newblock {In: }\emph{Eur. Phys. J. C} 82 (2022), Nr. 11, S. 988.
\newblock \url{http://dx.doi.org/10.1140/epjc/s10052-022-10785-0}. --
\newblock DOI 10.1140/epjc/s10052--022--10785--0

\bibitem[41]{CMS:2017ybg}
\textsc{Sirunyan}, Albert~M. u.\,a.:
\newblock {Search for Evidence of the Type-III Seesaw Mechanism in Multilepton
  Final States in Proton-Proton Collisions at $\sqrt{s}=13\text{ }\text{
  }\mathrm{TeV}$}.
\newblock {In: }\emph{Phys. Rev. Lett.} 119 (2017), Nr. 22, S. 221802.
\newblock \url{http://dx.doi.org/10.1103/PhysRevLett.119.221802}. --
\newblock DOI 10.1103/PhysRevLett.119.221802

\bibitem[42]{Giudice:2003jh}
\textsc{Giudice}, G.~F. ; \textsc{Notari}, A. ; \textsc{Raidal}, M. ;
  \textsc{Riotto}, A.  ; \textsc{Strumia}, A.:
\newblock {Towards a complete theory of thermal leptogenesis in the SM and
  MSSM}.
\newblock {In: }\emph{Nucl. Phys. B} 685 (2004), S. 89--149.
\newblock \url{http://dx.doi.org/10.1016/j.nuclphysb.2004.02.019}. --
\newblock DOI 10.1016/j.nuclphysb.2004.02.019

\bibitem[43]{Cirelli:2007xd}
\textsc{Cirelli}, Marco ; \textsc{Strumia}, Alessandro  ; \textsc{Tamburini},
  Matteo:
\newblock {Cosmology and Astrophysics of Minimal Dark Matter}.
\newblock {In: }\emph{Nucl. Phys. B} 787 (2007), S. 152--175.
\newblock \url{http://dx.doi.org/10.1016/j.nuclphysb.2007.07.023}. --
\newblock DOI 10.1016/j.nuclphysb.2007.07.023

\end{thebibliography}

\end{document}